\documentclass{article}

\usepackage{PRIMEarxiv}
\usepackage{natbib}
\usepackage[flushleft]{threeparttable}
\usepackage{algpseudocode}
\usepackage[utf8]{inputenc} 
\usepackage[T1]{fontenc}    
\usepackage{hyperref}       
\usepackage{url}            
\usepackage{booktabs}       
\usepackage{amsfonts}       
\usepackage{nicefrac}       
\usepackage{microtype}      
\usepackage{lipsum}
\usepackage{fancyhdr}       
\usepackage{graphicx}       
\graphicspath{{media/}}     

\graphicspath{{Fig/}}

\usepackage{amsmath}
\usepackage{graphicx}
\usepackage{comment}
\usepackage{enumerate}
\usepackage{siunitx}
\usepackage{booktabs}
\usepackage{mathtools}
\usepackage{amssymb}
\usepackage{comment}
\newcommand{\RNum}[1]{\uppercase\expandafter{\romannumeral #1\relax}}
\usepackage{float}
\newcommand{\indep}{\perp \!\!\! \perp}
\usepackage{algpseudocode}
\usepackage{algorithm}
\usepackage{lipsum}
\usepackage{stix}
\DeclareMathAlphabet\mathbfcal{LS2}{stixcal}{b}{n}
\usepackage{amsthm}

\usepackage{enumitem}
\setlist{nolistsep}

\usepackage[figuresright]{rotating}

\newtheorem{assumption}{Assumption}
\newtheorem{manualtheoreminner}{Proposition}
\newenvironment{manualtheorem}[1]{%
  \renewcommand\themanualtheoreminner{#1}%
  \manualtheoreminner
}{\endmanualtheoreminner}

\title{A Bayesian semi-parametric approach to causal mediation for longitudinal mediators and time-to-event outcomes with application to a 
cardiovascular disease cohort study
}

\author{
  Saurabh Bhandari \\
  Department of Statistics \\
  University of Florida \\
  s.bhandari@ufl.edu \\
   \And
  Michael J. Daniels \\
  Department of Statistics \\
  University of Florida \\
  daniels@ufl.edu \\
   \AND
  Maria Josefsson \\
  Department of Statistics, USBE \\
  Umeå University \\
  maria.josefsson@umu.se \\
   \AND
   Donald M. Lloyd-Jones \\
  Department of Preventive Medicine \\
  Northwestern University Feinberg School of Medicine \\
  dlj@northwestern.edu \\
   \AND
   Juned Siddique \\
  Department of Preventive Medicine \\
  Northwestern University Feinberg School of Medicine \\
  siddique@northwestern.edu 
}

\begin{document}
\maketitle

\begin{abstract}
Causal mediation analysis of observational data is an important tool for investigating the potential causal effects of medications on disease-related risk factors, and on time-to-death (or disease progression) through these risk factors. However, when analyzing data from a cohort study, such analyses are complicated by the longitudinal structure of the risk factors and the presence of time-varying confounders. Leveraging data from the Atherosclerosis Risk in Communities (ARIC) cohort study, we develop a causal mediation approach, using (semi-parametric) Bayesian Additive Regression Tree (BART) models for the longitudinal and survival data. Our framework allows for time-varying exposures, confounders, and mediators, all of which can either be continuous or binary. We also identify and estimate direct and indirect causal effects in the presence of a competing event. We apply our methods to assess how medication, prescribed to target cardiovascular disease (CVD) risk factors, affects the time-to-CVD death.
\end{abstract}


\section{Introduction}
\label{s:intro}
Medications targeting cardiovascular risk factors are essential tools for delaying and preventing the development of cardiovascular disease (CVD). A comprehensive study of CVD development requires identifying and analyzing longitudinal trajectories for CVD risk factors and other relevant covariates that may confound the relationship between risk factors and CVD outcomes. A key challenge in this setting is that some of these covariates are not measured in observational studies. Individuals with elevated CVD risk factors are typically prescribed medications that work to mitigate the harmful consequences of these risk factors. A natural question that arises in this setting is how these medications, designed to target the risk factors, go on to affect the time to CVD death. To investigate this question of interest, we develop a general causal mediation framework by modeling longitudinal and survival data. Our objective is to estimate the causal effects of two contrasting exposure regimes (on medication versus not on medication) on the observed time-to-death from CVD, mediated by changes in the trajectory of CVD risk factors in discrete time.

The existing causal mediation literature highlights two significant challenges when analyzing survival outcomes in the presence of longitudinal exposures, confounders, and mediators \citep{zeng2022causal, lin2017mediation, vansteelandt2019mediation}. First, when there are longitudinal confounders that are affected by prior exposure, mediator, and survival outcomes (also known as "recanting witnesses"), the natural direct and indirect effects defined using static interventions are not identified from the data \citep{avin2005identifiability}. Second, in the survival setting, if individuals die before the end of the study, their so-called cross-world counterfactual mediator values (potential mediator values under the counterfactual setting of exposure) may be poorly defined. 

To deal with the first complication, \cite{vanderweele2017mediation} proposed interventional effects to decompose the total effect into Interventional Direct Effect (IDE) and Interventional Indirect Effect (IIE). These "interventional effects" are considered random interventional analogues of natural direct and indirect effects. To define these effects, we fix the mediator for each individual not to the level it would have been for that individual under a specific exposure, but instead to a level that is randomly chosen from the distribution of the mediator among all of those with a particular exposure and other covariates. \cite{lin2017mediation} use these estimands in their work by setting the mediator for each exposed individual to a value randomly chosen from the distribution of what would be observed if everyone had been unexposed. 

Similarly, there are numerous techniques proposed in the literature to address the second complication of undefined counterfactual mediators. \cite{zeng2022causal} circumvent this issue by making the counterfactual mediator values exist in the entire span of time in a  framework that only includes longitudinal mediator and does not include longitudinal exposure. Additionally, their setup requires the assumption that these mediator values remain well-defined even after the subject's death.
\cite{lin2017mediation} address the complication by modifying the mediator model, making the mediator at time $t$ dependent on survival at age $t-1$. They also define nested counterfactual survival status to allow for hypothetical interventions on the survival of an individual. However, in doing so, they only consider random draws from the mediator distribution conditional on baseline---and not on longitudinal---confounders. Furthermore, \cite{lin2017mediation} only focus on the survival probability at the end of the study. An extension to this framework, albeit in a slightly different setting of causal pathways without time-varying exposures, is given in \cite{vansteelandt2019mediation}, who consider random draws from the mediator distribution conditional on longitudinal confounders. Their potential outcome framework is related to the work of \cite{zheng2017longitudinal}, whose mediation analysis framework addresses these complications in the presence of longitudinal exposures, confounders, and mediators, with applications to survival outcomes.

In this article, we approach these challenges following the arguments of \cite{zheng2017longitudinal}. Our causal mediation analysis technique is constructed upon a flexible and comprehensive modeling framework. It is flexible in the sense that we model the observed data using a Bayesian semi-parametric approach (using Bayesian additive regression trees (BART) models). It is comprehensive in the sense that it allows for time-varying exposures, mediators, and confounders, with the latter two being either continuous or binary. We also extend our work to estimate causal estimands in the presence of a competing event. Additionally, motivated by our longitudinal cohort setting where the history of risk factor levels prior to enrollment is unmeasured, we propose an approach that uses external data to address the presence of a known unmeasured confounder affecting the relationship between the mediator and the outcome.

 The advance beyond  \cite{zeng2022causal} and \cite{vansteelandt2019mediation} in our work is the presence of time-varying exposures. Similarly, the advance beyond \cite{lin2017mediation} is that we consider random draws from the mediator distribution conditional on longitudinal confounders. Another contribution of our work is to establish a general causal mediation framework by modeling longitudinal and survival data using BART. In doing so, we extend previous work in the literature, which predominantly focuses on analyzing survival probabilities through Structural Equation Models (SEMs)  \citep{zheng2017longitudinal,lin2017mediation}. These SEM frameworks often rely on parametric modeling assumptions, lack the flexibility to handle categorical or survival data, or assume linear relationships among variables. In this article, we use BART models as introduced by \cite{chipman2010bart} to model longitudinal confounders and mediators. Similarly, to model the time-to-event outcome, we use ideas from non-parametric discrete-time survival BART models described in \cite{sparapani2016nonparametric, sparapani2020nonparametric} with relevant modifications. 
 
\cite{josefsson2021bayesian} have previously developed a flexible G-computation approach to evaluate the causal effects using BART models for the observed data. However, their framework does not specifically focus on mediation analysis, and their final outcome of interest is a continuous variable rather than a survival outcome. An application of BART in survival causal inference without mediators was recently introduced in \cite{chen2024bayesian}, while its use with mediators for continuous and/or binary cross-sectional data was presented in \cite{linero2022mediation}. Nevertheless, a key difference in our work is that we are specifically interested in causal mediation analysis for survival data with time-varying covariates. Some other relevant work in mediation analysis with longitudinal and survival data are \cite{roy2024bayesian}, \cite{kim2019bayesian}, \cite{bind2016causal}, and  \cite{zhou2023causal}.
 
The article is structured as follows: In Section \ref{sec2}, we briefly introduce our motivating example. Section \ref{sec3} defines the potential outcome framework and the (interventional) causal estimands, providing the conditions under which these estimands can be identified from the observed data. In Section \ref{sec4}, we extend the framework to incorporate competing events. Section \ref{sec5} describes our proposed Bayesian semi-parametric modeling approach for longitudinal and survival data. In Section \ref{sec6}, we conduct simulation studies to compare the performance of our proposed model with a standard Bayesian parametric model. We then apply our proposed mediation technique to the Atherosclerosis Risk in Communities (ARIC) study and present our results in Section \ref{sec7}. Section \ref{sec8} outlines our proposed methodology for addressing unmeasured confounders using external data. We conclude our paper with a discussion in Section \ref{sec9}.

\section{Motivating example: the effect of blood pressure medication on time-to-death by CVD in ARIC}\label{sec2}

We develop our methods to examine the causal effect of blood pressure (BP) medication on time-to-death by CVD, taking into account time-varying mediators and confounders in ARIC. ARIC is a cohort study designed to investigate, among other factors, the variations in cardiovascular risk factors and disease by race, gender, location, and age (ARIC Investigators, 1989)\nocite{aric1989atherosclerosis}. A total of $15,792$ individuals aged $45-64$ at entry participated in the study from four geographically diverse communities. These participants were monitored for over 35 years, starting in $1987$, to understand the risks of heart attack and deaths from heart disease. In our analysis, we use a publicly available de-identified limited access version of the dataset from the NHLBI BioLINNC repository.

In this article, we specifically focus on individuals who were hypertensive at baseline in ARIC, defined as having a systolic blood pressure (SBP) greater than or equal to $140$ or a diastolic blood pressure (DBP) greater than or equal to $90$. We utilize data from the first three examinations, scheduled at three-year intervals, resulting in a total of six years of follow-up. The sample size of our study is $2230$, comprising $1038$ males and $1192$ females. The total number of deaths in the three examinations used in our study sample is $425$, with $172$ deaths attributed to CVD. 

Patients with high BP, a significant risk factor for CVD, are typically prescribed medications for lowering their BP. Consequently, we hypothesize that BP medications may influence the time-to-death by CVD and aim to quantify the extent of this impact through the reduction of BP. Specifically, we investigate the causal effect of time-varying hypertension medication on time to CVD deaths, considering non-CVD death as a competing event (as discussed in Section 4). The time-varying risk factor, BP, serves as a mediator, and the time-varying smoking status acts as a confounder of the relationship between treatment, mediator, and outcome. The outcome of interest in our analysis is an individual's survival status at a particular visit. Similarly, we include the following variables as baseline confounders: sex, race, BMI, education, multivariate smoking status (never, former, current), diabetes status, age, and total-to-HDL cholesterol ratio.

\section{Identification and estimation}\label{sec3}
\subsection{Notation and data structure}\label{sec31}
Suppose we have a longitudinal study with $N$ participants. For individual $i$, we record longitudinal risk factor (mediator) measurements $M_{i}(t_{j}) = M_{i}^{t_{j}}$ at visit time $ t_{j}, j \in \{1, \ldots, J \}$, along with time-to-event outcomes $T_{i}$. Throughout this article, we assume that all participants are seen at equidistant time points (e.g., risk factor assessments are performed once every 3 years for an individual).

Consider data in the following form: $T_{i}, \delta_{i}, \textbf{Z}_{i}^{t_{j}},\textbf{L}_{i}^{t_{j}},\textbf{M}_{i}^{t_{j}},\underline{L}^0_i$ with $i = 1,\ldots,N$ indexing participants in the study. In this representation, $T_i$ is the observed event/censoring time and $\delta_i$ is the event indicator that distinguishes event ($\delta_i =1$) from right censoring  ($\delta_i =0$) for individual $i$. We denote $J$ distinct visit times by $0< t_{1} < \ldots < t_{J} < \infty $,  with $t_{0} = 0$. Now, using the survival data pair ($T_{i}, \delta_{i}$), we define visit-specific survival status indicators $Y_{i}^{t_{j}} = \delta_{i}I(t_{j-1} < T_{i} \leq t_{j})$ for individual $i$ at each distinct time $t_{j}$ up to and including the subject's observation time $t_{i}$, $$Y_{i}^{t_{j}} = 
\begin{cases} 
    0  \quad\quad\quad\quad \text{  if } t_{j} < T_{i} \\ 
    \delta_{i} \quad\quad\quad\quad \text{if } t_{j-1} < T_{i} \leq t_{j} \\
    \text{undefined} \quad \text{otherwise}
\end{cases}$$  
for $i = 1,\ldots,N$; $j = 1,\ldots,J$. When defining the survival status indicator $Y_{i}^{t_j}$, we use $T_i$ to denote the observed event time for subject 
$i$ rather than the random variable representing the time-to-event outcome throughout the article. This slight abuse of notation avoids the need to use $t_i$ as the observation of $T_i$, thereby avoiding confusion between the observed event time and the $j^{th}$ visit time $t_j$.
We use the bold letters to denote random variables until visit: $\textbf{M}_{i}^{t_{j}} \equiv \{M_{i}^{t_{1}}, \ldots, M_{i}^{t_{j}}\}$.

In the presence of a competing event, we define an additional variable $d_{i} \in \{1,2\}$ which distinguishes the main event of interest ($d_{i} =1$) from the competing event ($d_{i} =2$). Similarly, we redefine $Y_{i}^{t_{j},(k)}$ as a sequence of multinomial (previously binary) events $Y_{i}^{t_{j},(k)} = I(t_{j-1} < T_{i} \leq t_{j},d_{i} = k)$, $i = 1,\ldots,N$; $j = 1,\ldots,J$; $k = 1,2$. Note that the outcome variables defined in \cite{sparapani2016nonparametric,sparapani2020nonparametric}, which are based on separate time indexes, are different from the ones defined here.

At each visit $t_{j}$, we assume the following order for the longitudinal variables: $Y_{i}^{t_{j}},Z_{i}^{t_{j}},L_{i}^{t_{j}}, M_{i}^{t_{j}}$. In other words, for $j \in \{1,\ldots,J\}$, we have the data structure
$$\underline{L}_{i}^{0},Y_{i}^{t_{1}},Z_{i}^{t_{1}}, L_{i}^{t_{1}}, M_{i}^{t_{1}},  \ldots, Y_{i}^{t_{j-1}}, Z_{i}^{t_{j-1}}, L_{i}^{t_{j-1}}, M_{i}^{t_{j-1}}, Y_{i}^{t_{j}}, \text{ where:}$$
\begin{itemize}
    \item the row vector $\underline{L}_{i}^0 \equiv \begin{bmatrix} V_{i}^{0} \quad W_{i}^{0} \end{bmatrix}$ denotes baseline confounders, with $V_{i}^{0}$ indicating the time-varying confounder measured at baseline and $W_{i}^{0}$ indicating time-invariant covariates measured at baseline, including age at entry (baseline age), sex, race, and education level,
    \item $Z_{i}^{t_{j}} $, $L_{i}^{t_{j}}$, and $M_{i}^{t_{j}}$  denote the time-varying exposure, confounder, and mediator, respectively, which are well-defined as long as individual $i$ is alive at $t_{j}$, and 
    \item $Y^{t_{j}}$ denotes the survival status indicator as defined previously.
\end{itemize}

\subsection{Potential outcome framework}\label{sec32}

Following the arguments from \cite{zheng2017longitudinal} and \cite{wang2023targeted}, we define the potential outcome by joint intervention on exposure and mediator history. As an example, let $Y_{i}^{t_{j}}(z_{i}^{t_{j-1}},m_{i}^{t_{j-1}})$ denote the counterfactual value of $Y_{i}^{t_{j}}$ when $Z_{i}^{t_{j-1}}$ and $M_{i}^{t_{j-1}}$ are set--possibly, contrary to the fact--to $z_{i}^{t_{j-1}}$ and $m_{i}^{t_{j-1}}$, respectively. This joint intervention is established by a static intervention on the exposure variables, followed by a random intervention on the mediator variables. Our objective is to estimate the causal effects of two contrasting exposure regimes, $\textbf{Z}_{i}^{t_{j}}$ and $\textbf{Z}_{i*}^{t_{j}}$, on the observed time-to-event, $T_{i}$, mediated by $\textbf{M}_{i}^{t_{j}}$. The exposure regime, denoted by $\textbf{Z}_{i}^{t_{j}} \equiv \{Z_{i}^{t_{1}}, \ldots, Z_{i}^{t_{j}}\}$, is defined on fixed time intervals (here, every 3 years).

 Suppose that $\textbf{z}^{t_{j}}$ and $\textbf{z}_{*}^{t_{j}}$ are two possible exposure regimes (pair of static treatment and control interventions). We assume that treatment assignment involves no randomness and is not based on a subject's past covariate history, meaning no dynamic treatment strategy is applied (see Chapter 19 of \cite{hernan2024causal} for a detailed discussion of static and dynamic treatment strategies in a longitudinal setting). For example, two exposure strategies of interest are “always treat” and “never treat” during the follow-up period. Now, for $j \in \{1,\ldots,J \}$, let 
\begin{equation}
    \begin{aligned}
        &\mathcal{G}_{z_{*}}^{t_{j}}(m^{t_{j}}|\boldsymbol{\ell}^{t_{j}},\textbf{m}^{t_{j-1}}, \underline{\boldsymbol{\ell}}^{0} ) 
        \\& =  p\Bigl(M^{t_{j}}(\textbf{z}^{t_{j}}_{*}) = m^{t_{j}}|\textbf{Y}^{t_{j}}(\textbf{z}^{t_{j-1}}_{*}) = \mathbb{0}_{j},\textbf{L}^{t_{j}}(\textbf{z}^{t_{j}}_{*}) = \boldsymbol{\ell}^{t_{j}},\textbf{M}^{t_{j-1}}(\textbf{z}^{t_{j-1}}_{*}) = \textbf{m}^{t_{j-1}},\underline{\textbf{L}}^{0} =  \underline{\boldsymbol{\ell}}^{0}\Bigr) 
    \end{aligned} 
\end{equation}
denote the conditional distribution of the control intervention counterfactual mediator given the history of covariates under $\textbf{Z}^{t_{j}} = \textbf{z}_{*}^{t_{j}}$. In this representation, $\mathbb{0}_{j}$ is a zero vector of size $j$. This conditional distribution provides a random draw $M^{t_{j}} \sim \mathcal{G}_{z_{*}}^{t_{j}}(m^{t_{j}}|\boldsymbol{\ell}^{t_{j}},\textbf{m}^{t_{j-1}}, \underline{\boldsymbol{\ell}}^{0} )$ within each stratum $( \boldsymbol{\ell}^{t_{j}},\textbf{m}^{t_{j-1}}, \underline{\boldsymbol{\ell}}^{0})$ at each visit $t_{j}$. Throughout this article, we will denote $\mathcal{G}_{z_{*}}^{t_{j}} = \mathcal{G}_{z_{*}}^{t_{j}}(m^{t_{j}}|\boldsymbol{\ell}^{t_{j}},\textbf{m}^{t_{j-1}},\underline{\boldsymbol{\ell}}^{0} )$ and $\mathbfcal{G}_{z_{*}}^{t_{j}} = \{ \mathcal{G}_{z_{*}}^{t_{1}}, \ldots , \mathcal{G}_{z_{*}}^{t_{j}}\}$ for convenience.

Now, consider a joint intervention to statically set $\textbf{Z}^{t_{j}} = \textbf{z}^{t_{j}}$ in the population and randomly draw $\textbf{M}^{t_{j}}$ from  $\mathbfcal{G}_{z_{*}}^{t_{j}}$ defined above. Resulting from this intervention, the counterfactual quantities $\mathcal{L}_{z}^{t_{j}}$ and $\mathcal{Y}_{z}^{t_{j}}$ are defined as $\mathcal{L}_{z}^{t_{j}} \equiv L^{t_{j}}(\textbf{z}^{t_{j}}, \mathbfcal{G}_{z_{*}}^{t_{j-1}}) \text{ and } \mathcal{Y}_{z}^{t_{j}} \equiv Y^{t_{j}}(\textbf{z}^{t_{j-1}}, \mathbfcal{G}_{z_{*}}^{t_{j-1}}).$
Similarly, we define $T_{i}(\textbf{z}^{t_{j-1}}, \textbf{m}^{t_{j-1}})$ to be the potential time to event under the exposure regime $\textbf{z}^{t_{j-1}}$, and mediator history $\textbf{m}^{t_{j-1}}$. Using this, we define the probability of an event at visit $t_{j}$ conditional on no previous event under the exposure process $\textbf{z}$ and the  mediator process taking the value as if the individual is under the regime $\textbf{z}_{*}$, that is, $\textbf{m}^{t_{j-1}} \sim \mathbfcal{G}_{z_{*}}^{t_{j-1}}$ as:
\begin{equation}
    \mathcal{P}_{\textbf{z},\textbf{z}_{*}}(t_{j}) = P\bigl[T_i(\textbf{z}^{t_{j-1}}, \textbf{m}^{t_{j-1}}) \in (t_{j-1},t_{j}]\bigr].
\end{equation}
Finally, we define the potential survival function at visit $t_{j}$    as:
\begin{equation}
    \mathcal{S}_{\textbf{z},\textbf{z}_{*}}(t_{j}) = P\big[T_i(\textbf{z}^{t_{j-1}}, \textbf{m}^{t_{j-1}}) > t_{j}\big] = \prod_{l=1}^{j}\big\{1-\mathcal{P}_{\textbf{z},\textbf{z}_{*}}(t_{l})\big\}.
\end{equation}

\subsection{Causal estimands and identification}\label{sec33}
Using the quantities defined in the previous section, we define the Interventional Direct Effects (IDE) and the  Interventional Indirect Effects (IIE) as:
\begin{equation}
    \begin{aligned}
        IDE(t_{j}) &= \mathcal{S}_{\textbf{z},\textbf{z}_{*}}(t_{j}) - \mathcal{S}_{\textbf{z}_{*},\textbf{z}_{*}}(t_{j}), \\
        IIE(t_{j}) &= \mathcal{S}_{\textbf{z},\textbf{z}}(t_{j}) - \mathcal{S}_{\textbf{z},\textbf{z}_{*}}(t_{j}).
    \end{aligned}
\end{equation}
This implies that for individual $i$ at visit $t_{j}$, the interventional direct and indirect effects can be expressed as differences between the following survival probabilities:
\begin{equation}    
    \begin{aligned}
    IDE(t_{j}) &=  P\big[T_i(\textbf{z}^{t_{j-1}}, \mathbfcal{G}_{z_{*}}^{t_{j-1}} ) > t_{j}\big] - 
    P\big[T_i(\textbf{z}_{*}^{t_{j-1}},  \mathbfcal{G}_{z_{*}}^{t_{j-1}} ) > t_{j}\big], \\
    IIE(t_{j}) &=  P\big[T_i(\textbf{z}^{t_{j-1}},  \mathbfcal{G}_{z}^{t_{j-1}} ) > t_{j}\big] - 
    P\big[T_i(\textbf{z}^{t_{j-1}},  \mathbfcal{G}_{z_{*}}^{t_{j-1}}) > t_{j}\big].
    \end{aligned}
\end{equation}
IIE captures the indirect effect of the exposure on the outcome resulting from changes in the mediator. IDE, on the other hand,  captures the direct effect of the exposure through pathways that do not involve the mediator. These two contrasts decompose the total effects (TE) as:
\begin{align*}
    TE(t_{j}) &= IDE(t_{j}) + IIE(t_{j}) = \mathcal{S}_{\textbf{z},\textbf{z}}(t_{j}) - \mathcal{S}_{\textbf{z}_{*},\textbf{z}_{*}}(t_{j}) 
    \\
    &= P\big[T_i(\textbf{z}^{t_{j-1}},  \mathbfcal{G}_{z}^{t_{j-1}} ) > t_{j}\big] - P\big[T_i(\textbf{z}_{*}^{t_{j-1}},  \mathbfcal{G}_{z_{*}}^{t_{j-1}} ) > t_{j}\big].
\end{align*}
In order to identify the mediation parameters $IDE(t_{j}) \text{ and } IIE(t_{j})$ from the observed data, we must rely on the following assumptions \citep{zheng2017longitudinal}: 

\begin{assumption}\label{assump1}
    \textbf{Consistency}
    \begin{enumerate}[label=(\roman*)] 
            \item $L^{t_{j}}(\textbf{z}^{t_{j}}_{*}) = L^{t_{j}}$ and $M^{t_{j}}(\textbf{z}^{t_{j}}_{*}) = M^{t_{j}}$ given $\textbf{Z}^{t_{j}} = \textbf{z}^{t_{j}}_{*}$ 
            \item  $Y^{t_{j}}(\textbf{z}^{t_{j-1}}_{*}) = Y^{t_{j}}$ given $\textbf{Z}^{t_{j-1}} = \textbf{z}^{t_{j-1}}_{*}$ 
            \item $L^{t_{j}}(\textbf{z}^{t_{j}}, \textbf{m}^{t_{j-1}}) = L^{t_{j}}$ given $\textbf{Z}^{t_{j}} = \textbf{z}^{t_{j}}$, and $\textbf{M}^{t_{j-1}} = \textbf{m}^{t_{j-1}}$
            \item $Y^{t_{j}}(\textbf{z}^{t_{j-1}},  \textbf{m}^{t_{j-1}}) = Y^{t_{j}}$ given $\textbf{Z}^{t_{j-1}} = \textbf{z}^{t_{j-1}}$, and $\textbf{M}^{t_{j-1}} = \textbf{m}^{t_{j-1}}$
            \item $T_i(\textbf{z}^{t_{j}}, \textbf{m}^{t_{j}}) = T_i$ given $\textbf{Z}^{t_{j}} = \textbf{z}^{t_{j}}$,   and $\textbf{M}^{t_{j}} = \textbf{m}^{t_{j}}$.
        \end{enumerate}
\end{assumption}
\begin{assumption}\label{assump2}
    \textbf{Positivity} \\
    For all visit $t_{j}$ and for all $\boldsymbol{\ell}\text{ and }\textbf{m}$, we assume that the following conditions hold:
   \begin{enumerate}[label=(\roman*)]
       \item if $p(\textbf{y}^{t_{j}}, \textbf{z}_{*}^{t_{j-1}},\boldsymbol{\ell}^{t_{j-1}}, \textbf{m}^{t_{j-1}} ) > 0 ,\text{ then } p(z_{*}^{t_{j}}|\textbf{y}^{t_{j}},\textbf{z}_{*}^{t_{j-1}},\boldsymbol{\ell}^{t_{j-1}}, \textbf{m}^{t_{j-1}} ) > 0$
       \item if $p(\textbf{y}^{t_{j}}, \textbf{z}^{t_{j-1}},\boldsymbol{\ell}^{t_{j-1}}, \textbf{m}^{t_{j-1}} ) > 0 ,\text{ then } p(z^{t_{j}}|\textbf{y}^{t_{j}}, \textbf{z}^{t_{j-1}},\boldsymbol{\ell}^{t_{j-1}}, \textbf{m}^{t_{j-1}} ) > 0$
       \item if $p(\ell^{t_{j}}|\textbf{y}^{t_{j}}, \textbf{z}^{t_{j}},\boldsymbol{\ell}^{t_{j-1}}, \textbf{m}^{t_{j-1}} ) > 0 ,\text{ then } p(\ell^{t_{j}}|\textbf{y}^{t_{j}}, \textbf{z}_{*}^{t_{j}},\boldsymbol{\ell}^{t_{j-1}}, \textbf{m}^{t_{j-1}} ) > 0$
       \item if $p(y^{t_{j}}|\textbf{y}^{t_{j-1}},\textbf{z}^{t_{j}}, \boldsymbol{\ell}^{t_{j}}, \textbf{m}^{t_{j}}) > 0 ,\text{ then } p(y^{t_{j}}|\textbf{y}^{t_{j-1}}, \textbf{z}_{*}^{t_{j}}, \boldsymbol{\ell}^{t_{j}}, \textbf{m}^{t_{j}}) > 0$
       \item if $p(\textbf{y}^{t_{j}}, \textbf{z}^{t_{j}},\boldsymbol{\ell}^{t_{j}}, \textbf{m}^{t_{j-1}} ) > 0$ and $p(m^{t_{j}}|\textbf{y}^{t_{j}}, \textbf{z}_{*}^{t_{j}},\boldsymbol{\ell}^{t_{j}}, \textbf{m}^{t_{j-1}} ) > 0 , \text{ then } \\ p(m^{t_{j}}|\textbf{y}^{t_{j}}, \textbf{z}^{t_{j}},\boldsymbol{\ell}^{t_{j}}, \textbf{m}^{t_{j-1}} ) > 0$.
   \end{enumerate}
\end{assumption}
Assumption \ref{assump2} (i) and (ii) require exposures to be observed in each confounder and mediator stratum.  Assumption \ref{assump2} (iii) and (iv) require confounder and survival outcomes supported under $\textbf{z}$  to also be supported under  $\textbf{z}_{*}$. Assumption \ref{assump2} (v) requires the mediator supported under $\textbf{z}_{*}$ to also be supported under $\textbf{z}$. We use $p(\cdot)$ to denote a density (or mass) function of a continuous (or discrete) variable.
    
\begin{assumption} \label{assump3}
    \textbf{No Unmeasured Confounding} 
    \begin{enumerate}[label=(\roman*)]
    \item $Z^{t_{j}}$ \textbf{is randomized conditional on the observed history:} \\
    There are no unmeasured confounders of the relationship between each $Z^{t_{j}}$ and all its subsequent covariates, conditional on the observed history. For any $l \geq j$: 
    \begin{enumerate}
        \item $\big\{ \mathbfcal{Y}(\textbf{z}_{*}^{t_{l-1}}),\mathbfcal{L}(\textbf{z}_{*}^{t_{l}}), \mathbfcal{M}(\textbf{z}_{*}^{t_{l}}), T(\textbf{z}_{*}^{t_{l-1}})\big\} \indep Z^{t_{j}}_{*}|\textbf{Y}^{t_{j}}= \mathbb{0}_{j}, \textbf{Z}^{t_{j-1}}_{*},\textbf{L}^{t_{j-1}},\textbf{M}^{t_{j-1}},\underline{L}^0$
        \item $\big\{\mathbfcal{Y}(\textbf{z}^{t_{l-1}},\textbf{m}^{t_{l-1}}),\mathbfcal{L}(\textbf{z}^{t_{l}},\textbf{m}^{t_{l-1}}),  T(\textbf{z}^{t_{l-1}},\textbf{m}^{t_{l}})\big\} \indep Z^{t_{j}}|
        \textbf{Y}^{t_{j}}= \mathbb{0}_{j}, \textbf{Z}^{t_{j-1}},\textbf{L}^{t_{j-1}},\textbf{M}^{t_{j-1}},\underline{L}^0$.
    \end{enumerate}
    
    \item  $M^{t_{j}}$ \textbf{is randomized conditional on the observed history:} \\
    There are no unmeasured confounders of the relationship between each $M^{t_{j}}$ and all its subsequent covariates, conditional on the observed history. For any $l \geq j$: \\
    $\big\{ \mathbfcal{Y}(\textbf{z}^{t_{l-1}},\textbf{m}^{t_{l-1}}),\mathbfcal{L}(\textbf{z}^{t_{l}},\textbf{m}^{t_{l-1}}), T(\textbf{z}^{t_{l-1}},\textbf{m}^{t_{l-1}})\big\} \indep M^{t_{j}}| \textbf{Y}^{t_{j}}= \mathbb{0}_{j},\textbf{Z}^{t_{j}},\textbf{L}^{t_{j}},\textbf{M}^{t_{j-1}},\underline{L}^0$.  
\end{enumerate}
\end{assumption}
Assumption \ref{assump3} (i) requires that, for an individual at visit $t_{j}$,  the history of medication, covariates, risk factor measurements, and survival indicators account for all the confounding in the relationship between current medication and current and future covariates, risk factor measurements, and survival outcomes. Similarly, Assumption \ref{assump3} (ii) requires that at visit $t_{j}$, current medication and covariates, along with the history of medication, covariates, risk factors, and survival indicators, account for all the confounding in the relationship between current CVD risk factor measurement and current and future covariates and survival outcomes.

These sequential randomization assumptions are widely used in causal mediation literature. However, these assumptions are strong and could be violated, for instance, if there is an unmeasured event that would affect an individual's current medication and current and future risk factor measurements and covariates. Positivity assumptions, necessary to ensure that the conditional densities used for the identification of the mediation parameter are well-defined, have been discussed in \cite{zheng2017longitudinal}.

\begin{manualtheorem}{1a}\label{prop1a}
Under the assumptions in this sub-section, we can identify $\mathcal{P}_{\textbf{z},\textbf{z}_{*}}(t_{j})$ using the observed data distribution as : \\
\begin{equation}\label{eq::prop1a}
    \begin{aligned}
        \mathcal{P}_{\textbf{z},\textbf{z}_{*}}(t_{j} ) &= \int_{\textbf{m}^{t_{j-1}}}\int_{\boldsymbol{\ell}^{t_{j-1}}}\int_{\textbf{y}^{t_{j-1}}}\int_{\underline{\ell}^{0}} 
      p\big(\underline{L}^0\big) \times
       P \Bigl[T_i \in  (t_{j-1},t_{j}]\big|\textbf{Y}^{t_{j-1}} = \mathbb{0}_{j-1}, 
       \textbf{Z}^{t_{j-1}} = \textbf{z}^{t_{j-1}}, 
       \textbf{L}^{t_{j-1}} = \boldsymbol{\ell}^{t_{j-1}},\textbf{M}^{t_{j-1}} = \textbf{m}^{t_{j-1}},  \underline{L}^0= \underline{\ell}^0 \Bigr] \times \\
      & \prod_{l=1}^{j-1} \Bigl\{  p\big(Y^{t_{l}}\big|\textbf{Y}^{t_{l-1}} = \mathbb{0}_{l-1}, \textbf{Z}^{t_{l-1}} = \textbf{z}^{t_{l-1}}, \textbf{L}^{t_{l-1}} = \boldsymbol{\ell}^{t_{l-1}}, \textbf{M}^{t_{l-1}} = \textbf{m}^{t_{l-1}}, 
      \underline{L}^0= \underline{\ell}^0 \big) \times
      \\& 
       p\big(L^{t_{l}}\big|\textbf{Y}^{t_{l}} = \mathbb{0}_{l}, \textbf{Z}^{t_{l}} = \textbf{z}^{t_{l}},  \textbf{L}^{t_{l-1}} = \boldsymbol{\ell}^{t_{l-1}},  \textbf{M}^{t_{l-1}} = \textbf{m}^{t_{l-1}}, 
        \underline{L}^0 = \underline{\ell}^0 \big) \times
        \\& 
      p\big( M^{t_{l}}\big| \textbf{Y}^{t_{l}} = \mathbb{0}_{l}, \textbf{Z}^{t_{l}} = \textbf{z}_{*}^{t_{l}},  \textbf{L}^{t_{l}} = \boldsymbol{\ell}^{t_{l}},  \textbf{M}^{t_{l-1}} = \textbf{m}^{t_{l-1}}, 
      \underline{L}^0= \underline{\ell}^0\big)    
          \Bigr\}  d\underline{\ell}^{0}d\textbf{y}^{t_{j-1}}d\boldsymbol{\ell}^{t_{j-1}}d\textbf{m}^{t_{j-1}}.
    \end{aligned}
\end{equation}
\end{manualtheorem}
\begin{manualtheorem}{1b}\label{prop1b}
For a known baseline age $A^{0} = a^{0}$, defining $\underline{L}_{i}^{0,*} = \underline{L}_{i}^{0} \setminus A_{i}^{0}$ as the vector of baseline confounders excluding the baseline age, we may define our causal quantity $\mathcal{P}_{\textbf{z},\textbf{z}_{*}}(t_{j})$ conditional on $a^{0}$ as:
\begin{equation}\label{eq::prop1b}
    \begin{aligned}
        \mathcal{P}_{\textbf{z},\textbf{z}_{*}}(t_{j}|a_{0} ) &= \int_{\textbf{m}^{t_{j-1}}}\int_{\boldsymbol{\ell}^{t_{j-1}}}\int_{\textbf{y}^{t_{j-1}}}\int_{\underline{\ell}^{0,*}} p\big(\underline{L}^{0,*}|a_{0}\big) \times
       P \Bigl[T_i \in  (t_{j-1},t_{j}]\big|\textbf{Y}^{t_{j-1}} = \mathbb{0}_{j-1}, 
       \textbf{Z}^{t_{j-1}} = \textbf{z}^{t_{j-1}}, 
       \textbf{L}^{t_{j-1}} = \boldsymbol{\ell}^{t_{j-1}},\textbf{M}^{t_{j-1}} = \textbf{m}^{t_{j-1}},  
       \\& 
      \underline{L}^{0,*}= \underline{\ell}^{0,*},a_{0}\Bigr] \times  \prod_{l=1}^{j-1} \Bigl\{  p\big(Y^{t_{l}}\big|\textbf{Y}^{t_{l-1}} = \mathbb{0}_{l-1}, \textbf{Z}^{t_{l-1}} = \textbf{z}^{t_{l-1}}, \textbf{L}^{t_{l-1}} = \boldsymbol{\ell}^{t_{l-1}}, \textbf{M}^{t_{l-1}} = \textbf{m}^{t_{l-1}}, 
      \underline{L}^{0,*}= \underline{\ell}^{0,*},a_{0} \big) 
       \times
       \\& 
      p\big(L^{t_{l}}\big|\textbf{Y}^{t_{l}} = \mathbb{0}_{l}, \textbf{Z}^{t_{l}} = \textbf{z}^{t_{l}},  \textbf{L}^{t_{l-1}} = \boldsymbol{\ell}^{t_{l-1}},  \textbf{M}^{t_{l-1}} = \textbf{m}^{t_{l-1}}, 
       \underline{L}^{0,*} = \underline{\ell}^{0,*},a_{0} \big) \times 
       \\& 
      p\big( M^{t_{l}}\big| \textbf{Y}^{t_{l}} = \mathbb{0}_{l}, \textbf{Z}^{t_{l}} = \textbf{z}_{*}^{t_{l}},  \textbf{L}^{t_{l}} = \boldsymbol{\ell}^{t_{l}},  \textbf{M}^{t_{l-1}} = \textbf{m}^{t_{l-1}}, 
      \underline{L}^{0,*}= \underline{\ell}^{0,*},a_{0}\big)    
          \Bigr\}   d\underline{\ell}^{0,*}d\textbf{y}^{t_{j-1}}d\boldsymbol{\ell}^{t_{j-1}}d\textbf{m}^{t_{j-1}}.
    \end{aligned}
\end{equation}
\end{manualtheorem}
The proof of Proposition \ref{prop1a} is available in Section 1 of the supplementary material. The proof of Proposition \ref{prop1b} can be written by repeating the steps in the proof of Proposition \ref{prop1a} conditional on baseline age $a_0$. Using $\mathcal{P}_{\textbf{z},\textbf{z}_{*}}(t_{j}|a_{0} )$, we can estimate the potential survival function given baseline age as $\mathcal{S}_{\textbf{z},\textbf{z}_{*}}(t_{j}|a_{0}) =  \prod_{l=1}^{j}\big(1-P_{\textbf{z},\textbf{z}_{*}}(t_{l}|a_{0})\big).$

Computational details on the numerical approximations of the integrals in Propositions (\ref{prop1a}) and (\ref{prop1b}) are available in Section \ref{sec5}. We use the Bayesian bootstrap \citep{rubin1981bayesian} to sample from the distributions $p\big(\underline{L}^0\big)$ and $p\big(\underline{L}^{0,*}|a_{0}\big)$
of the baseline confounders. The Bayesian bootstrap offers several advantages over its empirical counterpart. While the empirical distribution places a uniform weight of $1/N$ to each baseline confounder vector, the Bayesian bootstrap assigns an improper Dirichlet prior over the weights for observed confounder vectors. These weights are computationally easy to sample due to conjugacy and are agnostic to the choice of the confounder, mediator, and outcome models. The Bayesian bootstrap retains the flexibility of the empirical distribution while accounting for variability in the empirical estimate.
This approach has been widely used in Bayesian causal inference literature, as discussed in \cite{oganisian2024hierarchical,boatman2021borrowing,roy2017bayesian,taddy2016nonparametric}. We refer readers to these papers for additional details.

\section{An extension with competing events}\label{sec4}
 Competing events refer to events experienced by study participants that preclude them from experiencing the event of interest. In our motivating example, a participant's death due to non-CVD causes can be considered a competing event. Causal mediation analysis in the presence of competing events has garnered some recent attention \citep{young2020causal,vo2022longitudinal}. \cite{young2020causal} describe various counterfactual contrasts based on both risks and hazards of the event of interest when competing events exist. They summarize these contrasts in terms of the estimands defined in the literature, noting that the contrasts involving counterfactual hazards do not generally possess a causal interpretation. However, their formulation differs from our setting, as it is based on longitudinal event indicators for both the main and competing events, with the latter acting as a mediator between the exposure and the former, in the presence of a baseline exposure. \cite{vo2022longitudinal} define total, direct, and indirect effects using the cause-specific cumulative incidence scale. Their estimation relies on a weighting approach, but in a setting involving a point exposure. 
 
 In survival settings involving competing risks, the conventional approach is to model cause-specific hazards using separate Cox regression models or model sub-distribution hazards using the Fine and Gray model \citep{fine1999proportional}. Typically, a proportional hazards assumption is used in both scenarios. In practice, regression relationships between two competing risks often involve complex nonlinear relationships among covariates or non-proportional relationships among hazards. \cite{sparapani2020nonparametric} introduced a non-parametric approach for discrete-time-based competing risks data using BART. Their approach eliminates the need for restrictive proportional or sub-distribution hazards assumptions. In this article, we employ this framework for competing risk data (see Section 5 for the model specification). 
 
\subsection{Causal estimands and identification}\label{sec41}
Suppose we have two events: the main event of interest and the competing event. Recall that we define the event indicator $d_{i}$ to distinguish the main event ($d_{i} =1$) from the competing event ($d_{i} =2$). Our new outcome of interest is the pair $(T_i, d_i)$ where $T_{i}$ is defined as before. We define the corresponding potential outcome variables for the pair $(T_i, d_i)$ as:
$$ \big(T_i(\textbf{Z}^{t_{j-1}}, \mathbfcal{G}_{z_{*}}^{t_{j-1}}), \mathcal{D}_i(\textbf{Z}^{t_{j-1}}, \mathbfcal{G}_{z_{*}}^{t_{j-1}})\big)$$
where $T_i(\textbf{z}^{t_{j-1}},\textbf{m}^{t_{j-1}})$ denotes the potential time-to-event as before, and  $\mathcal{D}_i(\textbf{z}^{t_{j-1}},\textbf{m}^{t_{j-1}})$ denotes the counterfactual value of $d_i$ under the exposure and the mediator history  $\textbf{z}^{t_{j-1}}$, and $\textbf{m}^{t_{j-1}}$, respectively. 

Using these counterfactual outcomes, we now define the probability of event $k$ at visit $t_{j}$ given that the subject is still at risk (has not experienced either event yet). Under the exposure process $\textbf{z}$ and the mediator process taking the value as if the individual is under the regime $\textbf{z}_{*}$ ($\textbf{m}^{t_{j}} \sim \mathbfcal{G}_{z_{*}}^{t_{j}}$), this probability is given by $\mathcal{P}^{(k)}_{\textbf{z},\textbf{z}_{*}}(t_{j}) = P\Bigl[T_i(\textbf{z}^{t_{j-1}}, \textbf{m}^{t_{j-1}}) \in (t_{j-1},t_{j}],D_i(\textbf{z}^{t_{j-1}}, \textbf{m}^{t_{j-1}}) = k\Bigr]$.

Our causal estimand of interest is the Cumulative Incidence Function (CIF), which for the event type $k  \in \{1,2\}$, is defined as $$F^{(k)}(t_{j}) \equiv P\Bigl[T_i \leq t_{j}, d_i = k\Bigr].$$ 
Now, for potential outcomes $(T_i(\cdot),\mathcal{D}_i(\cdot))$ and event type $k$, we define the potential CIF at visit $t_{j}$ as:
\begin{equation}
    \begin{aligned}
        \mathcal{F}^{(k)}_{\textbf{z},\textbf{z}_{*}}(t_{j}) &= P\Bigl[T_i(\textbf{Z}^{t_{j-1}}, \mathbfcal{G}_{z_{*}}^{t_{j-1}})\leq t_{j} ,  \mathcal{D}_i(\textbf{Z}^{t_{j-1}} , \mathbfcal{G}_{z_{*}}^{t_{j-1}}) =k\Bigr] 
        = \sum_{l=1}^{j} \tilde{\mathcal{S}}_{\textbf{z},\textbf{z}}(t_{l-1})\mathcal{P}^{(k)}_{\textbf{z},\textbf{z}_{*}}(t_{l}),
    \end{aligned}
\end{equation}
where $\tilde{\mathcal{S}}_{\textbf{z},\textbf{z}}(t_{j}) = \prod_{l=1}^{j}\big(1-\mathcal{P}^{1}_{\textbf{z},\textbf{z}_{*}}(t_{l})\big)\big(1-\mathcal{P}^{2}_{\textbf{z},\textbf{z}_{*}}(t_{l})\big).$
Finally, we use the following contrasts to quantify the direct, indirect, and total effects of the exposure on event $k$ for an individual at visit $t_{j}$:
\begin{equation}
    \begin{aligned}
        IDE^{(k)}(t_{j}) &= \mathcal{F}^{(k)}_{\textbf{z},\textbf{z}_{*}}(t_{j}) - \mathcal{F}^{(k)}_{\textbf{z}_{*},\textbf{z}_{*}}(t_{j}) \\
        IIE^{(k)}(t_{j}) &= \mathcal{F}^{(k)}_{\textbf{z},\textbf{z}}(t_{j}) - \mathcal{F}^{(k)}_{\textbf{z},\textbf{z}_{*}}(t_{j}) \\
        TE^{(k)}(t_{j}) &= IDE^{(k)}(t_{j}) + IIE^{(k)}(t_{j})  = \mathcal{F}^{(k)}_{\textbf{z},\textbf{z}}(t_{j}) - \mathcal{F}^{(k)}_{\textbf{z}_{*},\textbf{z}_{*}}(t_{j}).
    \end{aligned}
\end{equation}
To identify $\mathcal{F}^{(k)}_{\textbf{z},\textbf{z}_{*}}(t_{j})$ from the observed data, we modify the assumptions from the previous section to accommodate the new outcome of interest $(T_i, d_i)$ as follows:
\begin{assumption}\label{assump4}
    \textbf{Consistency 1:} 
    \begin{enumerate}
        \item $Y^{t_{j},(k)}(\textbf{z}^{t_{j-1}},  \textbf{m}^{t_{j-1}}) = Y^{t_{j},(k)}$ given $\textbf{Z}^{t_{j-1}} = \textbf{z}^{t_{j-1}}$ and $\textbf{M}^{t_{j-1}} = \textbf{m}^{t_{j-1}}$
            \item $T_i(\textbf{z}^{t_{j}}, \textbf{m}^{t_{j}}) = T_i$ and $\mathcal{D}_i(\textbf{z}^{t_{j}},\textbf{m}^{t_{j}}) = d_i$ given $\textbf{Z}^{t_{j}} = \textbf{z}^{t_{j}}$   and $\textbf{M}^{t_{j}} = \textbf{m}^{t_{j}}$.
    \end{enumerate}
    If an individual's exposure and mediator histories are consistent with the trajectories of their interventions until visit $t_{j}$, then their counterfactual outcomes under the same interventions equal their observed outcomes until that visit.
\end{assumption}
\begin{assumption}\label{assump5}
    \textbf{Positivity 1:}
     $$ \text{If } p_{\textbf{Y}^{t_{j-1},(k)},\textbf{Z}^{t_{j-1}},\textbf{L}^{t_{j-1}},\textbf{M}^{t_{j-1}}}(\textbf{y}^{t_{j-1},(k)},\textbf{z}^{t_{j-1}},\boldsymbol{\ell}^{t_{j-1}},\textbf{m}^{t_{j-1}}) \neq 0, \text{ then }
    Pr( d_{i}=1|\textbf{y}^{t_{j-1},(k)},\textbf{z}^{t_{j-1}},\boldsymbol{\ell}^{t_{j-1}},\textbf{m}^{t_{j-1}}) >0.
    $$
    For any observed covariate history level among participants who remain uncensored (assuming they are free of competing events) and have not experienced the event of interest up to visit $t_{j-1}$, there are some participants who remain uncensored through visit $t_{j}$.
\end{assumption}

\begin{assumption}\label{assump6}
    \textbf{No Unmeasured Confounding 1:}
    \begin{enumerate}[label=(\roman*)]
    \item $Z^{t_{j}}$ \textbf{is randomized conditional on the observed history:} There are no unmeasured confounders of the relationship between each $Z^{t_{j}}$ and all its subsequent covariates, conditional on the observed history. For any $l \geq j$:
        $$\big\{\mathbfcal{Y}^{(k)}(\textbf{z}^{t_{l-1}},\textbf{m}^{t_{l-1}}),\mathbfcal{L}(\textbf{z}^{t_{l}},\textbf{m}^{t_{l-1}}),  T(\textbf{z}^{t_{l-1}},\textbf{m}^{t_{l-1}}),\mathcal{D}(\textbf{z}^{t_{l-1}},\textbf{m}^{t_{l-1}})\big\}  
        \indep Z^{t_{j}}| \textbf{Y}^{t_{j},(1)} = \textbf{Y}^{t_{j},(2)} = \mathbb{0}_{j},\textbf{Z}^{t_{j-1}},\textbf{L}^{t_{j-1}},\textbf{M}^{t_{j-1}},\underline{L}^0.$$
    
    \item  $M^{t_{j}}$ \textbf{is randomized conditional on the observed history:} There are no unmeasured confounders of the relationship between each $M^{t_{j}}$ and all its subsequent covariates, conditional on the observed history. For any $l \geq j$:
    $$\big\{\mathbfcal{Y}^{(k)}(\textbf{z}^{t_{l-1}},\textbf{m}^{t_{l-1}}),\mathbfcal{L}(\textbf{z}^{t_{l}},\textbf{m}^{t_{l-1}}),  T(\textbf{z}^{t_{l-1}},\textbf{m}^{t_{l-1}}),\mathcal{D}(\textbf{z}^{t_{l-1}},\textbf{m}^{t_{l-1}})\big\} 
        \indep M^{t_{j}}|\textbf{Y}^{t_{j},(1)} = \textbf{Y}^{t_{j},(2)} = \mathbb{0}_{j}, \textbf{Z}^{t_{j}},\textbf{L}^{t_{j}},\textbf{M}^{t_{j-1}},\underline{L}^0.$$
\end{enumerate}
\end{assumption}    
   
\begin{manualtheorem}{2}\label{prop2a}
    Under Assumptions \ref{assump1} - \ref{assump6}, we can identify $\mathcal{P}^{(1)}_{\textbf{z},\textbf{z}_{*}}(t_{j})$ and $\mathcal{P}^{(2)}_{\textbf{z},\textbf{z}_{*}}(t_{j})$ using the observed data distribution as:   
\begin{align*}
      \mathcal{P}^{(1)}_{\textbf{z},\textbf{z}_{*}}(t_{j}) &=  \int_{\textbf{m}^{t_{j-1}}}\int_{\boldsymbol{\ell}^{t_{j-1}}}\int_{\textbf{y}^{t_{j-1},(1)},\textbf{y}^{t_{j-1},(2)}}\int_{\underline{\ell}^{0}} p\big(\underline{L}^0\big) \times
      \\& 
      P \Bigl[T_i \in  (t_{j-1},t_{j}], d_i =1 \big| \textbf{Y}^{t_{j-1},(1)} 
       = \textbf{Y}^{t_{j-1},(2)} = \mathbb{0}_{j-1},
      \textbf{Z}^{t_{j-1}} = \textbf{z}^{t_{j-1}}, 
      \textbf{L}^{t_{j-1}} = \boldsymbol{\ell}^{t_{j-1}},\textbf{M}^{t_{j-1}} = \textbf{m}^{t_{j-1}},  \underline{L}^0= \underline{\ell}^0 \Bigr] \times \\
      & \prod_{l=1}^{j-1} \Bigl\{  p\big(Y^{t_{l},(1)}, Y^{t_{l},(2)}\big|\textbf{Y}^{t_{l-1},(1)} 
      = \textbf{Y}^{t_{l-1},(2)} = \mathbb{0}_{l-1},
      \textbf{Z}^{t_{l-1}} = \textbf{z}^{t_{l-1}}, \textbf{L}^{t_{l-1}} = \boldsymbol{\ell}^{t_{l-1}},
      \textbf{M}^{t_{l-1}} = \textbf{m}^{t_{l-1}},  \underline{L}^0= \underline{\ell}^0 \big) \times 
      \\&
      p\big(L^{t_{l}}\big|\textbf{Y}^{t_{l},(1)} 
      = \textbf{Y}^{t_{l},(2)} = \mathbb{0}_{l},
      \textbf{Z}^{t_{l}} = \textbf{z}^{t_{l}},
      \textbf{L}^{t_{l-1}} = \boldsymbol{\ell}^{t_{l-1}},
      \textbf{M}^{t_{l-1}} = \textbf{m}^{t_{l-1}},  \underline{L}^0 = \underline{\ell}^0 \big) \times 
      \\&
      p\big( M^{t_{l}}\big| \textbf{Y}^{t_{l},(1)} 
      = \textbf{Y}^{t_{l},(2)} = \mathbb{0}_{l},
      \textbf{Z}^{t_{l}} = \textbf{z}_{*}^{t_{l}},  \textbf{L}^{t_{l}} = \boldsymbol{\ell}^{t_{l}},
      \textbf{M}^{t_{l-1}} = \textbf{m}^{t_{l-1}}, \underline{L}^0= \underline{\ell}^0\big)    
          \Bigr\}    d\underline{\ell}^{0}d\textbf{y}^{t_{j-1},(k)}d\boldsymbol{\ell}^{t_{j-1}}d\textbf{m}^{t_{j-1}}, \text{ and }
          \\
    \mathcal{P}^{(2)}_{\textbf{z},\textbf{z}_{*}}(t_{j}) &=  \int_{\textbf{m}^{t_{j-1}}}\int_{\boldsymbol{\ell}^{t_{j-1}}}\int_{\textbf{y}^{t_{j},(1)},\textbf{y}^{t_{j-1},(2)}}\int_{\underline{\ell}^{0}} p\big(\underline{L}^0\big) \times
      \\& 
      P \Bigl[T_i \in  (t_{j-1},t_{j}], d_i =2\big| \textbf{Y}^{t_{j},(1)} = \mathbb{0}_{j},
       \textbf{Y}^{t_{j-1},(2)} = \mathbb{0}_{j-1},
      \textbf{Z}^{t_{j-1}} = \textbf{z}^{t_{j-1}}, 
      \textbf{L}^{t_{j-1}} = \boldsymbol{\ell}^{t_{j-1}},
      \textbf{M}^{t_{j-1}} = \textbf{m}^{t_{j-1}},  \underline{L}^0= \underline{\ell}^0 \Bigr] \times 
      \\& 
     p\big(Y^{t_{j},(1)}\big|\textbf{Y}^{t_{j-1},(1)} 
      = \textbf{Y}^{t_{j-1},(2)} = \mathbb{0}_{j-1},
      \textbf{Z}^{t_{j-1}} = \textbf{z}^{t_{j-1}}, 
       \textbf{L}^{t_{j-1}} = \boldsymbol{\ell}^{t_{j-1}},
      \textbf{M}^{t_{j-1}} = \textbf{m}^{t_{j-1}},  \underline{L}^0= \underline{\ell}^0 \big) \times
      \\& 
       \prod_{l=1}^{j-1} \Bigl\{  p\big(Y^{t_{l},(2)}, Y^{t_{l},(1)}\big|\textbf{Y}^{t_{l-1},(1)} = 
       \textbf{Y}^{t_{l-1},(2)} = \mathbb{0}_{l-1},
      \textbf{Z}^{t_{l-1}} = \textbf{z}^{t_{l-1}}, \textbf{L}^{t_{l-1}} = \boldsymbol{\ell}^{t_{l-1}},
      \textbf{M}^{t_{l-1}} = \textbf{m}^{t_{l-1}},  \underline{L}^0= \underline{\ell}^0 \big) \times
      \\&
      p\big(L^{t_{l}}\big|\textbf{Y}^{t_{l},(1)} 
      = \textbf{Y}^{t_{l},(2)} = \mathbb{0}_{l},
      \textbf{Z}^{t_{l}} = \textbf{z}^{t_{l}},
      \textbf{L}^{t_{l-1}} = \boldsymbol{\ell}^{t_{l-1}},
      \textbf{M}^{t_{l-1}} = \textbf{m}^{t_{l-1}},  \underline{L}^0 = \underline{\ell}^0 \big) \times
      \\&
      p\big( M^{t_{l}}\big| \textbf{Y}^{t_{l},(1)} 
      = \textbf{Y}^{t_{l},(2)} = \mathbb{0}_{l},
      \textbf{Z}^{t_{l}} = \textbf{z}_{*}^{t_{l}},  \textbf{L}^{t_{l}} = \boldsymbol{\ell}^{t_{l}},
      \textbf{M}^{t_{l-1}} = \textbf{m}^{t_{l-1}}, \underline{L}^0= \underline{\ell}^0\big)    
          \Bigr\}    d\underline{\ell}^{0}d\textbf{y}^{t_{j-1},(k)}d\boldsymbol{\ell}^{t_{j-1}}d\textbf{m}^{t_{j-1}}.
  \end{align*} 
\end{manualtheorem}
The proof of Proposition \ref{prop2a} is available in Section 1 of the supplementary material. For the binary survival indicators $Y^{t_{l},(1)}$ and $ Y^{t_{l},(2)}$, the joint mass function $p\big(Y^{t_{l},(1)}, Y^{t_{l},(2)}\big|\textbf{Y}^{t_{l-1},(1)} 
      = \textbf{Y}^{t_{l-1},(2)} = \mathbb{0}_{l-1},
      \textbf{Z}^{t_{l-1}} = \textbf{z}^{t_{l-1}}, \textbf{L}^{t_{l-1}} = \boldsymbol{\ell}^{t_{l-1}},
      \\
      \textbf{M}^{t_{l-1}} = \textbf{m}^{t_{l-1}},  \underline{L}^0= \underline{\ell}^0 \big)$
in Proposition \ref{prop2a} can be further decomposed into the mass function of $Y^{t_{l},(1)}$ given past covariates and the mass function of  $Y^{t_{l},(2)}$ given $Y^{t_{l},(1)}$ and past covariates as:
\begin{equation}
    \begin{aligned}
        & p\big(Y^{t_{l},(1)}, Y^{t_{l},(2)}\big|\textbf{Y}^{t_{l-1},(1)} = \textbf{Y}^{t_{l-1},(2)} = \mathbb{0}_{l-1},
      \textbf{Z}^{t_{l-1}} = \textbf{z}^{t_{l-1}}, \textbf{L}^{t_{l-1}} = \boldsymbol{\ell}^{t_{l-1}},
      \textbf{M}^{t_{l-1}} = \textbf{m}^{t_{l-1}},  \underline{L}^0= \underline{\ell}^0 \big) 
      \\&
      = p\big(Y^{t_{l},(1)}\big|\textbf{Y}^{t_{l-1},(1)} = \textbf{Y}^{t_{l-1},(2)} = \mathbb{0}_{l-1},
      \textbf{Z}^{t_{l-1}} = \textbf{z}^{t_{l-1}}, \textbf{L}^{t_{l-1}} = \boldsymbol{\ell}^{t_{l-1}},
      \textbf{M}^{t_{l-1}} = \textbf{m}^{t_{l-1}},  \underline{L}^0= \underline{\ell}^0 \big) \times
      \\& 
      p\big(Y^{t_{l},(2)}\big|\textbf{Y}^{t_{l-1},(1)} = \mathbb{0}_{l-1}, \textbf{Y}^{t_{l},(2)} = \mathbb{0}_{l},
      \textbf{Z}^{t_{l-1}} = \textbf{z}^{t_{l-1}}, \textbf{L}^{t_{l-1}} = \boldsymbol{\ell}^{t_{l-1}},
      \textbf{M}^{t_{l-1}} = \textbf{m}^{t_{l-1}},  \underline{L}^0= \underline{\ell}^0 \big)
    \end{aligned}
\end{equation}
This decomposition allows us to fit two models: one for the conditional probability of the main event at the current visit given no prior event, and another for the conditional probability of the competing event given no prior event and no main event occurring at that visit. Details on model specification in the presence of a competing event are provided in Section \ref{sec523}.

Suppose $A_{i}^{0}$ denotes the baseline age of individual $i$, and $\underline{L}_{i}^{0,*} = \underline{L}_{i}^{0} \setminus A_{i}^{0}$ denotes the vector of baseline confounders excluding the baseline age. As before, we can define the causal quantity $\mathcal{P}^{(1)}_{\textbf{z},\textbf{z}_{*}}(t_{j})$ and $\mathcal{P}^{(2)}_{\textbf{z},\textbf{z}_{*}}(t_{j})$ conditional on $a_{0}$ and compute causal effect estimates within different strata of baseline age groups.

\section{Model specification and implementation using BART}\label{sec5}
\subsection{Review of BART}\label{sec51}
BART is an ensemble of trees model. A detailed description of the BART model is available in \cite{chipman2010bart} and \cite{tan2019bayesian}. We provide a brief overview of the model below. 

Suppose we have a continuous outcome $Y$ and a covariate vector $\textbf{X}$. BART uses an ensemble of $R$ regression trees to estimate the unknown function $g(\textbf{x})$ in the model $y = g(\textbf{x}) + \epsilon$, with $\epsilon \sim N(0, \sigma^2_Y)$, capturing the complex relationship between $\textbf{x}$ and $y$. A BART model for predicting $Y$ using $\textbf{X}$ can be represented as:
\begin{equation}
    y = g(\textbf{x}) + \epsilon = \sum_{r =1}^{R}\text{Tree}(\textbf{x}; \mathcal{T}_{r}, \boldsymbol{\mu}_{r}) + \epsilon,
\end{equation}
where, for the $r^{\text{th}}$ binary decision tree, $r \in \{1, \ldots, R\}$, $\mathcal{T}_{r}$ denotes the tree topology and splitting rules of the tree and $\boldsymbol{\mu}_{r}$ denotes the vector of predicted responses for all leaf nodes in tree $r$. The function $\text{Tree}(\textbf{x}; \mathcal{T}_{r}, \boldsymbol{\mu}_{r})$ returns $\mu_{r,d}$ if covariate $x$ is associated with leaf node $d$ of tree $r$. The decision tree nodes are divided into a collection of leaf nodes $d \in \mathcal{D}$ and branch nodes $b \in \mathcal{B}$. Each branch $b$ has a corresponding \textit{decision rule}, denoted $[X_b \leq C_b]$, and each leaf $d$ is associated with a prediction $\mu_{r,d}$.

In BART models, posterior inference is carried out using Markov Chain Monte Carlo (MCMC) by incorporating the posterior sampling tools for single-tree models into Metropolis-within-Gibbs samplers. We assume that the parameters of the conditional distributions of $Z_{i}^{t_{j}}, L_{i}^{t_{j}}, M_{i}^{t_{j}}, p_{ij},  \text{ and } p_{ij}^{(k)}$ are independent, and therefore their posteriors can be sampled simultaneously. In this article, we implement BART for continuous, binary, and survival responses using the R package by \cite{sparapani2019bart}. We use default priors on all parameters of the sum-of-trees model. Next, we provide details of the BART models for our setting.

\subsection{Models}\label{sec52} 
\subsubsection{Models for time-varying confounders and  mediators}\label{sec521}
Let $\{t_{j}: j = 0,1,\ldots,J\}$ denote the discrete visit times (with $t_{0}$ indicating baseline). We assume the following models:
\begin{equation}
    \begin{aligned}
        L_{i}^{t_{j}}|\textbf{Y}_{i}^{t_{j}}= \mathbb{0}_{j}, \textbf{Z}_{i}^{t_{j}},\textbf{L}_{i}^{t_{j-1}}, \textbf{M}_{i}^{t_{j-1}}, \underline{L}_{i}^{0} &=   g_{L}\big( \textbf{Z}_{i}^{t_{j}},\textbf{L}_{i}^{t_{j-1}}, \textbf{M}_{i}^{t_{j-1}}, \underline{L}_{i}^{0}\big) + \epsilon_{i,L},\\
        M_{i}^{t_{j}}|\textbf{Y}_{i}^{t_{j}} = \mathbb{0}_{j}, \textbf{Z}_{i}^{t_{j}},\textbf{L}_{i}^{t_{j}}, \textbf{M}_{i}^{t_{j-1}},  \underline{L}_{i}^{0} &=   g_{M}\big( \textbf{Z}_{i}^{t_{j}},\textbf{L}_{i}^{t_{j}}, \textbf{M}_{i}^{t_{j-1}}, \underline{L}_{i}^{0}\big) + \epsilon_{i,M},
    \end{aligned}
\end{equation}
where the unknown function $g\big(\cdot\big) \sim BART$ \citep{chipman2010bart}, $\epsilon_{i, L} \sim N(0, \sigma^2_{L})$ and $\epsilon_{i, M} \sim N(0, \sigma^2_{M})$. For binary confounders and mediators, the corresponding models are:
\begin{equation}
    \begin{aligned}
        Pr(L_{i}^{t_{j}} =1|\textbf{Y}_{i}^{t_{j}}= \mathbb{0}_{j}, \textbf{Z}_{i}^{t_{j}},\textbf{L}_{i}^{t_{j-1}}, \textbf{M}_{i}^{t_{j-1}}, \underline{L}_{i}^{0}) &= \Phi\Big(g_{L}\big( \textbf{Z}_{i}^{t_{j}},\textbf{L}_{i}^{t_{j-1}}, \textbf{M}_{i}^{t_{j-1}}, \underline{L}_{i}^{0}\big)\Big), \\
        Pr(M_{i}^{t_{j}} =1|\textbf{Y}_{i}^{t_{j}} = \mathbb{0}_{j}, \textbf{Z}_{i}^{t_{j}},\textbf{L}_{i}^{t_{j}}, \textbf{M}_{i}^{t_{j-1}},  \underline{L}_{i}^{0}) &= \Phi\Big(g_{M}\big( \textbf{Z}_{i}^{t_{j}},\textbf{L}_{i}^{t_{j}}, \textbf{M}_{i}^{t_{j-1}},  \underline{L}_{i}^{0}\big)\Big)
    \end{aligned}
\end{equation}
where $\Phi(\cdot)$ denotes the cumulative distribution function of a standard normal distribution. Note that incorporating multiple longitudinal confounders and mediators into the proposed framework would require including them in the BART models according to their causal ordering. If multiple mediators are measured simultaneously, determining their causal ordering becomes challenging. Addressing this issue would require modifying the identification assumptions and observed data models presented in this article to account for simultaneous mediators at each visit. However, such extensions are beyond the scope of this work.

\subsubsection{Non-parametric survival model with no competing event}\label{sec522}
Suppose $p_{ij} = P\big(T_{i} \in (t_{j-1},t_{j}]|T_{i} > t_{j-1}, \textbf{Z}_{i}^{t_{j-1}},\textbf{L}_{i}^{t_{j-1}},\textbf{M}_{i}^{t_{j-1}}, \underline{L}^0_i\big)$  is the probability of an event occurring in the interval $(t_{j-1},t_{j}]$ conditional on no previous event and the history of the other variables. In other words, $p_{ij}$ represents the discrete-time hazard of death due to CVD in the interval $(t_{j-1},t_{j}]$ given that an individual has survived to that interval. Conditional on $p_{ij}$, we specify the following model for event indicators $Y_{i}^{t_{j}} = \delta_{i}I(t_{j-1} < T_{i} \leq t_{j}), j =1,.., n_{i}$:

\begin{equation}
    \begin{aligned}
        & Y_{i}^{t_{j}}|p_{ij} \sim Bernoulli(p_{ij}) \\
        & p_{ij}|g_{T} = \Phi\Big(g_{T}\big(t_{j},\textbf{Z}_{i}^{t_{j-1}},\textbf{L}_{i}^{t_{j-1}},\textbf{M}_{i}^{t_{j-1}},\underline{L}^0_i\big)\Big) \\
        & g_{T}\big(\cdot\big) \sim BART \\
    \end{aligned}
\end{equation}
Hence, we have the likelihood $L(p|\textbf{y}) = \prod_{i=1}^{N}\prod_{j=1}^{n_{i}} p_{ij}^{Y_{i}^{t_{j}}}(1-p_{ij})^{1-Y_{i}^{t_{j}}}.$

\subsubsection{Non-parametric survival model with competing event}\label{sec523}
Suppose $p_{ij}^{(k)} = P\big(T_{i}\in (t_{j-1},t_{j}],d_{i} = k|T_{i} > t_{j-1}, \textbf{Z}_{i}^{t_{j-1}},\textbf{L}_{i}^{t_{j-1}},\textbf{M}_{i}^{t_{j-1}},\underline{L}^0_i\big)$ is the probability of an event occurring at interval $\in (t_{j-1},t_{j}]$, given the subject is still at risk (not experienced either main or competing event), and conditional on the history of the other variables. Thus, we have the likelihood $L(p|\textbf{y}) = \prod_{i=1}^{N}\prod_{j=1}^{n_{i}} \big(p_{ij}^{(1)}\big)^{y_{i}^{t_{j},(1)}} \times \big(p_{ij}^{(2)}\big)^{y_{i}^{t_{j},(2)}} \times \big(1-p_{ij}^{(1)}-p_{ij}^{(2)}\big)^{1-y_{i}^{t_{j},(1)} -y_{i}^{t_{j},(2)}}$,
where $y_{i}^{t_{j},(k)} = I\big(t_{j-1} < T_{i} \leq t_{j},d_{i} = k\big), k \in \{1,2\}$ is the observed event  indicator for individual $i$ at visit $j$. These indicators form a sequence of multinomial events with corresponding conditional probabilities $p_{ij}^{(k)}$.
Following the arguments of \cite{sparapani2020nonparametric}, we define $\tilde{p}_{ij}^{(2)} = \frac{p_{ij}^{(2)}}{1-p_{ij}^{(1)}}$ as the conditional probability of the competing event at visit $j$ given no main event occuring at that visit, and  rewrite this as:
\begin{align*}
        L(p|\textbf{y}) &= \prod_{i=1}^{N}\prod_{j=1}^{n_{i}} \big(p_{ij}^{(1)}\big)^{y_{i}^{t_{j},(1)}} \times  \big[\tilde{p}_{ij}^{(2)}\big(1-p_{ij}^{(1)}\big)\big]^{y_{i}^{t_{j},(2)}} \times 
        \big[1- p_{ij}^{(1)} - \tilde{p}_{ij}^{(2)}\big(1-p_{ij}^{(1)}\big) \big]^{1- y_{i}^{t_{j},(1)} - y_{i}^{t_{j},(2)}} \\
        &= \prod_{i=1}^{N}\prod_{j=1}^{n_{i}} \big(p_{ij}^{(1)}\big)^{y_{i}^{t_{j},(1)}} \big(1-p_{ij}^{(1)}\big)^{1-y_{i}^{t_{j},(1)}} \times 
        \prod_{i=1}^{N} \prod_{j:y_{i}^{t_{j},(1)} = 0}  \big(\tilde{p}_{ij}^{(2)}\big)^{y_{i}^{t_{j},(2)}} \big(1-\tilde{p}_{ij}^{(2)}\big)^{1-y_{i}^{t_{j},(2)}}.
\end{align*}
 With this, we can fit two separate probit BART models corresponding to $p_{ij}^{(1)}$ and $\tilde{p}_{ij}^{(2)}$.
Specifically, we assume the following  model for $p_{ij}^{(1)}$:
\begin{equation}
    \begin{aligned}
        & y_{i}^{t_{j},(1)} =  I\big(t_{j-1} < T_{i} \leq t_{j},d_{i} = 1\big), \quad j =1,.., n_{i}, \\
        & y_{i}^{t_{j},(1)}|p_{ij}^{(1)} \sim Bernoulli(p_{ij}^{(1)}), \\
        & p_{ij}^{(1)}|g_{Y^{(1)}} = \Phi\Big(g_{Y^{(1)}}\big(t_{j},\textbf{Z}_{i}^{t_{j-1}},\textbf{L}_{i}^{t_{j-1}},\textbf{M}_{i}^{t_{j-1}},\underline{L}^0_i\big)\Big), 
    \end{aligned}
\end{equation}
where $g_{Y^{(1)}}\big(\cdot\big) \sim BART$. This model can be thought of as a BART survival model for the main event at visit $j$, given the subject is still at risk prior to $j$. 
Similarly, we assume the following  model for  $\tilde{p}_{ij}^{(2)}$:
\begin{equation}
    \begin{aligned}
        & y_{i}^{t_{j},(2)} =  I\big(t_{j-1} < T_{i} \leq t_{j},d_{i} = 2\big), \quad j =1,.., n_{i}, \\
        & y_{i}^{t_{j},(2)}|\tilde{p}_{ij}^{(2)} \sim Bernoulli(\tilde{p}_{ij}^{(2)}), \\
        & \tilde{p}_{ij}^{(2)}|g_{Y^{(2)}} = \Phi\Big(g_{Y^{(2)}}\big(t_{j},\textbf{Z}_{i}^{t_{j-1}},\textbf{L}_{i}^{t_{j-1}},\textbf{M}_{i}^{t_{j-1}},\underline{L}^0_i\big)\Big),
    \end{aligned}
\end{equation}
where $g_{Y^{(2)}}\big(\cdot\big) \sim BART$. This model is for the conditional probability of the competing event at visit $j$, given that the individual is still at risk prior to $j$, and has not yet experienced the main event. 

\subsection{Computation of causal estimands}\label{sec53}

The baseline confounders $\underline{\ell}^0_i$ for participants are observed on entry into the study. Recall that at each visit $t_{j}$, we assume the following order for the longitudinal variables: $y_{i}^{t_{j}},z_{i}^{t_{j}},\ell_{i}^{t_{j}}, m_{i}^{t_{j}}$. Using samples from the posterior distributions of the observed data model parameters, we can generate samples from the posterior distributions of the causal quantities $\mathcal{P}_{\textbf{z},\textbf{z}_{*}}(t_{j})$ and $\mathcal{P}^{(k)}_{\textbf{z},\textbf{z}_{*}}(t_{j})$ using G-computation algorithms, \textbf{Algorithm 1} and \textbf{Algorithm 2}, respectively. Given our assumption of independent BART models for the time-varying confounders, mediators, and outcomes in our setup, we achieve high computational efficiency by parallelizing both the (BART) model fitting and Monte Carlo integration in the G-computation algorithms. We extend the \textit{GcompBART} package \citep{maria_josefsson_2023_10277820} to implement these algorithms. For a given baseline age $a_{0}$, the conditional causal estimands $\mathcal{P}_{\textbf{z},\textbf{z}_{*}}(t|a_{0})$ and $\mathcal{P}^{(k)}_{\textbf{z},\textbf{z}_{*}}(t|a_{0})$ can be calculated by repeating the steps outlined in \textbf{Algorithm 1} and \textbf{Algorithm 2}, respectively, conditional on baseline age $a_{0}$ . 

\section{Simulation}\label{sec6}
To evaluate the performance of the proposed BART models, we conduct simulation studies, comparing them with a standard parametric model. First, we fit parametric generalized linear models (GLMs) for the observed ARIC data to compute the 'ground truth' for interventional direct, indirect, and total effects. Specifically, for continuous data at visit $j$, we use the R Stats package to fit a linear model based on all preceding longitudinal and baseline covariates in the causal ordering. For binary data at visit $j$, we fit a GLM with a logit link using the same package. The coefficients from these models allow us to compute the ground truth for causal effect estimates using the G-computation algorithms proposed in the previous section.

Using the observed baseline confounders in ARIC and the fitted GLMs, we generate $1,000$ data replications for our simulation studies. Each replication has a sample size matching the baseline hypertensive sub-population in ARIC. On these replicated datasets, we fit the following parametric model in Stan for $j \in 1,\ldots, J$:
\begin{equation}\label{eq13}
    \begin{aligned}
         & Y_{i}^{t_{j}}|\textbf{Y}_{i}^{t_{j-1}} = \mathbb{0}, \textbf{M}_{i}^{t_{j}}, \textbf{L}_{i}^{t_{j}},\textbf{Z}_{i}^{t_{j}},\underline{L}^0_i 
         \sim Bern\big(logit(\boldsymbol{\gamma^{t_j}_{M}}\textbf{m}_{i}^{t_{j}}+\boldsymbol{\gamma^{t_j}_{L}}\boldsymbol{\ell}_{i}^{t_{j}} + \boldsymbol{\gamma^{t_j}_{Z}}\textbf{z}_{i}^{t_{j}}+\boldsymbol{\gamma^{t_j}_{\underline{L}^0}}\boldsymbol{\underline{\ell}}^0_{i} )\big), 
         \\&
        M_{i}^{t_{j}}|\textbf{Y}_{i}^{t_{j}} = \mathbb{0}, \textbf{M}_{i}^{t_{j-1}}, \textbf{L}_{i}^{t_{j}},\textbf{Z}_{i}^{t_{j}},\underline{L}^0_i 
        \sim Normal\big(\boldsymbol{\xi^{t_j}_{M}}\textbf{m}_{i}^{t_{j-1}}+\boldsymbol{\xi^{t_j}_{L}}\boldsymbol{\ell}_{i}^{t_{j-1}} + \boldsymbol{\xi^{t_j}_{Z}}\textbf{z}_{i}^{t_{j}}+ \boldsymbol{\xi^{t_j}_{\underline{L}^0}}\boldsymbol{\underline{\ell}}^0_{i}, \sigma^2_{M^{t_j}} \big), 
        \\&
        L_{i}^{t_{j}}|\textbf{Y}_{i}^{t_{j}} = \mathbb{0}, \textbf{M}_{i}^{t_{j-1}}, \textbf{L}_{i}^{t_{j-1}},\textbf{Z}_{i}^{t_{j}},\underline{L}^0_i 
        \sim Bern\big(logit(\boldsymbol{\beta^{t_j}_{M}}\textbf{m}_{i}^{t_{j-1}}+\boldsymbol{\beta^{t_j}_{L}}\boldsymbol{\ell}_{i}^{t_{j-1}} + \boldsymbol{\beta^{t_j}_{Z}}\textbf{z}_{i}^{t_{j}}+ \boldsymbol{\beta^{t_j}_{\underline{L}^0}}\boldsymbol{\underline{\ell}}^0_{i} )\big),
    \end{aligned}
\end{equation}
where $\boldsymbol{\gamma}^{t_j}$, $\boldsymbol{\xi}^{t_j}$, and $\boldsymbol{\beta}^{t_j}$ denote the regression coefficients for the time-varying outcome, mediator, and confounder variables at visit $j$, respectively. Similarly, the parameter $\sigma^2_{M^{t_j}}$ denotes the variance component of the mediator at visit $j$. We assign uninformative priors to these parameters.

Table \ref{tab1} shows results comparing the performance of the proposed BART model with that of the parametric model in (\ref{eq13}). The results indicate that causal effect estimators from both models are approximately unbiased for visits 2 and 3 in ARIC. We also observe that the MSEs are similar for both models, with the BART model achieving closer to the normal coverage rate for visit 2 and some over-coverage for visit 3.

We then assess the performance of the two models when the parametric model is misspecified. Specifically, we add several log terms and an interaction term among baseline covariates to the data-generating model. We then fit the model as specified in (\ref{eq13}) and compare its performance with the BART model.

Table \ref{tab2} displays the results comparing the performance of the two models. As expected, the bias for the causal effect estimates across visits increases for the misspecified parametric model in Table \ref{tab2} relative to the corresponding results in Table \ref{tab1}. The $95\%$ coverage rates at visit 3 are particularly lower with coverage as low as $57\%$. Between the two models fitted to replicated data generated from the misspecified GLM models, we observe that the BART model shows lower absolute bias and a substantial reduction in MSE. Overall, the BART model performs well for data with models that include interactions and/or log terms, as indicated by the results in Table \ref{tab2}.

\begin{table*}[!h]
\caption{Simulation results for comparing the bias, MSE, and coverage of 95\% credible intervals of the true parametric model versus BART.\label{tab1}}
\tabcolsep=0pt
\begin{tabular*}{\textwidth}{@{\extracolsep{\fill}}lccccccc@{\extracolsep{\fill}}}
\toprule%
& &\multicolumn{3}{@{}c@{}}{Bayesian parametric model} & \multicolumn{3}{@{}c@{}}{BART model} \\
\cline{3-5}\cline{6-8}%
Visit & & Bias & 95\% Coverage & MSE & Bias & 95\% Coverage & MSE \\
\midrule
2  & IDE & -6.2e-04 &  0.93 &  6.7e-05 & -2.2e-04  &  0.95 &   5.4e-05\\
 & IIE & 7.9e-04 & 1.00 & 1.01e-06  & 1.2e-03 &   1.00 &   1.6e-06\\
 & TE & 1.7e-04 & 0.93 & 6.8e-05  & 9.6e-04  &  0.94  &  5.5e-05\\
\midrule
3  & IDE & -2.4e-03  & 0.94 &  1.8e-04 & -2.4e-03   & 0.99   & 1.9e-04\\
 & IIE & 3.1e-03  & 0.96  & 1.3e-05 & 3.4e-03  &  0.99  &  1.5e-05\\
 & TE & 7.4e-04  & 0.94  & 1.7e-04  & 1.1e-03   & 0.99   & 1.9e-04\\
\bottomrule
\end{tabular*}
\end{table*}

\begin{table*}[!h]
\caption{Simulation results for comparing the bias, MSE, and coverage of 95\% credible intervals of the misspecified parametric model versus BART.\label{tab2}}
\tabcolsep=0pt
\begin{tabular*}{\textwidth}{@{\extracolsep{\fill}}lccccccc@{\extracolsep{\fill}}}
\toprule%
& &\multicolumn{3}{@{}c@{}}{Bayesian parametric model} & \multicolumn{3}{@{}c@{}}{BART model} \\
\cline{3-5}\cline{6-8}%
Visit & & Bias & 95\% Coverage & MSE & Bias & 95\% Coverage & MSE \\
\midrule
2  & IDE & -1.4e-02 &  0.95  & 3.4e-03 & -3.4e-03  &  0.97   & 4.6e-04\\
 & IIE & 3.0e-03 &  0.95 &  3.8e-05 & 2.9e-03   & 1.00   & 1.3e-05\\
 & TE & -1.1e-02  & 0.95  & 3.4e-03  & -5.0e-04  &  0.97   & 4.6e-04\\
\midrule
3  & IDE & -6.8e-03 &  0.57  & 1.5e-03 
& 3.6e-03  &  0.91   & 2.5e-05\\
 & IIE & -4.5e-03  & 0.97  & 1.2e-04 & 7.3e-04   & 0.98   & 7.3e-07\\
 & TE & -1.1e-02  & 0.57  & 1.5e-03 & 4.4e-03   & 0.88   & 3.1e-05\\
\bottomrule
\end{tabular*}
\end{table*}

\section{Analysis of the ARIC data}\label{sec7}
Our goal is to estimate the direct and indirect causal effects of BP medication on time-to-death by CVD, mediated by mean blood pressure (the unweighted average of SBP and DBP) among participants with hypertension at baseline in ARIC. Note that we restrict our estimands to individuals classified as hypertensive at baseline because our estimands are the effects of BP medication among those for whom BP medication is indicated.  The analysis includes individuals whose information is recorded in the study either with complete information or with monotone missingness due to death or dropout. We model the probability of each variable conditional on all preceding variables in the temporal order as given in Section 5.3, as well as on baseline covariates.  

BP medications can impact time-to-death by CVD as a result of a change in smoking status and/or decreased mean blood pressure (MBP). The difference, $P[T_i(\textbf{z}^{t_{j-1}},  \mathbfcal{G}_{z}^{t_{j-1}} ) > t_{j}] - P[T_i(\textbf{z}^{t_{j-1}},  \mathbfcal{G}_{z_{*}}^{t_{j-1}}) > t_{j}]$, captures the indirect effect (on time to CVD death) of BP medications due to increased/decreased MBP, but not due to the medications' effect on MBP as a result of continued/discontinued smoking status. In other words, this \textbf{indirect effect} only captures the paths from exposures into mediators, but not from exposures into confounders into mediators. On the other hand, the difference, $P[T_i(\textbf{z}^{t_{j-1}}, \mathbfcal{G}_{z_{*}}^{t_{j-1}} ) > t_{j}] - P[T_i(\textbf{z}_{*}^{t_{j-1}},  \mathbfcal{G}_{z_{*}}^{t_{j-1}} ) > t_{j}]$, captures the effect of BP medications on time to CVD death due to increased/decreased MBP as a result of smoking status. In other words, the \textbf{direct effect} captures the paths from exposure to the outcome, both from exposure into confounder into outcome, and from exposure into confounder into mediator into outcome. These two effects provide a decomposition of the \textbf{total effect} $P[T_i(\textbf{z}^{t_{j-1}},  \mathbfcal{G}_{z}^{t_{j-1}} ) > t_{j}] - P[T_i(\textbf{z}_{*}^{t_{j-1}},  \mathbfcal{G}_{z_{*}}^{t_{j-1}} ) > t_{j}]$.

\subsection{Results}\label{sec71}
Table \ref{tableone} and Table \ref{tabletwo} summarize the results of our analysis in the absence and presence of a competing event, respectively. These results are based on our analysis of the first three exams of the ARIC dataset. We calculated the causal effects of two contrasting exposure regimes, $\textbf{Z}_{i}^{t_j} = \{1,1\}$ (indicating medication at visits 1 and 2) and $\textbf{Z}_{i*}^{t_j} = \{0,0\}$ (indicating no medication at visits 1 and 2), on survival from death by CVD. Recall that the total causal effect at each visit in Table \ref{tableone} is the overall causal impact of BP medications on the survival status indicator corresponding to time-to-death by CVD, taking into account the mediation pathway through MBP measurements, as well as any confounding effect of the individual's smoking status up to that visit. A positive total causal effect (in Table \ref{tableone}) indicates that taking BP medication has a beneficial impact, increasing the survival probability for time-to-death by CVD. If medication is effective, we would expect all the causal effects in Table \ref{tableone} to be positive, and the effects in Table \ref{tabletwo} to be negative (i.e., medication to result in better survival outcomes). Table \ref{tableone} shows that the causal effects at visit 3 are positive, indicating improved survival outcomes for individuals on medication among baseline hypertensive individuals in ARIC. However, the Bayesian credible intervals suggest that none of these effects are significant. Additionally, we observe that the point estimates in Table \ref{tableone} remain relatively stable across the baseline age groups. Table \ref{tabletwo} indicates that the indirect effect point estimates for the main event at visits 2 and 3 are negative, as expected. The results in the opposite direction of our expectation may be related to the short duration on BP medication. Similarly, the lack of significance may also be related to short follow-up and/or a small number of events (death). The effects on total CVD (fatal and non-fatal) can be quick but the effect on death alone takes longer to be observed. Tables 7-10 (in Section 3 of the supplementary material) present the results separately by men and women. Consistent with the findings in Tables \ref{tableone} and \ref{tabletwo}, the results for baseline hypertensive men and women do not provide significant evidence of the effect of BP medication in improving survival probability, either in the presence or absence of a competing event.
\begin{center}
\begin{table}[!h]
\caption{Posterior means and 95 \% Bayesian credible intervals of causal effects in the absence of a competing event. The reported results indicate the differences in survival probabilities between (hypertensive at baseline) participants who were on BP medication and those who were not on BP medication in ARIC.\label{tableone}}
\begin{tabular}{llccccc}
\toprule
 \multicolumn{4}{c}{\bfseries \normalsize Hypertensive at baseline population in ARIC}  \\
\cline{1-4}
&  &   Visit 2 & Visit 3  \\
\cline{1-4}
\hline 
Age 45-49       & $IDE$                        &  -0.0053(-0.068, 0.045)             & 0.0084(-0.093, 0.13)             \\
at baseline        & $IIE$                        &  0.0005(-0.0032, 0.0071)             & 0.0014(-0.012, 0.016)              \\
& $TE$              &  -0.0049(-0.067, 0.046)             & 0.0098(-0.096, 0.13)             \\
\hline
Age 50-54       & $IDE$                        &  -0.0051(-0.064, 0.043)             & 0.0080(-0.090, 0.12)              \\
at baseline        & $IIE$                        &  0.0004(-0.0037, 0.0075)             & 0.0012(-0.012, 0.015)             \\
& $TE$              &  -0.0047(-0.064, 0.043)             & 0.0093(-0.091, 0.12)             \\
\hline
Age 55-59        & $IDE$                        &  -0.0051(-0.062, 0.041)             & 0.0077(-0.089, 0.12)              \\
at baseline        & $IIE$                        &  0.0004(-0.0041, 0.0064)             & 0.0013(-0.010, 0.016)              \\
& $TE$              &  -0.0047(-0.063, 0.042)             & 0.0090(-0.089, 0.12)            \\
\hline
Age 60-64        & $IDE$                        &  -0.0052(-0.060, 0.039)             & 0.0066(-0.089, 0.11)             \\
at baseline        & $IIE$                        &  0.0004(-0.0040, 0.0067)             & 0.0013(-0.010, 0.015)             \\
& $TE$              &  -0.0048(-0.058, 0.040)             & 0.0079(-0.085, 0.12)            \\
\hline
\end{tabular}
\begin{tablenotes}%
\item \textbf{Note:} $IDE$ = Interventional Direct Effects, $IIE$ = Interventional Indirect Effects, and $TE$ = Total Effects.\vspace*{6pt}
\end{tablenotes}
\end{table}
\end{center}

\begin{center}
\begin{table}[!h]
\caption{Posterior means and 95 \% Bayesian credible intervals of causal effects in the presence of a  competing event. The reported results indicate the differences in potential cumulative incidence functions (CIFs) between (hypertensive at baseline) participants who were on BP medication and those who were not on BP medication in ARIC.\label{tabletwo}}
\begin{tabular}{llccccc}
\toprule
 \multicolumn{4}{c}{\bfseries \normalsize Hypertensive at baseline population in ARIC}  \\
\cline{1-4}
&  &   Visit 2 & Visit 3  \\
\cline{1-4}
\hline 
Age 45-49        & $IDE^{(1)}$                     &  0.0047(-0.040, 0.063)             & 0.014(-0.028, 0.067)              \\
at baseline        & $IIE^{(1)}$                    &  -0.0011(-0.011, 0.0047)             & -0.0009(-0.010, 0.0054)              \\
& $TE^{(1)}$           &  0.0036(-0.041, 0.064)             & 0.013(-0.032, 0.067)              \\
\hline
 & $IDE^{(2)}$                    &  0.0017(-0.016, 0.026)             & 0.011(-0.016, 0.044)              \\
 & $IIE^{(2)}$                    &  0.0000(-0.0042, 0.0041)             & 0.0002(-0.0063, 0.0061)              \\
& $TE^{(2)}$          &  0.0017(-0.015, 0.026)             & 0.011(-0.016, 0.045)              \\
\hline
\hline
Age 50-54        & $IDE^{(1)}$                     &  0.0046(-0.038, 0.060)             & 0.014(-0.029, 0.064)              \\
at baseline        & $IIE^{(1)}$                   &  -0.0011(-0.011, 0.0043)             & -0.0010(-0.011, 0.0059)              \\
          & $TE^{(1)}$           &  0.0035(-0.040, 0.061)             & 0.013(-0.030, 0.063)              \\
\hline
& $IDE^{(2)}$                    &  0.0019(-0.015, 0.026)             & 0.011(-0.015, 0.043)              \\
& $IIE^{(2)}$                    &  0.0000(-0.0049, 0.0051)             & 0.0001(-0.0078, 0.0070)              \\ 
& $TE^{(2)}$           &  0.0019(-0.014, 0.027)             & 0.011(-0.017, 0.043)              \\
\hline
\hline
Age 55-59        & $IDE^{(1)}$                     &  0.0046(-0.037, 0.059)             & 0.014(-0.025, 0.062)              \\
at baseline        & $IIE^{(1)}$                    &  -0.0011(-0.011, 0.0045)             & -0.0009(-0.011, 0.0063)              \\
& $TE^{(1)}$           &  0.0035(-0.037, 0.059)             & 0.013(-0.027, 0.063)              \\
\hline
& $IDE^{(2)}$                    &  0.0020(-0.014, 0.026)             & 0.011(-0.014, 0.043)              \\
& $IIE^{(2)}$                    &  0.0000(-0.0049, 0.0044)             & 0.0000(-0.0071, 0.0069)              \\
& $TE^{(2)}$           &  0.0020(-0.014, 0.026)             & 0.011(-0.015, 0.044)              \\
\hline
\hline
  Age 60-64        & $IDE^{(1)}$                     &  0.0045(-0.035, 0.056)             & 0.014(-0.024, 0.062)              \\
 at baseline        & $IIE^{(1)}$                    &  -0.0011(-0.011, 0.0050)             & -0.0010(-0.011, 0.0063)              \\
         & $TE^{(1)}$           &  0.0034(-0.037, 0.055)             & 0.013(-0.025, 0.062)              \\
\hline
& $IDE^{(2)}$                    &  0.0020(-0.014, 0.026)             & 0.011(-0.016, 0.042)              \\
& $IIE^{(2)}$                    &  0.0000(-0.0050, 0.0051)             & 0.0002(-0.0073, 0.0077)              \\
& $TE^{(2)}$           &  0.0020(-0.014, 0.027)             & 0.011(-0.014, 0.044)              \\
\hline
\end{tabular}
\begin{tablenotes}%
\item \textbf{Note:} $IDE^{(1)}$ and $IDE^{(2)}$ represent direct effects on the main event and the competing event, respectively. We similarly define the indirect effects ($IIE^{(1)}$, $IIE^{(2)}$) and the total effects  ($TE^{(1)}$, $TE^{(2)}$) for both events.\vspace*{6pt}
\end{tablenotes}
\end{table}
\end{center}

\section{An approach to address unmeasured confounding using external data}\label{sec8}
In Tables \ref{tableone} and \ref{tabletwo}, we observe that certain causal effects in our analysis are in the opposite direction of our expectation. More precisely, some causal effects are negative when there is no competing event, and some are positive in the presence of a competing event, suggesting that taking BP medication may have detrimental effects on survival outcomes. We suspect that the reason for obtaining such results may be that the strong ignorability assumptions are not satisfied in our application. Previous work \citep{whelton20182017} has shown that the history of uncontrolled blood pressure serves as a confounder for individuals who are already on BP medications at the start of the study. However, this history (or the absence of it) is not measured in ARIC as ARIC participants are already aged 45-64 at baseline. In this section, we develop an approach to estimate the past history of MBP, referred to as predicted cumulative MBP, for those in ARIC using data from a different cohort study. 

\subsection{Methods for imputing cumulative MBP (unmeasured confounder)}\label{sec81}
In this subsection, we briefly explain how we estimate the unmeasured confounder cumulative mean blood pressure (CMBP) for individuals in ARIC. For individual $i$, we define the CMBP for the 20 years up to baseline age as: $CMBP_i = \int_{a_{i1}^{ARIC} -20}^{a_{i1}^{ARIC}} MBP(a_{ij})da_{ij}$. The idea is to use a mixed-effects model for mean blood pressure (MBP) based on data from the CARDIA cohort study \citep{friedman1988cardia} and utilize the model's coefficients to predict the cumulative MBP for participants in ARIC. In the CARDIA study, participants aged 18-30 at baseline were followed for 35 years, providing adequate information on blood pressure trends in young adulthood. Similar to ARIC, we use a
de-identified version of CARDIA from the NHLBI BioLINNC repository. Note that we cannot use CARDIA for our primary analysis as the study has very few events.

First, in CARDIA, we fit a mixed-effects model of the form:
\begin{equation}
    \begin{aligned}
        MBP(a_{ij}) &= \beta_0 + \beta_1 RACE_{i} + \beta_2 SEX_{i} +  \beta_3 BMI_{i} + \beta_4  a_{ij}  +  \beta_5 a_{ij}^2 +  \\
        & \beta_6 RACE_{i} * a_{ij} + \beta_7 SEX_{i} * a_{ij} + \beta_8 BMI_{i} * a_{ij} + b_{0i} +  b_{1i} a_{ij} + \epsilon_{ij}
    \end{aligned}
\end{equation}
where the pair $(b_{0i},b_{1i})$ represents the collection of random effects, and $a_{ij}$ denotes the age of individual $i$ at visit $j$. An implicit assumption for using coefficients from the CARDIA model above for prediction in ARIC is that the coefficients are transportable. The transportability assumption is formally stated as: 
\begin{assumption}\label{assump7}
    \textbf{Transportability} \\
    For two cohort studies, $cohort_1$ and $cohort_2$, the predicted (CMBP) is transportable given baseline covariates $\underline{L}_i^0$ (here, race and sex), that is,
    $P_{cohort_1}(CMBP_i|\underline{L}^0_i) =  P_{cohort_2}(CMBP_i|\underline{L}^0_i).$
\end{assumption} 
Since the random effects in the model are individual-specific in CARDIA, we use the following steps to assign each ARIC participant $i$ with one pair $(b_{0i},b_{1i})$:
\begin{enumerate}
    \item For each participant \(i\) in ARIC, we identify all matches (individuals) in CARDIA with the same features \((RACE_{i}, SEX_{i})\). We index these matches as \(\{cp_{i,1}, cp_{i,2}, \ldots, cp_{i,n_i}\}\), where \(cp\) abbreviates CARDIA participants, and \(n_i\) denotes the number of matches in CARDIA for individual \(i\) in ARIC.

    \item As each of the \(n_i\) matches has its corresponding \(n_i\) random slopes and intercepts, we randomly sample one pair of random effects, denoted as \((b_{0, cp_i}^{*},b_{1, cp_i}^{*})\), from the \(n_i\) pairs, $\{(b_{0, cp_{i,1}},b_{1, cp_{i,1}}), \ldots, (b_{0, cp_{i,n_i}},b_{1, cp_{i,n_i}})\}$, and set \((b_{0i},b_{1i}) = (b_{0, cp_i}^{*},b_{1, cp_i}^{*})\).

    \item For each participant \(i\) in ARIC, we use the randomly sampled pair \((b_{0i},b_{1i})\) and the fixed-effects coefficients from the model in (24) to compute their CMBP.
\end{enumerate}
Finally, using Assumption \ref{assump7} (transportability between ARIC and CARDIA cohorts given baseline covariates $RACE_i,\text{ and } SEX_i$), we include $CMBP_i$ as a baseline confounder. 

\subsection{Results including CMBP (unmeasured confounder)}\label{sec82}

\begin{center}
\begin{table}[!h]
\caption{Posterior means and 95 \% Bayesian credible intervals of causal effects in the absence of a competing event. The reported results indicate the differences in survival probabilities including CMBP as a baseline confounder between (hypertensive at baseline) participants who were on BP medication and those who were not on BP medication in ARIC. \label{tablethree}}
\begin{tabular}{llccccc}
\toprule
 \multicolumn{4}{c}{\bfseries \normalsize Hypertensive at baseline population in ARIC}  \\
\cline{1-4}
&  &   Visit 2 & Visit 3  \\
\cline{1-4}
\hline 
Age 45-49       & $IDE$                        & -0.0077(-0.071, 0.042)             & 0.0014(-0.095, 0.12)             \\
 at baseline        & $IIE$                        &  0.0004(-0.0036, 0.0059)             & 0.0007(-0.012, 0.015)             \\
           & $TE$              &  -0.0073(-0.071, 0.043)             & 0.0021(-0.093, 0.11)            \\
\hline
 Age 50-54       & $IDE$                        &  -0.0074(-0.067, 0.039)             & 0.0011(-0.091, 0.11)            \\
 at baseline        & $IIE$                        &  0.0003(-0.0036, 0.0062)             & 0.0006(-0.011, 0.015)             \\
           & $TE$              &  -0.0071(-0.067, 0.040)             & 0.0017(-0.089, 0.11)            \\
\hline
 Age 55-59        & $IDE$                        &  -0.0076(-0.065, 0.039)             & 0.0001(-0.087, 0.10)            \\
 at baseline        & $IIE$                        &  0.0004(-0.0036, 0.0069)             & 0.0010(-0.011, 0.016)             \\
           & $TE$              &  -0.0072(-0.066, 0.039)             & 0.0011(-0.090, 0.10)           \\
\hline
 Age 60-64        & $IDE$                        &  -0.0074(-0.062, 0.036)             & -0.0001(-0.089, 0.10)             \\
 at baseline        & $IIE$                        &  0.0003(-0.0043, 0.0066)             & 0.0008(-0.013, 0.015)             \\
           & $TE$              &  -0.0072(-0.061, 0.036)             & 0.0007(-0.088, 0.10)             \\

\hline
\end{tabular}
\begin{tablenotes}%
\item \textbf{Note:} $IDE$ = Interventional Direct Effects, $IIE$ = Interventional Indirect Effects, and $TE$ = Total Effects.\vspace*{6pt}
\end{tablenotes}
\end{table}
\end{center}

\begin{center}
\begin{table}[!h]
\caption{Posterior means and 95 \% Bayesian credible intervals of causal effects in the presence of a competing event. The reported results indicate the differences in potential cumulative incidence functions (CIFs) including CMBP as a baseline confounder between (hypertensive at baseline) participants who were on BP medication and those who were not on BP medication in ARIC. \label{tablefour}}
\begin{tabular}{llccccc}
\toprule
 \multicolumn{4}{c}{\bfseries \normalsize Hypertensive at baseline population in ARIC}  \\
\cline{1-4}
&  &   Visit 2 & Visit 3  \\
\cline{1-4}
\hline 
Age 45-49        & $IDE^{(1)}$                     &  0.0083(-0.055, 0.080)             & 0.018(-0.034, 0.05)              \\
 at baseline        & $IIE^{(1)}$                     &  -0.0006(-0.0087, 0.0053)             & -0.0001(-0.0082, 0.0066)              \\
& $TE^{(1)}$           &  0.0077(-0.056, 0.076)             & 0.018(-0.036, 0.083)              \\
\hline
& $IDE^{(2)}$                    &  0.0022(-0.013, 0.024)             & 0.0097(-0.017, 0.041)              \\
& $IIE^{(2)}$                    &  -0.0001(-0.0036, 0.0032)             & 0.0000(-0.0047, 0.0044)              \\
 & $TE^{(2)}$           &  0.0021(-0.013, 0.025)             & 0.0097(-0.017, 0.040)              \\
\hline
\hline
 Age 50-54        & $IDE^{(1)}$                     &  0.0081(-0.049, 0.077)             & 0.018(-0.033, 0.082)              \\
 at baseline        & $IIE^{(1)}$                    &  -0.0006(-0.0090, 0.0048)             & -0.0002(-0.0084, 0.0069)              \\
& $TE^{(1)}$           &  0.0075(-0.050, 0.075)             & 0.018(-0.034, 0.081)              \\
\hline
& $IDE^{(2)}$                    &  0.0026(-0.013, 0.025)             & 0.0095(-0.016, 0.040)              \\
& $IIE^{(2)}$                    &  0.0000(-0.0037, 0.0043)             & 0.0002(-0.0048, 0.0060)              \\
& $TE^{(2)}$           &  0.0026(-0.013, 0.025)             & 0.0097(-0.016, 0.041)              \\
\hline
\hline
 Age 55-59        & $IDE^{(1)}$                     &  0.0081(-0.048, 0.074)             & 0.018(-0.034, 0.078)              \\
 at baseline        & $IIE^{(1)}$                    &  -0.0005(-0.0088, 0.0055)             & -0.0001(-0.0088, 0.0076)              \\
& $TE^{(1)}$           &  0.0076(-0.051, 0.072)             & 0.018(-0.034, 0.076)              \\
\hline
& $IDE^{(2)}$                    &  0.0027(-0.012, 0.025)             & 0.0095(-0.016, 0.040)              \\
& $IIE^{(2)}$                    &  -0.0002(-0.0045, 0.0039)             & -0.0001(-0.0055, 0.0053)              \\ 
& $TE^{(2)}$           &  0.0025(-0.013, 0.025)             & 0.0094(-0.016, 0.039)              \\
\hline
\hline
  Age 60-64        & $IDE^{(1)}$                     &  0.0081(-0.042, 0.072)             & 0.017(-0.030, 0.076)              \\
 at baseline        & $IIE^{(1)}$                    &  -0.0006(-0.0088, 0.0053)             & -0.0003(-0.0099, 0.0070)              \\
& $TE^{(1)}$           &  0.0075(-0.045, 0.070)             & 0.017(-0.033, 0.073)              \\
\hline
& $IDE^{(2)}$                    &  0.0030(-0.012, 0.026)             & 0.0094(-0.015, 0.038)              \\
& $IIE^{(2)}$                   &  0.0000(-0.0047, 0.0044)             & 0.0001(-0.0058, 0.0058)              \\
& $TE^{(2)}$           &  0.0030(-0.012, 0.026)             & 0.0095(-0.015, 0.038)              \\
\hline
\end{tabular}
\begin{tablenotes}%
\item \textbf{Note:} $IDE^{(1)}$ and $IDE^{(2)}$ represent direct effects on the main event and the competing event, respectively. We similarly define the indirect effects ($IIE^{(1)}$, $IIE^{(2)}$) and the total effects  ($TE^{(1)}$, $TE^{(2)}$) for both events.\vspace*{6pt}
\end{tablenotes}
\end{table}
\end{center}

Tables \ref{tablethree} and \ref{tablefour} summarize the results of the analysis after including $CMBP_i$ as a baseline confounder. When compared to the results in Tables \ref{tableone} and \ref{tabletwo}, the causal effects in Tables \ref{tablethree} and \ref{tablefour} do not show any significant changes after adding predicted CMBP as a baseline confounder. The effect estimates remain insignificant, with no impact on our previous conclusions. Thus, assuming Assumption \ref{assump7} holds, the results are not very sensitive to this known, potential unmeasured confounder. As such, we do not pursue incorporating uncertainty in estimating CMBP or more complex models in our analysis.

\section{Discussion}\label{sec9}
This paper proposes a random intervention approach to causal mediation analysis using (semi-parametric) BART models for longitudinal and survival data and introduces an extension that allows for the presence of a competing event. A similar data structure, using random interventions for the mediator, has been implemented in recent (unpublished) work by \cite{wang2023targeted}  using Targeted Maximum Likelihood Estimation (TMLE). Some advantages of our proposed framework are: i) computational efficiency, allowing parallelization during both (BART) model fitting and G-computation, ii) semi-parametric models for observed continuous random variables, and (essentially) non-parametric models for binary random variables, and iii) a novel approach to address a known, unmeasured confounder. Additionally, we provide the R code to implement our methods (https://github.com/SBstats/DiscreteTimeMediationBART.git).

Our results provide weak support for the notion that BP medication has a beneficial impact on time-to-death by CVD through the reduction of MBP during short-term follow-up. The results are mostly non-significant. For baseline hypertensive women, we observe more evidence of positive direct and indirect effects on survival from time-to-death by CVD, both in the presence and absence of a competing event. For baseline hypertensive men, however, we did not obtain evidence to assert a beneficial direct impact of the medication on time-to-death by CVD, even after including predicted cumulative MBP as an unmeasured baseline confounder. Our analysis relies on strong ignorability assumptions, which may be suspect given other potential unmeasured confounders besides CMBP, such as dietary factors, physical activity, and a family history of high BP, that were not available for this analysis. Both diet and exercise changes could well covary with the initiation of BP medication since the participants are often advised to make lifestyle modifications for these two. Additionally, we lack data on the duration of anti-hypertensive (HTN) therapy and/or adherence to it, which are factors that certainly influence the effectiveness of BP medication and the strength of its association with our outcome.  This is a limitation in pharmacoepidemiology studies.

In future work, we will further evaluate the sensitivity of inferences to the violation of ignorability assumptions. In a setting involving baseline covariates and cross-sectional treatment, mediator, and outcome, \cite{linero2022mediation} note that machine learning tools like BART can yield biased estimates of treatment effects in the presence of moderate to high dimensional covariates. They refer to this phenomenon as "prior dogmatism." In order to overcome the challenge of prior dogmatism, they propose to include the predicted mediator values conditional on baseline covariates---the so-called "clever covariates"---instead of the observed mediator values as predictors into the model of the outcome. Incorporating this idea into our methods could also be an interesting future direction. Similarly, extending our proposed methods to include multiple mediators will be a component of future work. Finally, a further extension will be handling missing data (particularly dropout). The current approach handles MAR dropout but not MNAR dropout. This will be further explored in future work.

\section{Competing interests}
No competing interest is declared.

\section{Acknowledgments}
Bhandari, Daniels, Lloyd-Jones, and Siddique were partially supported by NIH grant R01 HL158963. Josefsson was partially supported by Forte Dnr 2019-01064. Daniels was also partially supported by NIH grant R01 HL166324. The authors acknowledge Dr. Hongyan Ning for help with ARIC and CARDIA datasets.

\bibliographystyle{apalike}
\bibliography{references}

\begin{thebibliography}{}

\bibitem[Avin et~al., 2005]{avin2005identifiability}
Avin, C., Shpitser, I., and Pearl, J. (2005).
\newblock Identifiability of path-specific effects.

\bibitem[Bind et~al., 2016]{bind2016causal}
Bind, M.-A., Vanderweele, T., Coull, B., and Schwartz, J. (2016).
\newblock Causal mediation analysis for longitudinal data with exogenous exposure.
\newblock {\em Biostatistics}, 17(1):122--134.

\bibitem[Boatman et~al., 2021]{boatman2021borrowing}
Boatman, J.~A., Vock, D.~M., and Koopmeiners, J.~S. (2021).
\newblock Borrowing from supplemental sources to estimate causal effects from a primary data source.
\newblock {\em Statistics in Medicine}, 40(24):5115--5130.

\bibitem[Chen et~al., 2024]{chen2024bayesian}
Chen, X., Harhay, M.~O., Tong, G., and Li, F. (2024).
\newblock A bayesian machine learning approach for estimating heterogeneous survivor causal effects: applications to a critical care trial.
\newblock {\em The annals of applied statistics}, 18(1):350.

\bibitem[Chipman et~al., 2010]{chipman2010bart}
Chipman, H.~A., George, E.~I., and McCulloch, R.~E. (2010).
\newblock {BART}: Bayesian additive regression trees.
\newblock {\em The Annals of Applied Statistics}, 4(1):266--298.

\bibitem[Fine and Gray, 1999]{fine1999proportional}
Fine, J.~P. and Gray, R.~J. (1999).
\newblock A proportional hazards model for the subdistribution of a competing risk.
\newblock {\em Journal of the American statistical association}, 94(446):496--509.

\bibitem[Friedman et~al., 1988]{friedman1988cardia}
Friedman, G.~D., Cutter, G.~R., Donahue, R.~P., Hughes, G.~H., Hulley, S.~B., Jacobs~Jr, D.~R., Liu, K., and Savage, P.~J. (1988).
\newblock {CARDIA}: study design, recruitment, and some characteristics of the examined subjects.
\newblock {\em Journal of clinical epidemiology}, 41(11):1105--1116.

\bibitem[Hernan and Robins, 2024]{hernan2024causal}
Hernan, M. and Robins, J. (2024).
\newblock {\em Causal Inference: What If}.
\newblock Chapman \& Hall/CRC Monographs on Statistics \& Applied Probab. CRC Press.

\bibitem[Investigators, 1989]{aric1989atherosclerosis}
Investigators, A. (1989).
\newblock The atherosclerosis risk in communit ({ARIC}) study: design and objectives.
\newblock {\em American journal of epidemiology}, 129(4):687--702.

\bibitem[Josefsson, 2023]{maria_josefsson_2023_10277820}
Josefsson, M. (2023).
\newblock m4ryjo/{GcompBART}: {GcomBART}-v.1.0.0.

\bibitem[Josefsson and Daniels, 2021]{josefsson2021bayesian}
Josefsson, M. and Daniels, M.~J. (2021).
\newblock Bayesian semi-parametric {G}-computation for causal inference in a cohort study with {MNAR} dropout and death.
\newblock {\em Journal of the Royal Statistical Society. Series C, Applied statistics}, 70(2):398.

\bibitem[Kim et~al., 2019]{kim2019bayesian}
Kim, C., Zigler, C.~M., Daniels, M.~J., Choirat, C., and Roy, J.~A. (2019).
\newblock Bayesian longitudinal causal inference in the analysis of the public health impact of pollutant emissions.
\newblock {\em arXiv preprint arXiv:1901.00908}.

\bibitem[Lin et~al., 2017]{lin2017mediation}
Lin, S.-H., Young, J.~G., Logan, R., and VanderWeele, T.~J. (2017).
\newblock Mediation analysis for a survival outcome with time-varying exposures, mediators, and confounders.
\newblock {\em Statistics in medicine}, 36(26):4153--4166.

\bibitem[Linero and Zhang, 2022]{linero2022mediation}
Linero, A.~R. and Zhang, Q. (2022).
\newblock Mediation analysis using {B}ayesian tree ensembles.
\newblock {\em Psychological Methods}.

\bibitem[Oganisian et~al., 2024]{oganisian2024hierarchical}
Oganisian, A., Mitra, N., and Roy, J.~A. (2024).
\newblock Hierarchical bayesian bootstrap for heterogeneous treatment effect estimation.
\newblock {\em The International Journal of Biostatistics}, 20(1):93--106.

\bibitem[Roy et~al., 2017]{roy2017bayesian}
Roy, J., Lum, K.~J., and Daniels, M.~J. (2017).
\newblock A bayesian nonparametric approach to marginal structural models for point treatments and a continuous or survival outcome.
\newblock {\em Biostatistics}, 18(1):32--47.

\bibitem[Roy et~al., 2024]{roy2024bayesian}
Roy, S., Daniels, M.~J., and Roy, J. (2024).
\newblock A bayesian nonparametric approach for multiple mediators with applications in mental health studies.
\newblock {\em Biostatistics}, page kxad038.

\bibitem[Rubin, 1981]{rubin1981bayesian}
Rubin, D.~B. (1981).
\newblock The {B}ayesian bootstrap.
\newblock {\em The annals of statistics}, pages 130--134.

\bibitem[Sparapani et~al., 2020]{sparapani2020nonparametric}
Sparapani, R., Logan, B.~R., McCulloch, R.~E., and Laud, P.~W. (2020).
\newblock Nonparametric competing risks analysis using {B}ayesian additive regression trees.
\newblock {\em Statistical methods in medical research}, 29(1):57--77.

\bibitem[Sparapani et~al., 2019]{sparapani2019bart}
Sparapani, R., Spanbauer, C., and McCulloch, R. (2019).
\newblock The {BART} {R} package.
\newblock {\em Accessed on Aug}, 21:2019.

\bibitem[Sparapani et~al., 2016]{sparapani2016nonparametric}
Sparapani, R.~A., Logan, B.~R., McCulloch, R.~E., and Laud, P.~W. (2016).
\newblock Nonparametric survival analysis using {B}ayesian additive regression trees ({BART}).
\newblock {\em Statistics in medicine}, 35(16):2741--2753.

\bibitem[Taddy et~al., 2016]{taddy2016nonparametric}
Taddy, M., Gardner, M., Chen, L., and Draper, D. (2016).
\newblock A nonparametric bayesian analysis of heterogenous treatment effects in digital experimentation.
\newblock {\em Journal of Business \& Economic Statistics}, 34(4):661--672.

\bibitem[Tan and Roy, 2019]{tan2019bayesian}
Tan, Y.~V. and Roy, J. (2019).
\newblock Bayesian additive regression trees and the general {BART} model.
\newblock {\em Statistics in medicine}, 38(25):5048--5069.

\bibitem[VanderWeele and Tchetgen~Tchetgen, 2017]{vanderweele2017mediation}
VanderWeele, T.~J. and Tchetgen~Tchetgen, E.~J. (2017).
\newblock Mediation analysis with time varying exposures and mediators.
\newblock {\em Journal of the Royal Statistical Society: Series B (Statistical Methodology)}, 79(3):917--938.

\bibitem[Vansteelandt et~al., 2019]{vansteelandt2019mediation}
Vansteelandt, S., Linder, M., Vandenberghe, S., Steen, J., and Madsen, J. (2019).
\newblock Mediation analysis of time-to-event endpoints accounting for repeatedly measured mediators subject to time-varying confounding.
\newblock {\em Statistics in medicine}, 38(24):4828--4840.

\bibitem[Vo et~al., 2022]{vo2022longitudinal}
Vo, T.-T., Davies-Kershaw, H., Hackett, R., and Vansteelandt, S. (2022).
\newblock Longitudinal mediation analysis of time-to-event endpoints in the presence of competing risks.
\newblock {\em Lifetime Data Analysis}, pages 1--21.

\bibitem[Wang et~al., 2023]{wang2023targeted}
Wang, Z., van~der Laan, L., Petersen, M., Gerds, T., Kvist, K., and van~der Laan, M. (2023).
\newblock Targeted maximum likelihood based estimation for longitudinal mediation analysis.
\newblock {\em arXiv preprint arXiv:2304.04904}.

\bibitem[Whelton et~al., 2018]{whelton20182017}
Whelton, P.~K., Carey, R.~M., Aronow, W.~S., Casey, D.~E., Collins, K.~J., Dennison~Himmelfarb, C., DePalma, S.~M., Gidding, S., Jamerson, K.~A., Jones, D.~W., et~al. (2018).
\newblock 2017 acc/aha/aapa/abc/acpm/ags/apha/ash/aspc/nma/pcna guideline for the prevention, detection, evaluation, and management of high blood pressure in adults: a report of the american college of cardiology/american heart association task force on clinical practice guidelines.
\newblock {\em Journal of the American College of Cardiology}, 71(19):e127--e248.

\bibitem[Young et~al., 2020]{young2020causal}
Young, J.~G., Stensrud, M.~J., Tchetgen~Tchetgen, E.~J., and Hern{\'a}n, M.~A. (2020).
\newblock A causal framework for classical statistical estimands in failure-time settings with competing events.
\newblock {\em Statistics in Medicine}, 39(8):1199--1236.

\bibitem[Zeng et~al., 2022]{zeng2022causal}
Zeng, S., Lange, E.~C., Archie, E.~A., Campos, F.~A., Alberts, S.~C., and Li, F. (2022).
\newblock A causal mediation model for longitudinal mediators and survival outcomes with an application to animal behavior.
\newblock {\em Journal of Agricultural, Biological and Environmental Statistics}, pages 1--22.

\bibitem[Zheng and van~der Laan, 2017]{zheng2017longitudinal}
Zheng, W. and van~der Laan, M. (2017).
\newblock Longitudinal mediation analysis with time-varying mediators and exposures, with application to survival outcomes.
\newblock {\em Journal of causal inference}, 5(2).

\bibitem[Zhou and Song, 2023]{zhou2023causal}
Zhou, X. and Song, X. (2023).
\newblock Causal mediation analysis for multivariate longitudinal data and survival outcomes.
\newblock {\em Structural Equation Modeling: A Multidisciplinary Journal}, 30(5):749--760.

\end{thebibliography}

\newpage

\begin{algorithm}
\caption{Algorithm  to sample from the posterior of $\mathcal{P}_{\textbf{z}_{1},\textbf{z}_{2}}(t_{j})$ where $(\textbf{z}_{1},\textbf{z}_{2}) \in \bigl\{ (\textbf{z},\textbf{z}_{*}),  (\textbf{z}_{*},\textbf{z}_{*}), (\textbf{z},\textbf{z})\bigr\}$:}\label{alg:cap}
\begin{enumerate}
            \item Fit the BART models for confounders $\textbf{L}^{t_{j}}$, mediators  $\textbf{M}^{t_{j}}$, and outcomes $\textbf{Y}^{t_{j}}$ as specified in Section 4.2.  For each model, burn in the first $B$ MCMC iterations and keep the next $Q$ iterations (i.e. keep iterations $B+1,\ldots, B+Q$). For each of the  $Q$ iterations, we have a sample from the posterior of observed data parameters $(\mathbfcal{T}^{L^{t_{j}},q}, \boldsymbol{\mu}^{L^{t_{j}},q},\sigma^{2}_{L,q})$, $(\mathbfcal{T}^{M^{t_{j}},q}, \boldsymbol{\mu}^{M^{t_{j}},q},\sigma^{2}_{M,q})$, and $(\mathbfcal{T}^{Y^{t_{j}},q}, \boldsymbol{\mu}^{Y^{t_{j}},q})$ for $L^{t_{j}}, M^{t_{j}}, \text{ and } Y^{t_{j}},$ respectively, with $j \in \{ 1, \ldots, J\}$.
            \item Compute a posterior sample for $\mathcal{P}_{\textbf{z}_{1},\textbf{z}_{2}}(t_{j})$ where $(\textbf{z}_{1},\textbf{z}_{2}) \in \bigl\{ (\textbf{z},\textbf{z}_{*}),  (\textbf{z}_{*},\textbf{z}_{*}),  (\textbf{z},\textbf{z})\bigr\}$ as follows:
                \begin{enumerate}
                    \item At the  $(B+q)^{th}$ iteration, randomly draw $C^{*}$ (row) vectors as samples for $\underline{L}^{0}$, say $\underline{\tilde{\ell}}^{0,c}$, $c \in \{1,\ldots, C^{*} \}$, using the following steps:
                            \begin{enumerate}
                                \item  Sample the Bayesian Bootstrap \citep{rubin1981bayesian} weights $$\boldsymbol{\omega}^{q} = (\omega_{1}^{q}, \ldots, \omega_{N}^{q}) \sim Dirichlet(\underbrace{1,\ldots,1}_{N \text{  times}}).$$
                                \item Generate $C^{*}$ indices by drawing  a sample from $ Multinomial( 1,\omega^{q})$ $C^{*}$ times. 
                                Use these indices to sample $C^{*}$ (row) vectors $\{\underline{\tilde{\ell}}^{0,c}\}_{c=1}^{C^{*}}$ from the collection $\{\underline{\tilde{\ell}}_{i}^{0}\}_{i=1}^{N}$.
                            \end{enumerate}
                    \item Sequentially repeat the following $C^{*}$ times for all $j = \{1,.., J-2 \}$: 
                        \begin{enumerate}
                            \item For a fixed regime $\textbf{Z}^{t_{j}} = \textbf{z}_{1}^{t_{j}} $, conditional on $\tilde{\textbf{y}}^{t_{j},c} = \mathbb{0}_{j}$, $\underline{\tilde{\boldsymbol{\ell}}}^{0,c}$,  $\tilde{\boldsymbol{\ell}}^{t_{j-1},c}$, and $\tilde{\textbf{m}}^{t_{j-1},c}$,  sample one observation of $L^{t_{j}}$, say $\tilde{\ell}^{t_{j},c}$, from: 
                            \begin{align*}
                                & N\Bigl(g_{L}( \textbf{z}_{1}^{t_{j}},\tilde{\boldsymbol{\ell}}^{t_{j-1},c}, \tilde{\textbf{m}}^{t_{j-1},c}, \underline{\tilde{\boldsymbol{\ell}}}^{0,c};\mathbfcal{T}^{L^{t_{j}},q}, \boldsymbol{\mu}^{L^{t_{j}},q}), \sigma^{2}_{L,q}\Bigr) \quad \text{if } L^{t_{j}} \text{ is continuous} \\
                                & Bern\Bigl(\Phi \bigl(g_{L}( \textbf{z}_{1}^{t_{j}},\tilde{\boldsymbol{\ell}}^{t_{j-1},c}, \tilde{\textbf{m}}^{t_{j-1},c}, \underline{\tilde{\boldsymbol{\ell}}}^{0,c};\mathbfcal{T}^{L^{t_{j}},q}, \boldsymbol{\mu}^{L^{t_{j}},q})\bigr)\Bigr)  \quad \text{if } L^{t_{j}} \text{ is binary.} 
                            \end{align*}
                            \item For a fixed regime $\textbf{Z}^{t_{j}} = \textbf{z}_{2}^{t_{j}} $, conditional on $\tilde{\textbf{y}}^{t_{j},c} = \mathbb{0}_{j}$,  $\underline{\tilde{\boldsymbol{\ell}}}^{0,c}$, $\tilde{\boldsymbol{\ell}}^{t_{j},c}$, and $\tilde{\textbf{m}}^{t_{j-1},c}$,   sample one observation of $M^{t_{j}}$, say $\tilde{m}^{t_{j},c}$, from:
                            \begin{align*}
                                &  N\Bigl(g_{M}( \textbf{z}_{2}^{t_{j}},\tilde{\boldsymbol{\ell}}^{t_{j},c}, \tilde{\textbf{m}}^{t_{j-1},c}, \underline{\tilde{\boldsymbol{\ell}}}^{0,c};\mathbfcal{T}^{M^{t_{j}},q}, \boldsymbol{\mu}^{M^{t_{j}},q}), \sigma^{2}_{M,q}\Bigr) \quad \text{if } M^{t_{j}} \text{ is continuous} \\
                                & Bern\Bigl(\Phi\bigl(g_{M}(  \textbf{z}_{2}^{t_{j}},\tilde{\boldsymbol{\ell}}^{t_{j},c}, \tilde{\textbf{m}}^{t_{j-1},c}, \underline{\tilde{\boldsymbol{\ell}}}^{0,c};\mathbfcal{T}^{M^{t_{j}},q}, \boldsymbol{\mu}^{M^{t_{j}},q})\bigr)\Bigr) \quad \text{if } M^{t_{j}} \text{ is binary.} 
                            \end{align*}
                            \item Compute $\mathcal{P}_{\textbf{z}_{1},\textbf{z}_{2}}^{ (B+q),c}(t_{j+1}) = \Phi\bigl(g_{Y}( \textbf{z}_{1}^{t_{j}},\tilde{\boldsymbol{\ell}}^{t_{j},c}, \tilde{\textbf{m}}^{t_{j},c}, \underline{\tilde{\boldsymbol{\ell}}}^{0,c};\mathbfcal{T}^{Y^{t_{j+1}},q}, \boldsymbol{\mu}^{Y^{t_{j+1}},q})\bigr)$.
                            \item For a fixed regime $\textbf{Z}^{t_{j}} = \textbf{z}_{1}^{t_{j}} $, conditional on $\tilde{\textbf{y}}^{t_{j},c} = \mathbb{0}_{j}$, $\underline{\tilde{\boldsymbol{\ell}}}^{0,c}$,  $\tilde{\boldsymbol{\ell}}^{t_{j},c}$, and $\tilde{\textbf{m}}^{t_{j},c}$,  sample one observation of $Y^{t_{j+1}}$, say $\tilde{y}^{t_{j+1},c}$, from:  
                            \begin{align*}
                               & Bern\Bigl(\mathcal{P}_{\textbf{z}_{1},\textbf{z}_{2}}^{ (B+q),c}(t_{j+1})\Bigr) .
                            \end{align*}
                        \end{enumerate}
                    \item At $j = J-1$, sequentially repeat Steps b. (i) - (iii)  $C^{*}$ times to compute 
    $$\mathcal{P}_{\textbf{z}_{1},\textbf{z}_{2}}^{ (B+q),c}(t_{J}) = \Phi\bigl(g_{Y}( \textbf{z}_{1}^{t_{J-1}},\tilde{\boldsymbol{\ell}}^{t_{J-1},c}, \tilde{\textbf{m}}^{t_{J-1},c}, \underline{\tilde{\boldsymbol{\ell}}}^{0,c};\mathbfcal{T}^{Y^{t_{J}},q}, \boldsymbol{\mu}^{Y^{t_{J}},q})\bigr)$$
                    \item For all $j \in {1, \ldots, J }$ Compute  $\mathcal{P}_{\textbf{z}_{1},\textbf{z}_{2}}^{ (B+q)}(t_{j})$ using $$\mathcal{P}_{\textbf{z}_{1},\textbf{z}_{2}}^{ (B+q)}(t_{j}) \approx \frac{1}{C^{*}} \sum_{c=1}^{C^{*}} \mathcal{P}_{\textbf{z}_{1},\textbf{z}_{2}}^{ (B+q),c}(t_{j}).$$
                \end{enumerate}

            \item Repeat Step 2 for each of the $Q$ posterior samples of the parameters from Step 1.
        \end{enumerate}
\end{algorithm}

\begin{algorithm}
\caption{Algorithm to sample from the posterior of $\mathcal{P}^{(k)}_{\textbf{z}_{1},\textbf{z}_{2}}(t_{j})$ where $(\textbf{z}_{1},\textbf{z}_{2}) \in \bigl\{ (\textbf{z},\textbf{z}_{*}),  (\textbf{z}_{*},\textbf{z}_{*}),  (\textbf{z},\textbf{z})\bigr\}$:}\label{alg:cap}
\begin{enumerate}
             \item Fit the BART models for confounders $\textbf{L}^{t_{j}}$, mediators  $\textbf{M}^{t_{j}}$, and outcomes $\textbf{Y}^{t_{j},(k)}, k \in \{1,2\},$ as specified in Section 4.2. For each model, burn in the first $B$ MCMC iterations and keep the next $Q$ iterations (i.e. keep iterations $B+1, \ldots, B+Q$). For each of the  $Q$ iterations, we have a sample from the posterior of observed data parameters $(\mathbfcal{T}^{L^{t_{j}},q}, \boldsymbol{\mu}^{L^{t_{j}},q},\sigma^{2}_{L,q})$, $(\mathbfcal{T}^{M^{t_{j}},q}, \boldsymbol{\mu}^{M^{t_{j}},q},\sigma^{2}_{M,q})$, and $(\mathbfcal{T}^{Y^{t_{j}},q}, \boldsymbol{\mu}^{Y^{t_{j},(k)},q})$ for $L^{t_{j}}, M^{t_{j}}, \text{ and } Y^{t_{j},(k)},$ respectively, with $j \in \{ 1, \ldots, J\}$. 
            \item Compute a posterior sample for $\mathcal{P}^{(k)}_{\textbf{z}_{1},\textbf{z}_{2}}(t_{j})$ where $k \in \{1,2\}$  and $(\textbf{z}_{1},\textbf{z}_{2}) \in \bigl\{ (\textbf{z},\textbf{z}_{*}),  (\textbf{z}_{*},\textbf{z}_{*}),  (\textbf{z},\textbf{z})\bigr\}$ as follows:
                \begin{enumerate}
                    \item At the  $(B+q)^{th}$ iteration, randomly draw $C^{*}$ (row) vectors as samples for $\underline{L}^{0}$, say $\underline{\tilde{\ell}}^{0,c}$, $c \in \{1,\ldots, C^{*} \}$, using the following steps:
                            \begin{enumerate}
                                \item  Sample the Bayesian Bootstrap \citep{rubin1981bayesian} weights $$\boldsymbol{\omega}^{q} = (\omega_{1}^{q}, \ldots, \omega_{N}^{q}) \sim Dirichlet(\underbrace{1,\ldots,1}_{N \text{  times}}).$$
                                \item Generate $C^{*}$ indices by drawing  a sample from $ Multinomial( 1,\omega^{q})$ $C^{*}$ times. 
                                Use these indices to sample $C^{*}$ (row) vectors $\{\underline{\tilde{\ell}}^{0,c}\}_{c=1}^{C^{*}}$ from the collection $\{\underline{\tilde{\ell}}_{i}^{0}\}_{i=1}^{N}$.
                            \end{enumerate}
                    \item Sequentially repeat the following $C^{*}$ times for all $j = \{1,.., J-2 \}$:    
                        \begin{enumerate}
                            \item For a fixed regime $\textbf{Z}^{t_{j}} = \textbf{z}_{1}^{t_{j}} $, conditional on $\tilde{\textbf{y}}^{t_{j},(1),c} =
                        \tilde{\textbf{y}}^{t_{j},(2),c} =\mathbb{0}_{j}$, $\underline{\tilde{\boldsymbol{\ell}}}^{0,c}$,  $\tilde{\boldsymbol{\ell}}^{t_{j-1},c}$, and $\tilde{\textbf{m}}^{t_{j-1},c}$,   sample one observation of $L^{t_{j}}$, say $\tilde{\ell}^{t_{j},c}$, from: 
                            \begin{align*}
                                & N\Bigl(g_{L}( \textbf{z}_{1}^{t_{j}},\tilde{\boldsymbol{\ell}}^{t_{j-1},c}, \tilde{\textbf{m}}^{t_{j-1},c}, \underline{\tilde{\boldsymbol{\ell}}}^{0,c};\mathbfcal{T}^{L^{t_{j}},q}, \boldsymbol{\mu}^{L^{t_{j}},q}), \sigma^{2}_{L,q}\Bigr) \quad \text{if } L^{t_{j}} \text{ is continuous} \\
                                & Bern\Bigl(\Phi \bigl(g_{L}( \textbf{z}_{1}^{t_{j}},\tilde{\boldsymbol{\ell}}^{t_{j-1},c}, \tilde{\textbf{m}}^{t_{j-1},c}, \underline{\tilde{\boldsymbol{\ell}}}^{0,c};\mathbfcal{T}^{L^{t_{j}},q}, \boldsymbol{\mu}^{L^{t_{j}},q})\bigr)\Bigr)  \quad \text{if } L^{t_{j}} \text{ is binary.} 
                            \end{align*}
                            \item For a fixed regime $\textbf{Z}^{t_{j}} = \textbf{z}_{2}^{t_{j}} $, conditional on $\tilde{\textbf{y}}^{t_{j},(1),c} =
                        \tilde{\textbf{y}}^{t_{j},(2),c} =\mathbb{0}_{j}$, $\underline{\tilde{\boldsymbol{\ell}}}^{0,c}$,  $\tilde{\boldsymbol{\ell}}^{t_{j},c}$, and $\tilde{\textbf{m}}^{t_{j-1},c}$,  sample one observation of $M^{t_{j}}$, say $\tilde{m}^{t_{j},c}$, from:
                            \begin{align*}
                                &  N\Bigl(g_{M}(  \textbf{z}_{2}^{t_{j}},\tilde{\boldsymbol{\ell}}^{t_{j},c}, \tilde{\textbf{m}}^{t_{j-1},c}, \underline{\tilde{\boldsymbol{\ell}}}^{0,c};\mathbfcal{T}^{M^{t_{j}},q}, \boldsymbol{\mu}^{M^{t_{j}},q}), \sigma^{2}_{M,q}\Bigr) \quad \text{if } M^{t_{j}} \text{ is continuous} \\
                                & Bern\Bigl(\Phi(g_{M}\bigl(\textbf{z}_{2}^{t_{j}},\tilde{\boldsymbol{\ell}}^{t_{j},c}, \tilde{\textbf{m}}^{t_{j-1},c}, \underline{\tilde{\boldsymbol{\ell}}}^{0,c};\mathbfcal{T}^{M^{t_{j}},q}, \boldsymbol{\mu}^{M^{t_{j}},q})\bigr)\Bigr) \quad \text{if } M^{t_{j}} \text{ is binary.} 
                            \end{align*}
                            \item  Compute $\mathcal{P}_{\textbf{z}_{1},\textbf{z}_{2}}^{ (1),(B+q),c}(t_{j+1}) = \Phi\bigl(g_{Y^{(1)}}(\textbf{z}_{1}^{t_{j}},\tilde{\boldsymbol{\ell}}^{t_{j},c}, \tilde{\textbf{m}}^{t_{j},c}, \underline{\tilde{\boldsymbol{\ell}}}^{0,c};\mathbfcal{T}^{Y^{t_{j+1},(1)},q}, \boldsymbol{\mu}^{Y^{t_{j+1},(1)},q})\bigr)$.
                                \item For a fixed regime $\textbf{Z}^{t_{j}} = \textbf{z}_{1}^{t_{j}} $, conditional on $\tilde{\textbf{y}}^{t_{j},(1),c} =
                        \tilde{\textbf{y}}^{t_{j},(2),c} =\mathbb{0}_{j}$, $\underline{\tilde{\boldsymbol{\ell}}}^{0,c}$,  $\tilde{\boldsymbol{\ell}}^{t_{j},c}$, and $\tilde{\textbf{m}}^{t_{j},c}$, sample one observation of $Y^{t_{j+1},(1)}$, say $\tilde{y}^{t_{j+1},(1),c}$, from:  
                            \begin{align*}
                               & Bern\Bigl(\mathcal{P}_{\textbf{z}_{1},\textbf{z}_{2}}^{ (1),(B+q),c}(t_{j+1})\Bigr) .
                            \end{align*}
                                \item For a fixed regime $\textbf{Z}^{t_{j}} = \textbf{z}_{1}^{t_{j}} $, conditional on $\tilde{\textbf{y}}^{t_{j+1},(1),c} =\mathbb{0}_{j+1}$, $\tilde{\textbf{y}}^{t_{j},(2),c} =\mathbb{0}_{j}$,  $\underline{\tilde{\boldsymbol{\ell}}}^{0,c}$,  $\tilde{\boldsymbol{\ell}}^{t_{j},c}$, and $\tilde{\textbf{m}}^{t_{j},c}$, compute $$\mathcal{P}_{\textbf{z}_{1},\textbf{z}_{2}}^{ (2),(B+q),c}(t_{j+1}) = \Phi\bigl(g_{Y}(\textbf{z}_{1}^{t_{j}},\tilde{\boldsymbol{\ell}}^{t_{j},c}, \tilde{\textbf{m}}^{t_{j},c}, \underline{\tilde{\boldsymbol{\ell}}}^{0,c};\mathbfcal{T}^{Y^{t_{j+1},(2)},q}, \boldsymbol{\mu}^{Y^{t_{j+1},(2)},q})\bigr).$$
                                \item For a fixed regime $\textbf{Z}^{t_{j}} = \textbf{z}_{1}^{t_{j}} $, conditional on $\tilde{\textbf{y}}^{t_{j+1},(1),c} =\mathbb{0}_{j+1}$  $\tilde{\textbf{y}}^{t_{j},(2),c} =\mathbb{0}_{j}$, $\underline{\tilde{\boldsymbol{\ell}}}^{0,c}$,  $\tilde{\boldsymbol{\ell}}^{t_{j},c}$, and $\tilde{\textbf{m}}^{t_{j},c}$, sample one observation of $Y^{t_{j+1},(1)}$, say $\tilde{y}^{t_{j+1},(1),c}$, from:  
                            \begin{align*}
                               & Bern\Bigl(\mathcal{P}_{\textbf{z}_{1},\textbf{z}_{2}}^{ (2),(B+q),c}(t_{j+1})\Bigr) .
                            \end{align*}
                \end{enumerate}
                    \item At $j = J-1$, sequentially repeat Steps b. (i) - (v) $C^{*}$ times to compute:
                    \begin{align*}
                        \mathcal{P}_{\textbf{z}_{1},\textbf{z}_{2}}^{ (1),(B+q),c}(t_{J}) &= \Phi\bigl(g_{Y^{(1)}}(\textbf{z}_{1}^{t_{J-1}},\tilde{\boldsymbol{\ell}}^{t_{J-1},c}, \tilde{\textbf{m}}^{t_{J-1},c}, \underline{\tilde{\boldsymbol{\ell}}}^{0,c};\mathbfcal{T}^{Y^{t_{J},(1)},q}, \boldsymbol{\mu}^{Y^{t_{J},(1)},q})\bigr)
                        \\
                        \mathcal{P}_{\textbf{z}_{1},\textbf{z}_{2}}^{ (2),(B+q),c}(t_{J}) &= \Phi\bigl(g_{Y}(\textbf{z}_{1}^{t_{J-1}},\tilde{\boldsymbol{\ell}}^{t_{J-1},c}, \tilde{\textbf{m}}^{t_{J-1},c}, \underline{\tilde{\boldsymbol{\ell}}}^{0,c};\mathbfcal{T}^{Y^{t_{J},(2)},q}, \boldsymbol{\mu}^{Y^{t_{J},(2)},q})\bigr)
                    \end{align*}
                    \item For all $j = \{1, \ldots J\}$ and $k = 1,2$ , compute $\mathcal{P}_{\textbf{z}_{1},\textbf{z}_{2}}^{ (k),(B+q)}(t_{j})$ using $$\mathcal{P}_{\textbf{z}_{1},\textbf{z}_{2}}^{ (k),(B+q)}(t_{j}) \approx \frac{1}{C^{*}} \sum_{c=1}^{C^{*}} \mathcal{P}_{\textbf{z}_{1},\textbf{z}_{2}}^{(k), (B+q),c}(t_{j}).$$ 
                \end{enumerate}
            \item Repeat Step 2 for each of the $Q$ posterior samples of the parameters from Step 1.
        \end{enumerate}
\end{algorithm}

\newpage
\appendix

\section{Supplementary Section 1: Proofs}
Recall that the random draw $M^{t_{j}} \sim \mathcal{G}_{z_{*}}^{t_{j}}(m^{t_{j}}| \boldsymbol{\ell}^{t_{j}},\textbf{m}^{t_{j-1}}, \underline{\boldsymbol{\ell}}^{0} )$ where:
\begin{equation}
    \begin{aligned}
        & \mathcal{G}_{z_{*}}^{t_{j}}(m^{t_{j}}|\boldsymbol{\ell}^{t_{j}},\textbf{m}^{t_{j-1}}, \underline{\boldsymbol{\ell}}^{0} )  \\
        &=  p(M^{t_{j}}(\textbf{z}^{t_{j}}_{*}) = m^{t_{j}}|\textbf{Y}^{t_{l}}(\textbf{z}^{t_{l}}_{*}) = \mathbb{0}_j,\textbf{L}^{t_{j}}(\textbf{z}^{t_{j}}_{*}) = \boldsymbol{\ell}^{t_{j}},\textbf{M}^{t_{j-1}}(\textbf{z}^{t_{j-1}}_{*}) = \textbf{m}^{t_{j-1}},\underline{\textbf{L}}^{0} =  \underline{\boldsymbol{\ell}}^{0}) 
    \end{aligned} 
\end{equation}
denotes the conditional probability of mediators for $j \in \{1,\ldots,J \}$ in the world where the exposure is set to $\textbf{Z}^{t_{j}} = \textbf{z}_{*}^{t_{j}}$. In an ideal experiment, the data would be generated as follows: 
\begin{itemize}
    \item \textbf{At baseline:} Measure the covariates $\underline{\textbf{L}}^{0} = \underline{\boldsymbol{\ell}}^{0}$.
    \item  \textbf{At $t_{1}$:} Since all participants are alive at $t_{1}$, set $\mathcal{Y}_{z}^{t_{1}} = Y^{t_{1}} (z^{t_{0}},\mathcal{G}_{z_{*}}^{t_{0}} ) = Y^{t_{1}}= y^{t_{1}} (\text{say}) =   0$. Intervene on the exposure to set $Z^{t_{1}} = z^{t_{1}}$. Then, sequentially,  measure the covariate $\mathcal{L}_{z}^{t_{1}} = L^{t_{1}}(z^{t_{1}},\mathcal{G}_{z_{*}}^{t_{0}} ) = L^{t_{1}}(z^{t_{1}})  $ (say $\ell^{t_{1}}$) and intervene to randomly draw $M^{t_{1}} \sim \mathcal{G}_{z_{*}}^{t_{1}}(.|\ell^{t_{1}},\underline{\boldsymbol{\ell}}^{0})$ (say $m^{t_{1}}$). 
    \item  \textbf{At $t_{2}$:} Measure the outcome variable $\mathcal{Y}_{z}^{t_{2}} = Y^{t_{2}}(z^{t_{1}},\mathcal{G}_{z_{*}}^{t_{1}} )$ (say $y^{t_{2}}$). Intervene on the exposure to set $Z^{t_{2}} = z^{t_{2}}$. Then, sequentially, measure the covariate $\mathcal{L}_{z}^{t_{2}} = L^{t_{2}}(\textbf{z}^{t_{2}},\mathcal{\textbf{G}}_{z_{*}}^{t_{1}} )$ (say $\ell^{t_{2}}$) and intervene to randomly draw $M^{t_{2}} \sim \mathcal{G}_{z_{*}}^{t_{2}}(.| \ell^{t_{2}},m^{t_{1}},\underline{\boldsymbol{\ell}}^{0})$ (say $m^{t_{2}}$). 
    \item  Continue in a similar way for $t_{j}: j \in \{3,\ldots, J \}$ and for each individual $i \in \{1, \ldots, N \}$ until their death or the end of the study, whichever occurs first. 
    \item Measure the outcome $Y_{i}^{t_{(J)}}(\textbf{z}_{i}^{t_{J-1}}, \mathcal{\textbf{G}}_{z_{*}}^{t_{J-1}})$ for each individual $i$.
\end{itemize}

\subsection{Identification in the absence of Competing Event (Proposition 1a)}
Under the assumptions specified in Section 2.3, we can identify $\mathcal{P}_{\textbf{z},\textbf{z}_{*}}(t_{j})$  using the observed data distribution as : \\
\begin{align*}
       \mathcal{P}_{\textbf{z},\textbf{z}_{*}}(t_{j} ) &= \int_{\textbf{m}^{t_{j-1}}}\int_{\boldsymbol{\ell}^{t_{j-1}}}\int_{\textbf{y}^{t_{j-1}}}\int_{\underline{\ell}^{0}} 
      p\big(\underline{L}^0\big) \times
       P \Bigl[T_i \in  (t_{j-1},t_{j}]\big|\textbf{Y}^{t_{j-1}} = \mathbb{0}_{j-1}, 
       \textbf{Z}^{t_{j-1}} = \textbf{z}^{t_{j-1}}, 
       \textbf{L}^{t_{j-1}} = \boldsymbol{\ell}^{t_{j-1}},\textbf{M}^{t_{j-1}} = \textbf{m}^{t_{j-1}},  \underline{L}^0= \underline{\ell}^0 \Bigr] \times 
       \\
      & \prod_{l=1}^{j-1} \Bigl\{  p\big(Y^{t_{l}}\big|\textbf{Y}^{t_{l-1}} = \mathbb{0}_{l-1}, \textbf{Z}^{t_{l-1}} = \textbf{z}^{t_{l-1}}, \textbf{L}^{t_{l-1}} = \boldsymbol{\ell}^{t_{l-1}}, \textbf{M}^{t_{l-1}} = \textbf{m}^{t_{l-1}}, 
      \underline{L}^0= \underline{\ell}^0 \big)\times
      \\&
       p\big(L^{t_{l}}\big|\textbf{Y}^{t_{l}} = \mathbb{0}_{l}, \textbf{Z}^{t_{l}} = \textbf{z}^{t_{l}},  \textbf{L}^{t_{l-1}} = \boldsymbol{\ell}^{t_{l-1}},  \textbf{M}^{t_{l-1}} = \textbf{m}^{t_{l-1}}, 
       \underline{L}^0 = \underline{\ell}^0 \big) \times 
       \\&
         p\big( M^{t_{l}}\big| \textbf{Y}^{t_{l}} = \mathbb{0}_{l}, \textbf{Z}^{t_{l}} = \textbf{z}_{*}^{t_{l}},  \textbf{L}^{t_{l}} = \boldsymbol{\ell}^{t_{l}},  \textbf{M}^{t_{l-1}} = \textbf{m}^{t_{l-1}}, 
      \underline{L}^0= \underline{\ell}^0\big)    
          \Bigr\}  d\underline{\ell}^{0}d\textbf{y}^{t_{j-1}}d\boldsymbol{\ell}^{t_{j-1}}d\textbf{m}^{t_{j-1}}.     
  \end{align*}

\begin{proof}
    By the definition of $\mathcal{P}_{\textbf{z},\textbf{z}_{*}}(t_{j})$, we have:
    \begin{equation}
        \mathcal{P}_{\textbf{z},\textbf{z}_{*}}(t_{j}) =
        Pr(T_i(\textbf{Z}^{t_{j-1}} = \textbf{z}^{t_{j-1}}, \mathcal{\textbf{G}}_{z_{*}}^{t_{j-1}} = \textbf{m}^{t_{j-1}} ) \in  (t_{j-1},t_{j}] )      
    \end{equation}
    
     Using the law of total probability ($p(C) = \int_{D} p(C|D)p(D)$) iteratively for all $t_{j} \in \{t_{1}, \ldots , t_{J} \}$, we get:
    \begin{align*}
        \mathcal{P}_{\textbf{z},\textbf{z}_{*}}(t_{j}) &=
        \int_{\textbf{m}^{t_{j-1}}}\int_{\boldsymbol{\ell}^{t_{j-1}}}\int_{\textbf{y}^{t_{j-1}}}\int_{\underline{\ell}^{0}}  P \Bigl[T_i(\textbf{Z}^{t_{j-1}} = \textbf{z}^{t_{j-1}}, \mathcal{\textbf{G}}_{z_{*}}^{t_{j-1}} = \textbf{m}^{t_{j-1}} ) \in  (t_{j-1},t_{j}]|\mathcal{\textbf{Y}}_{z}^{t_{j-1}} = \mathbb{0}_{j-1}, \mathcal{\textbf{L}}_{z}^{t_{j-1}} = \boldsymbol{\ell}^{t_{j-1}},  \\& \mathcal{\textbf{G}}_{z_{*}}^{t_{j-1}} = \textbf{m}^{t_{j-1}},  \underline{L}^0= \underline{\ell}^0 \Bigr]   \times
        \prod_{l=1}^{j-1} \{p(\mathcal{Y}_{z}^{t_{l}} = y^{t_{l}}|\mathcal{\textbf{Y}}_{z}^{t_{l-1}} = \mathbb{0}_{l-1}, \mathcal{\textbf{L}}_{z}^{t_{l-1}}= \boldsymbol{\ell}^{t_{l-1}}, \mathcal{\textbf{G}}_{z_{*}}^{t_{l-1}} =\textbf{m}^{t_{l-1}}, \underline{L}^0= \underline{\ell}^0) \times  \\&
        p(\mathcal{L}_{z}^{t_{l}} = \ell^{t_{l}}|\mathcal{\textbf{Y}}_{z}^{t_{l}} = \mathbb{0}_{l}, \mathcal{\textbf{L}}_{z}^{t_{l-1}}= \boldsymbol{\ell}^{t_{l-1}}, \mathcal{\textbf{G}}_{z_{*}}^{t_{l-1}} =\textbf{m}^{t_{l-1}}, \underline{L}^0= \underline{\ell}^0) \times  \\& 
         p(\mathcal{G}_{z_{*}}^{t_{l}} = m^{t_{l}}| \mathcal{\textbf{Y}}_{z}^{t_{l}} = \mathbb{0}_{l},\mathcal{\textbf{L}}_{z}^{t_{l}} = \boldsymbol{\ell}^{t_{l}}, \mathcal{\textbf{G}}_{z_{*}}^{t_{l-1}} =  \textbf{m}^{t_{l-1}},\underline{L}^0= \underline{\ell}^0)\} \times  \\&
         p(\underline{L}^0) \text{ } d\underline{\ell}^{0}d\textbf{y}^{t_{j-1}}d\boldsymbol{\ell}^{t_{j-1}}d\textbf{m}^{t_{j-1}}
    \end{align*}
Then, observe the following:
\begin{enumerate}[label=(\roman*)]
    \item 
    \begin{align*}
        & p(\mathcal{G}_{z_{*}}^{t_{l}} = m^{t_{l}}| \mathcal{\textbf{Y}}_{z}^{t_{l}} = \mathbb{0}_{l},\mathcal{\textbf{L}}_{z}^{t_{l}} = \boldsymbol{\ell}^{t_{l}}, \mathcal{\textbf{G}}_{z_{*}}^{t_{l-1}} =  \textbf{m}^{t_{l-1}},\underline{L}^0= \underline{\ell}^0)\} \\
        =& p(M^{t_{l}}(\textbf{z}^{t_{l}}_{*}) = m^{t_{l}}|\textbf{Y}^{t_{l}}(\textbf{z}^{t_{l-1}}_{*}) = \mathbb{0}_{l},\textbf{L}^{t_{l}}(\textbf{z}^{t_{l}}_{*}) = \boldsymbol{\ell}^{t_{l}},\textbf{M}^{t_{l-1}}(\textbf{z}^{t_{l-1}}_{*}) = \textbf{m}^{t_{l-1}},\underline{L}^{0} =  \underline{\ell}^{0}) \quad \quad  \{\text{By the definition of } \mathcal{G}_{z_{*}}^{t_{j}} \}\\
        =& p(M^{t_{l}} = m^{t_{l}}| \textbf{Y}^{t_{l}} =\mathbb{0}_{l},\textbf{Z}^{t_{l}} = \textbf{z}_{*}^{t_{l}}, \textbf{L}^{t_{l}} = \boldsymbol{\ell}^{t_{l}}, \textbf{M}^{t_{l-1}} =\textbf{m}^{t_{l-1}}, \underline{L}^0= \underline{\ell}^0) \\& \{ \text{By using unmeasured confounding assumptions \textbf{III}(i)(a)  and Consistency (i) } \} \\
    \end{align*}
    \item Now, to identify the conditional density $p(\mathcal{L}_{z}^{t_{l}} = \ell^{t_{l}}|\mathcal{\textbf{Y}}_{z}^{t_{l}} = \mathbb{0}_{l}, \mathcal{\textbf{L}}_{z}^{t_{l-1}}= \boldsymbol{\ell}^{t_{l-1}}, \mathcal{\textbf{G}}_{z_{*}}^{t_{l-1}} =\textbf{m}^{t_{l-1}}, \underline{L}^0= \underline{\ell}^0)$, we demonstrate the steps for  $p(\mathcal{L}_{z}^{t_{1}} = \ell^{t_{1}}| \mathcal{Y}_{z}^{t_{1}} = 0, \underline{L}^0= \underline{\ell}^0)$ and the same arguments can be used for subsequent covariates by induction. Observe that:
    \begin{align*}
        & p(\mathcal{L}_{z}^{t_{1}}= \ell^{t_{1}}|\mathcal{Y}_{z}^{t_{1}} = 0,  \underline{L}^0= \underline{\ell}^0) 
        \\
        = & p(L^{t_{1}}({z}^{t_{1}},\mathcal{\textbf{G}}_{z_{*}}^{t_{0}})  = \ell^{t_{1}}| \mathcal{Y}_{z}^{t_{1}} = 0, \underline{L}^0= \underline{\ell}^0)  \quad \quad  \{\text{definition of } \mathcal{L}_{z}^{t_{1}}  \} \\
        =& p(L^{t_{1}}({z}^{t_{1}},m^{t_{0}})  = \ell^{t_{1}}| \mathcal{Y}_{z}^{t_{1}} = 0,   \underline{L}^0= \underline{\ell}^0)    \\
        =& p(L^{t_{1}}({z}^{t_{1}},m^{t_{0}})  = \ell^{t_{1}}|\mathcal{Y}_{z}^{t_{1}} = 0, Z^{t_{0}}=z^{t_{0}},M^{t_{0}}=m^{t_{0}},\underline{L}^0= \underline{\ell}^0) \\& \{\text{ Assuming \textbf{III}(i)(b), \textbf{III}(ii) and }  Z^{t_{-1}} = L^{t_{-1}} = M^{t_{-1}}= \emptyset\} \\
        =& p(L^{t_{1}}({z}^{t_{1}},m^{t_{0}})  = \ell^{t_{1}}|\mathcal{Y}_{z}^{t_{1}} = 0,  Z^{t_{1}} = z^{t_{1}}, M^{t_{0}}=m^{t_{0}}, \underline{L}^0= \underline{\ell}^0)  \\& \{\text{ By Unmeasured Confounding assumption \textbf{III}(i)(b)}\} \\
        =& p(L^{t_{1}} = \ell^{t_{1}} | Y^{t_{1}} = 0, Z^{t_{1}} = z^{t_{1}}, M^{t_{0}}=m^{t_{0}},\underline{L}^0= \underline{\ell}^0   ) \quad  \{\text{By  Consistency (ii), (iii)}\}
    \end{align*}

Using the argument above for any $l \in \{2,3, \ldots , J \}$, we get $p(\mathcal{L}_{z}^{t_{l}} = \ell^{t_{l}}|\mathcal{\textbf{Y}}_{z}^{t_{l}} = \mathbb{0}_{l}, \mathcal{\textbf{L}}_{z}^{t_{l-1}}= \boldsymbol{\ell}^{t_{l-1}}, \mathcal{\textbf{G}}_{z_{*}}^{t_{l-1}} =\textbf{m}^{t_{l-1}}, \underline{L}^0= \underline{\ell}^0) =
        p(L^{t_{l}} = \ell^{t_{l}} | \textbf{Y}^{t_{l}} = \mathbb{0}_{l}, \textbf{Z}^{t_{l}} = \textbf{z}^{t_{l}}, \textbf{L}^{t_{l-1}}= \boldsymbol{\ell}^{t_{l-1}}, \textbf{M}^{t_{l-1}} = \textbf{m}^{t_{l-1}}, \underline{L}^0 = \underline{\ell}^0 ).$
        
Specifically, we could use the same reasons as above to write:
    \begin{equation}
        \begin{aligned}
        &p(\mathcal{L}_{z}^{t_{j}} = \ell^{t_{j}}|\mathcal{\textbf{Y}}_{z}^{t_{j}} = \mathbb{0}_{j}, \mathcal{\textbf{L}}_{z}^{t_{j-1}}= \boldsymbol{\ell}^{t_{j-1}}, \mathcal{\textbf{G}}_{z_{*}}^{t_{j-1}} =\textbf{m}^{t_{j-1}}, \underline{L}^0= \underline{\ell}^0) \\
        =& p(L^{t_{j}}(\textbf{z}^{t_{j}},\textbf{m}^{t_{j-1}})  = \ell^{t_{j}}| \textbf{Y}^{t_{j}}(\textbf{z}^{t_{j-1}},\textbf{m}^{t_{j-1}}) = \mathbb{0}_{j},\textbf{L}^{t_{j-1}}(\textbf{z}^{t_{j-1}},\textbf{m}^{t_{j-2}})  = \boldsymbol{\ell}^{t_{j-1}}, \mathcal{\textbf{G}}_{z_{*}}^{t_{j-1}} =\textbf{m}^{t_{j-1}},\underline{L}^0= \underline{\ell}^0)   \\
        =& p(L^{t_{j}}(\textbf{z}^{t_{j}},\textbf{m}^{t_{j-1}})  = \ell^{t_{j}}|\textbf{Y}^{t_{j}}(\textbf{z}^{t_{j-1}},\textbf{m}^{t_{j-1}}) = \mathbb{0}_{j},\textbf{L}^{t_{j-1}}(\textbf{z}^{t_{j-1}},\textbf{m}^{t_{j-2}})  = \boldsymbol{\ell}^{t_{j-1}},  \underline{L}^0= \underline{\ell}^0)  \\ & \{ \text{conditional on }  \boldsymbol{\ell}^{t_{j-1}},   \text{ and }  \underline{\ell}^0, \text{ } \textbf{m}^{t_{j-1}} \text{ is a random draw }  \text{from } \mathcal{\textbf{G}}_{z_{*}}^{t_{j-1}} 
        \text{(randomized intervention)} \\& \text{and does not affect confounder }  L^{t_{j}}(\textbf{z}^{t_{j}},\textbf{m}^{t_{j-1}}) \} \\
        =& p(L^{t_{j}}(\textbf{z}^{t_{j}},\textbf{m}^{t_{j-1}})  = \ell^{t_{j}}| \textbf{Y}^{t_{j}}(\textbf{z}^{t_{j-1}},\textbf{m}^{t_{j-1}}) = \mathbb{0}_{j}, \textbf{Z}^{t_{j}} = \textbf{z}^{t_{j}}, \textbf{L}^{t_{j-1}}(\textbf{z}^{t_{j-1}},\textbf{m}^{t_{j-2}})  = \boldsymbol{\ell}^{t_{j-1}},  \underline{L}^0= \underline{\ell}^0) \\ & \{ \text{ By Unmeasured Confounding assumption \textbf{III}(i)(b)}\}  \\
        =& p(L^{t_{j}}(\textbf{z}^{t_{j}},\textbf{m}^{t_{j-1}})  = \ell^{t_{j}}| \textbf{Y}^{t_{j}}(\textbf{z}^{t_{j-1}},\textbf{m}^{t_{j-1}}) = \mathbb{0}_{j}, \textbf{Z}^{t_{j}} = \textbf{z}^{t_{j}}, \textbf{L}^{t_{j-1}}(\textbf{z}^{t_{j-1}},\textbf{m}^{t_{j-2}})  = \boldsymbol{\ell}^{t_{j-1}},\textbf{M}^{t_{j-1}} = \textbf{m}^{t_{j-1}},   \underline{L}^0= \underline{\ell}^0) \\ & \{ \text{ By Unmeasured Confounding assumption \textbf{III}(ii)}\} \\
        =& p(L^{t_{j}} = \ell^{t_{j}} |\textbf{Y}^{t_{j}} = \mathbb{0}_{j}, \textbf{Z}^{t_{j}} = \textbf{z}^{t_{j}},\textbf{L}^{t_{j-1}} = \boldsymbol{\ell}^{t_{j-1}}, \textbf{M}^{t_{j-1}} = \textbf{m}^{t_{j-1}},  \underline{L}^0= \underline{\ell}^0   ) \quad \quad \{\text{ By Consistency (ii), (iii)} \}
        \end{aligned}
    \end{equation}

    \item Similarly, to identify the conditional density $p(\mathcal{Y}_{z}^{t_{l}} = y^{t_{l}}|\mathcal{\textbf{Y}}_{z}^{t_{l-1}} = \mathbb{0}_{l-1}, \mathcal{\textbf{L}}_{z}^{t_{l-1}}= \boldsymbol{\ell}^{t_{l-1}}, \mathcal{\textbf{G}}_{z_{*}}^{t_{l-1}} =\textbf{m}^{t_{l-1}}, \underline{L}^0= \underline{\ell}^0)$, we demonstrate the steps for  $p(\mathcal{Y}_{z}^{t_{2}} = y^{t_{2}}| \mathcal{Y}_{z}^{t_{1}} = 0, \mathcal{L}_{z}^{t_{1}}= \ell^{t_{1}}, \mathcal{G}_{z_{*}}^{t_{1}} =m^{t_{1}},\underline{L}^0= \underline{\ell}^0)$ (since we assume that $Y_i^{t_{1}} = 0$ for every individual) and the same arguments can be used for subsequent covariates by induction. Observe that: 
    \begin{align*}
        & p(\mathcal{Y}_{z}^{t_{2}} = y^{t_{2}}| \mathcal{Y}_{z}^{t_{1}} = 0, \mathcal{L}_{z}^{t_{1}}= \ell^{t_{1}}, \mathcal{G}_{z_{*}}^{t_{1}} =m^{t_{1}},\underline{L}^0= \underline{\ell}^0) \\
        =& p(Y^{t_{2}}(z^{t_{1}},m^{t_{1}})  = y^{t_{2}}| Y^{t_{1}}(z^{t_{0}},m^{t_{0}})  = 0, L^{t_{1}}({z}^{t_{1}},m^{t_{0}})= \ell^{t_{1}}, \mathcal{G}_{z_{*}}^{t_{1}} =m^{t_{1}}, \underline{L}^0= \underline{\ell}^0)  \\&
        \{\text{definition of } \mathcal{Y}_{z}^{t_{1}}, \mathcal{Y}_{z}^{t_{2}} \text{ and } \mathcal{L}_{z}^{t_{1}}  \} \\
        =& p(Y^{t_{2}}(z^{t_{1}},m^{t_{1}})  = y^{t_{2}}|Y^{t_{1}}(z^{t_{0}},m^{t_{0}})  = 0, L^{t_{1}}({z}^{t_{1}},m^{t_{0}})= \ell^{t_{1}},  \underline{L}^0= \underline{\ell}^0)  \\&
        \{\text{Conditional on }  \ell^{t_{1}}, \text{ and } \underline{\ell}^0, m^{t_{1}} \text{ is a random draw from } \mathcal{G}_{z_{*}}^{t_{1}} \text{(randomized intervention)} \\&
        \text{and does not affect } Y^{t_{2}}(z^{t_{1}},m^{t_{1}}) \}\\
        =& p(Y^{t_{2}}(z^{t_{1}},m^{t_{1}})  = y^{t_{2}}|Y^{t_{1}}(z^{t_{0}},m^{t_{0}})  = 0, Z^{t_{0}} = z^{t_{0}}, L^{t_{1}}({z}^{t_{1}},m^{t_{0}})= \ell^{t_{1}}, M^{t_{0}} = m^{t_{0}}, \underline{L}^0= \underline{\ell}^0)    \\& \{ \text{ Assuming \textbf{III}(i)(b), \textbf{III}(ii) and  }  Z^{t_{-1}} = L^{t_{-1}} = M^{t_{-1}}= \emptyset\} \\
        =& p(Y^{t_{2}}(z^{t_{1}},m^{t_{1}})  = y^{t_{2}}|Y^{t_{1}}  = 0,Z^{t_{0}} = z^{t_{0}}, L^{t_{1}}({z}^{t_{1}},m^{t_{0}})= \ell^{t_{1}},M^{t_{0}} = m^{t_{0}},  \underline{L}^0= \underline{\ell}^0)  \\&
        \{Y^{t_{1}}(z^{t_{0}},m^{t_{0}}) =  Y^{t_{1}}  \text{ by Consistency (iii) } \}  \\
        =& p(Y^{t_{2}}(z^{t_{1}},m^{t_{1}})  = y^{t_{2}}|Y^{t_{1}}  = 0, Z^{t_{1}} = z^{t_{1}}, L^{t_{1}}({z}^{t_{1}},m^{t_{0}})= \ell^{t_{1}}, M^{t_{0}} = m^{t_{0}}, \underline{L}^0= \underline{\ell}^0)    \\& \{ \text{ By Unmeasured Confounding assumption \textbf{III}(i)(b)}\} \\
        =& p(Y^{t_{2}}(z^{t_{1}},m^{t_{1}})  = y^{t_{2}}|Y^{t_{1}}  = 0, Z^{t_{1}} = z^{t_{1}}, L^{t_{1}}= \ell^{t_{1}},M^{t_{0}} = m^{t_{0}},  \underline{L}^0= \underline{\ell}^0)    \\& \{L^{t_{1}}({z}^{t_{1}},m^{t_{0}}) = L^{t_{1}} \text{ by Consistency (ii))}\} \\
        =& p(Y^{t_{2}}(z^{t_{1}},m^{t_{1}})  = y^{t_{2}}|Y^{t_{1}}  = 0, Z^{t_{1}} = z^{t_{1}}, L^{t_{1}}= \ell^{t_{1}}, M^{t_{1}} = m^{t_{1}},  \underline{L}^0= \underline{\ell}^0)    \\& \{ \text{ By Unmeasured Confounding assumption \textbf{III}(ii)}\} \\
        =& p(Y^{t_{2}}  = y^{t_{2}}|Y^{t_{1}}  = 0, Z^{t_{1}} = z^{t_{1}}, L^{t_{1}}= \ell^{t_{1}}, M^{t_{1}} = m^{t_{1}},  \underline{L}^0= \underline{\ell}^0) \\&
        \{Y^{t_{2}}(z^{t_{1}},m^{t_{1}})  =Y^{t_{2}} \text{ by  Consistency (iii)} \}
    \end{align*}
    Using the argument above for any $l \in \{3, \ldots , J \}$, we get $p(\mathcal{Y}_{z}^{t_{l}} = y^{t_{l}}|\mathcal{\textbf{Y}}_{z}^{t_{l-1}} = \mathbb{0}_{l-1}, \mathcal{\textbf{L}}_{z}^{t_{l-1}}= \boldsymbol{\ell}^{t_{l-1}}, \mathcal{\textbf{G}}_{z_{*}}^{t_{l-1}} =\textbf{m}^{t_{l-1}}, \underline{L}^0= \underline{\ell}^0) = p(Y^{t_{l}} = y^{t_{l}} | \textbf{Y}^{t_{l-1}} = \mathbb{0}_{l-1}, \textbf{Z}^{t_{l-1}} = \textbf{z}^{t_{l-1}}, \textbf{L}^{t_{l-1}}= \boldsymbol{\ell}^{t_{l-1}}, \textbf{M}^{t_{l-1}} = \textbf{m}^{t_{l-1}}, \underline{L}^0 = \underline{\ell}^0 )$.
    \item Finally, to identify the conditional density $$P\Bigl[T_i(\textbf{Z}^{t_{j-1}} = \textbf{z}^{t_{j-1}}, \mathcal{\textbf{G}}_{z_{*}}^{t_{j-1}} = \textbf{m}^{t_{j-1}} ) \in  (t_{j-1},t_{j}]|\mathcal{\textbf{Y}}_{z}^{t_{j-1}} = \mathbb{0}_{j-1}, \mathcal{\textbf{L}}_{z}^{t_{j-1}} = \boldsymbol{\ell}^{t_{j-1}}, \mathcal{\textbf{G}}_{z_{*}}^{t_{j-1}} = \textbf{m}^{t_{j-1}}, \underline{L}^0= \underline{\ell}^0 \Bigr]$$ from the observed data, we demonstrate the steps for the second visit (since all participants are alive at the first visit) and the same argument applies for the subsequent covariates by induction. Observe that:
    \begin{align*}
        & P\Bigl[T_i(Z^{t_{1}} = z^{t_{1}}, \mathcal{G}_{z_{*}}^{t_{1}} = m^{t_{1}} ) \in   (t_{1},t_{2}]  | \mathcal{Y}_{z}^{t_{1}} = 0,\mathcal{L}_{z}^{t_{1}} = \ell^{t_{1}}, \mathcal{G}_{z_{*}}^{t_{1}} = m^{t_{1}}, \underline{L}^0= \underline{\ell}^0 \Bigr]
        \\
        &= P\Bigl[T_i(Z^{t_{1}} = z^{t_{1}}, \mathcal{G}_{z_{*}}^{t_{1}} = m^{t_{1}} ) \in   (t_{1},t_{2}]  |Y^{t_{1}}(z^{t_{0}}, m^{t_{0}}) = 0, L^{t_{1}}(z^{t_{1}}, m^{t_{0}}) = \ell^{t_{1}}, \mathcal{G}_{z_{*}}^{t_{1}} = m^{t_{1}}, \underline{L}^0= \underline{\ell}^0 \Bigr] \\& \{ \text{By the definition of } \mathcal{L}_{z}^{t_{1}}  \text{ and } \mathcal{Y}_{z}^{t_{1}}\}
        \\
        &= P\Bigl[T_i(Z^{t_{1}} = z^{t_{1}}, \mathcal{G}_{z_{*}}^{t_{1}} = m^{t_{1}} ) \in   (t_{1},t_{2}]  |Y^{t_{1}}(z^{t_{0}}, m^{t_{0}}) = 0, L^{t_{1}}(z^{t_{1}}, m^{t_{0}}) = \ell^{t_{1}},  \underline{L}^0= \underline{\ell}^0 \Bigr] \\ &  \{ \text{conditional on }  \ell^{t_{1}},  \text{ and } \underline{L}^0= \underline{\ell}^0 , m^{ t_{1}} \text{ is a random draw from  } \mathcal{G}_{z_{*}}^{t_{1}} 
        \text{(randomized intervention)} \\&
        \text{and does not affect } T_i(Z^{t_{1}} = z^{t_{1}}, \mathcal{G}_{z_{*}}^{t_{1}} = m^{t_{1}} ) \}
        \\
        &= P\Bigl[T_i(Z^{t_{1}} = z^{t_{1}}, \mathcal{G}_{z_{*}}^{t_{1}} = m^{t_{1}} ) \in   (t_{1},t_{2}]  |Y^{t_{1}}(z^{t_{0}}, m^{t_{0}}) = 0, Z^{t_{0}} = z^{t_{0}}, L^{t_{1}}(z^{t_{1}}, m^{t_{0}}) = \ell^{t_{1}}, M^{t_{0}} = m^{t_{0}}, \underline{L}^0= \underline{\ell}^0 \Bigr] \\& \{ \text{ Assuming \textbf{III}(i)(b), \textbf{III}(ii) and  }  Z^{t_{-1}} = L^{t_{-1}} = M^{t_{-1}}= \emptyset\} \\
        &= P\Bigl[T_i(Z^{t_{1}} = z^{t_{1}}, \mathcal{G}_{z_{*}}^{t_{1}} = m^{t_{1}} ) \in   (t_{1},t_{2}]  |Y^{t_{1}} = 0, L^{t_{1}}(z^{t_{1}}, m^{t_{0}}) = \ell^{t_{1}},  \underline{L}^0= \underline{\ell}^0 \Bigr]  \\&
        \{Y^{t_{1}}(z^{t_{0}},m^{t_{0}}) =  Y^{t_{1}}  \text{ by Consistency (iii) } \} 
        \\
        &= P\Bigl[T_i(Z^{t_{1}} = z^{t_{1}}, \mathcal{G}_{z_{*}}^{t_{1}} = m^{t_{1}} ) \in   (t_{1},t_{2}]  |Y^{t_{1}} = 0, Z^{t_{1}} = z^{t_{1}}, L^{t_{1}}(z^{t_{1}}, m^{t_{0}}) = \ell^{t_{1}}, M^{t_{0}} = m^{t_{0}}, \underline{L}^0= \underline{\ell}^0 \Bigr] \\ & \{ \text{By Unmeasured Confounding Assumption \textbf{III}(i)(b)  } \}
        \\
        &= P\Bigl[T_i(Z^{t_{1}} = z^{t_{1}}, \mathcal{G}_{z_{*}}^{t_{1}} = m^{t_{1}} ) \in   (t_{1},t_{2}]  |Y^{t_{1}} = 0, Z^{t_{1}} = z^{t_{1}}, L^{t_{1}} = \ell^{t_{1}}, M^{t_{0}} = m^{t_{0}}, \underline{L}^0= \underline{\ell}^0 \Bigr]  \\&
        \{L^{t_{1}}({z}^{t_{1}},m^{t_{0}}) = L^{t_{1}} \text{ by Consistency (ii))}\} 
        \\
        &= P\Bigl[T_i(Z^{t_{1}} = z^{t_{1}}, \mathcal{G}_{z_{*}}^{t_{1}} = m^{t_{1}} ) \in   (t_{1},t_{2}]  |Y^{t_{1}} = 0, Z^{t_{1}} = z^{t_{1}}, L^{t_{1}} = \ell^{t_{1}}, M^{t_{1}} = m^{t_{1}}, \underline{L}^0= \underline{\ell}^0 \Bigr]  \\&
        \{ \text{By Unmeasured Confounding Assumption \textbf{III}(ii)  } \} 
        \\
        &= P\Bigl[T_i \in  (t_{1},t_{2}] |Y^{t_{1}} = 0, Z^{t_{1}} = z^{t_{1}}, L^{t_{1}} = \ell^{t_{1}}, M^{t_{1}} = m^{t_{1}}, \underline{L}^0= \underline{\ell}^0 \Bigr]  \quad \quad  \{ \text{By Consistency (iv)}\}  
    \end{align*}
  Hence, the result follows.
\end{enumerate}    
\end{proof}

The positivity assumptions specified for the proposition above assure that the conditional densities used to identify the mediation parameter are well-defined.

\subsection{Identification in the presence of a competing event (Proposition 2a)}
\subsubsection{Indentification for the main event}
Under the assumptions specified in Section 3.1, we can identify $\mathcal{P}^{(1)}_{\textbf{z},\textbf{z}_{*}}(t_{j})$ using the observed data distribution as: \\
\begin{align*}
      \mathcal{P}^{(1)}_{\textbf{z},\textbf{z}_{*}}(t_{j}) &=  \int_{\textbf{m}^{t_{j-1}}}\int_{\boldsymbol{\ell}^{t_{j-1}}}\int_{\textbf{y}^{t_{j-1},(1)},\textbf{y}^{t_{j-1},(2)}}\int_{\underline{\ell}^{0}} p\big(\underline{L}^0\big) \times
      \\& 
      P \Bigl[T_i \in  (t_{j-1},t_{j}], d_i =1\big| \textbf{Y}^{t_{j-1},(1)} 
       = \textbf{Y}^{t_{j-1},(2)} = \mathbb{0}_{j-1},
      \textbf{Z}^{t_{j-1}} = \textbf{z}^{t_{j-1}}, 
      \textbf{L}^{t_{j-1}} = \boldsymbol{\ell}^{t_{j-1}},\textbf{M}^{t_{j-1}} = \textbf{m}^{t_{j-1}},  \underline{L}^0= \underline{\ell}^0 \Bigr] \times  
      \\& 
       \prod_{l=1}^{j-1} \Bigl\{  p\big(Y^{t_{l},(1)}, Y^{t_{l},(2)}\big|\textbf{Y}^{t_{l-1},(1)} 
      = \textbf{Y}^{t_{l-1},(2)} = \mathbb{0}_{l-1},
      \textbf{Z}^{t_{l-1}} = \textbf{z}^{t_{l-1}}, \textbf{L}^{t_{l-1}} = \boldsymbol{\ell}^{t_{l-1}},
      \textbf{M}^{t_{l-1}} = \textbf{m}^{t_{l-1}},  \underline{L}^0= \underline{\ell}^0 \big) \times 
      \\&
      p\big(L^{t_{l}}\big|\textbf{Y}^{t_{l},(1)} 
      = \textbf{Y}^{t_{l},(2)} = \mathbb{0}_{l},
      \textbf{Z}^{t_{l}} = \textbf{z}^{t_{l}},
      \textbf{L}^{t_{l-1}} = \boldsymbol{\ell}^{t_{l-1}},
      \textbf{M}^{t_{l-1}} = \textbf{m}^{t_{l-1}},  \underline{L}^0 = \underline{\ell}^0 \big) \times
      \\&
      p\big( M^{t_{l}}\big| \textbf{Y}^{t_{l},(1)} 
      = \textbf{Y}^{t_{l},(2)} = \mathbb{0}_{l},
      \textbf{Z}^{t_{l}} = \textbf{z}_{*}^{t_{l}},  \textbf{L}^{t_{l}} = \boldsymbol{\ell}^{t_{l}},
      \textbf{M}^{t_{l-1}} = \textbf{m}^{t_{l-1}}, \underline{L}^0= \underline{\ell}^0\big)    
          \Bigr\}    d\underline{\ell}^{0}d\textbf{y}^{t_{j-1},(k)}d\boldsymbol{\ell}^{t_{j-1}}d\textbf{m}^{t_{j-1}}
  \end{align*}

\begin{proof}
 By definition,
   \begin{equation}
       \mathcal{P}^{(1)}_{\textbf{z},\textbf{z}_{*}}(t_{j}) = P\Bigl[T_i(\textbf{Z}^{t_{j-1}} = \textbf{z}^{t_{j-1}}, \mathcal{\textbf{G}}_{z_{*}}^{t_{j-1}}= \textbf{m}^{t_{j-1}}) \in  (t_{j-1},t_{j}], \quad \mathcal{D}_i(\textbf{Z}^{t_{j-1}} = \textbf{z}^{t_{j-1}}, \mathcal{\textbf{G}}_{z_{*}}^{t_{j-1}}= \textbf{m}^{t_{j-1}}) =1\Bigr]
   \end{equation}

Using the law of total probability iteratively for all $t_{j} \in \{t_{1},  \ldots, t_{J} \}$, we get:
\begin{align*}
    \mathcal{P}^{(1)}_{\textbf{z},\textbf{z}_{*}}(t_{j}) 
    &= \int_{\textbf{m}^{t_{j-1}}}\int_{\boldsymbol{\ell}^{t_{j-1}}}\int_{\textbf{y}^{t_{j-1},(1)},\textbf{y}^{t_{j-1},(2)}}\int_{\underline{\ell}^{0}} \\&  P\Bigl[T_i(\textbf{Z}^{t_{j-1}} = \textbf{z}^{t_{j-1}}, \mathcal{\textbf{G}}_{z_{*}}^{t_{j-1}}= \textbf{m}^{t_{j-1}}) \in  (t_{j-1},t_{j}], \mathcal{D}_i(\textbf{Z}^{t_{j-1}} = \textbf{z}^{t_{j-1}}, \mathcal{\textbf{G}}_{z_{*}}^{t_{j-1}}= \textbf{m}^{t_{j-1}}) =1 \big| 
    \\& 
    \mathcal{\textbf{Y}}_{z}^{t_{j-1},(1)} = \mathcal{\textbf{Y}}_{z}^{t_{j-1},(2)} = \mathbb{0}_{j-1}, \mathcal{\textbf{L}}_{z}^{t_{j-1}} = \boldsymbol{\ell}^{t_{j-1}}, \mathcal{\textbf{G}}_{z_{*}}^{t_{j-1}} = \textbf{m}^{t_{j-1}}, \underline{L}^0= \underline{\ell}^0 \Bigr]  \\& \times
        \prod_{l=1}^{j-1} \Big\{p\big(\mathcal{Y}_{z}^{t_{l},(1)} = y^{t_{l},(1)}, \mathcal{Y}_{z}^{t_{l},(2)} = y^{t_{l},(2)}\big|\mathcal{\textbf{Y}}_{z}^{t_{l-1},(1)} = \mathcal{\textbf{Y}}_{z}^{t_{l-1},(2)} = \mathbb{0}_{l-1}, \mathcal{\textbf{L}}_{z}^{t_{l-1}}= \boldsymbol{\ell}^{t_{l-1}}, \mathcal{\textbf{G}}_{z_{*}}^{t_{l-1}} =\textbf{m}^{t_{l-1}}, \underline{L}^0= \underline{\ell}^0\big) \times  
        \\&
        p\big(\mathcal{L}_{z}^{t_{l}} = \ell^{t_{l}}\big|\mathcal{\textbf{Y}}_{z}^{t_{l},(1)} = \mathcal{\textbf{Y}}_{z}^{t_{l},(2)} = \mathbb{0}_{l}, \mathcal{\textbf{L}}_{z}^{t_{l-1}}= \boldsymbol{\ell}^{t_{l-1}}, \mathcal{\textbf{G}}_{z_{*}}^{t_{l-1}} =\textbf{m}^{t_{l-1}}, \underline{L}^0= \underline{\ell}^0\big) \times  
        \\& 
         p\big(\mathcal{G}_{z_{*}}^{t_{l}} = m^{t_{l}}\big| \mathcal{\textbf{Y}}_{z}^{t_{l},(1)} = \mathcal{\textbf{Y}}_{z}^{t_{l},(2)} = \mathbb{0}_{l},\mathcal{\textbf{L}}_{z}^{t_{l}} = \boldsymbol{\ell}^{t_{l}}, \mathcal{\textbf{G}}_{z_{*}}^{t_{l-1}} =  \textbf{m}^{t_{l-1}},\underline{L}^0= \underline{\ell}^0\big)\Big\} \times  \\&
         p(\underline{L}^0) \text{ } d\underline{\ell}^{0}d\textbf{y}^{t_{j-1},(k)}d\boldsymbol{\ell}^{t_{j-1}}d\textbf{m}^{t_{j-1}}
    \end{align*}

In  \textbf{Section 1.1}, we have already shown the identification steps for the following conditional densities:
\begin{equation}\label{eq:identG}
    \begin{aligned}
        & p(\mathcal{G}_{z_{*}}^{t_{l}} = m^{t_{l}}| \mathcal{\textbf{Y}}_{z}^{t_{l}} = \mathbb{0}_{l},\mathcal{\textbf{L}}_{z}^{t_{l}} = \boldsymbol{\ell}^{t_{l}}, \mathcal{\textbf{G}}_{z_{*}}^{t_{l-1}} =  \textbf{m}^{t_{l-1}},\underline{L}^0= \underline{\ell}^0) \\
        =& p(M^{t_{l}} = m^{t_{l}}| \textbf{Y}^{t_{l}} =\mathbb{0}_{l},  \textbf{Z}^{t_{l}} = \textbf{z}_{*}^{t_{l}}, \textbf{L}^{t_{l}} = \boldsymbol{\ell}^{t_{l}},  \textbf{M}^{t_{l-1}} =\textbf{m}^{t_{l-1}},    \underline{L}^0= \underline{\ell}^0)
    \end{aligned}
\end{equation}

\begin{equation}\label{eq:identL}
    \begin{aligned}
        & p(\mathcal{L}_{z}^{t_{l}} = \ell^{t_{l}}|\mathcal{\textbf{Y}}_{z}^{t_{l}} = \mathbb{0}_{l}, \mathcal{\textbf{L}}_{z}^{t_{l-1}}= \boldsymbol{\ell}^{t_{l-1}}, \mathcal{\textbf{G}}_{z_{*}}^{t_{l-1}} =\textbf{m}^{t_{l-1}},  \underline{L}^0= \underline{\ell}^0)\\
            =& p(L^{t_{l}} = \ell^{t_{l}}| \textbf{Y}^{t_{l}} =\mathbb{0}_{l},  \textbf{Z}^{t_{l}} = \textbf{z}^{t_{l}}, \textbf{L}^{t_{l-1}} = \boldsymbol{\ell}^{t_{l-1}},  \textbf{M}^{t_{l-1}} =\textbf{m}^{t_{l-1}},    \underline{L}^0= \underline{\ell}^0)
    \end{aligned}
\end{equation}

\begin{equation}\label{eq:identY}
    \begin{aligned}
         & p(\mathcal{Y}_{z}^{t_{l}} = \ell^{t_{l}}|\mathcal{\textbf{Y}}_{z}^{t_{l-1}} = \mathbb{0}_{l}, \mathcal{\textbf{L}}_{z}^{t_{l-1}}= \boldsymbol{\ell}^{t_{l-1}}, \mathcal{\textbf{G}}_{z_{*}}^{t_{l-1}} =\textbf{m}^{t_{l-1}},  \underline{L}^0= \underline{\ell}^0)\\
            =& p(Y^{t_{l}} = y^{t_{l}}|\textbf{Y}^{t_{l-1}} =\mathbb{0}_{l},  \textbf{Z}^{t_{l-1}} = \textbf{z}^{t_{l-1}}, \textbf{L}^{t_{l-1}} = \boldsymbol{\ell}^{t_{l-1}},  \textbf{M}^{t_{l-1}} =\textbf{m}^{t_{l}},   \underline{L}^0= \underline{\ell}^0).
    \end{aligned}
\end{equation}
The same arguments can be used for the identification of the conditional densities $p\big(\mathcal{G}_{z_{*}}^{t_{l}} = m^{t_{l}}\big| \mathcal{\textbf{Y}}_{z}^{t_{l},(1)} = \mathcal{\textbf{Y}}_{z}^{t_{l},(2)} = \mathbb{0}_{l},\mathcal{\textbf{L}}_{z}^{t_{l}} = \boldsymbol{\ell}^{t_{l}}, \mathcal{\textbf{G}}_{z_{*}}^{t_{l-1}} =  \textbf{m}^{t_{l-1}},\underline{L}^0= \underline{\ell}^0\big)$, and $p\big(\mathcal{L}_{z}^{t_{l}} = \ell^{t_{l}}\big|\mathcal{\textbf{Y}}_{z}^{t_{l},(1)} = \mathcal{\textbf{Y}}_{z}^{t_{l},(2)} = \mathbb{0}_{l}, \mathcal{\textbf{L}}_{z}^{t_{l-1}}= \boldsymbol{\ell}^{t_{l-1}}, \mathcal{\textbf{G}}_{z_{*}}^{t_{l-1}} =\textbf{m}^{t_{l-1}}, \underline{L}^0= \underline{\ell}^0\big)$ by adding both main and competing events in the conditioning sets in Equations \ref{eq:identG} and \ref{eq:identL}.
The identification of the joint conditional mass function  $p\big(\mathcal{Y}_{z}^{t_{l},(1)} = y^{t_{l},(1)}, \mathcal{Y}_{z}^{t_{l},(2)} = y^{t_{l},(2)}\big|\mathcal{\textbf{Y}}_{z}^{t_{l-1},(1)} = \mathcal{\textbf{Y}}_{z}^{t_{l-1},(2)} = \mathbb{0}_{l-1}, \mathcal{\textbf{L}}_{z}^{t_{l-1}}= \boldsymbol{\ell}^{t_{l-1}}, \mathcal{\textbf{G}}_{z_{*}}^{t_{l-1}} =\textbf{m}^{t_{l-1}}, \underline{L}^0= \underline{\ell}^0\big)$ also follows the same arguments as the identification steps for the conditional mass function in Equation \ref{eq:identY}. However, the resulting joint conditional mass function of the observed main and competing events can be further decomposed as the conditional mass function of the observed main event at the current visit given no past events and other covariates, and the conditional mass function of the observed competing event at the current visit given no main event at the current visit, no other events in the past and other covariates:
\begin{align*}\label{eq:identYtoY1Y2}
        & p\big(\mathcal{Y}_{z}^{t_{l},(1)} = y^{t_{l},(1)}, \mathcal{Y}_{z}^{t_{l},(2)} = y^{t_{l},(2)}\big|\mathcal{\textbf{Y}}_{z}^{t_{l-1},(1)} = \mathcal{\textbf{Y}}_{z}^{t_{l-1},(2)} = \mathbb{0}_{l-1}, \mathcal{\textbf{L}}_{z}^{t_{l-1}}= \boldsymbol{\ell}^{t_{l-1}}, \mathcal{\textbf{G}}_{z_{*}}^{t_{l-1}} =\textbf{m}^{t_{l-1}}, \underline{L}^0= \underline{\ell}^0\big)
        \\&
        = p\big(Y^{t_{l},(1)} = y^{t_{l},(1)}, Y^{t_{l},(2)} = y^{t_{l},(2)}\big|\mathcal{\textbf{Y}}_{z}^{t_{l-1},(1)} = \mathcal{\textbf{Y}}_{z}^{t_{l-1},(2)} = \mathbb{0}_{l-1}, \mathcal{\textbf{L}}_{z}^{t_{l-1}}= \boldsymbol{\ell}^{t_{l-1}}, \mathcal{\textbf{G}}_{z_{*}}^{t_{l-1}} =\textbf{m}^{t_{l-1}}, \underline{L}^0= \underline{\ell}^0\big)
        \\&
        = p\big(Y^{t_{l},(1)} = y^{t_{l},(1)}\big|\mathcal{\textbf{Y}}_{z}^{t_{l-1},(1)} = \mathcal{\textbf{Y}}_{z}^{t_{l-1},(2)} = \mathbb{0}_{l-1}, \mathcal{\textbf{L}}_{z}^{t_{l-1}}= \boldsymbol{\ell}^{t_{l-1}}, \mathcal{\textbf{G}}_{z_{*}}^{t_{l-1}} =\textbf{m}^{t_{l-1}}, \underline{L}^0= \underline{\ell}^0\big) \times 
        \\&
        p\big(Y^{t_{l},(2)} = y^{t_{l},(2)}\big|\mathcal{\textbf{Y}}_{z}^{t_{l},(1)} = \mathbb{0}_{l}, \mathcal{\textbf{Y}}_{z}^{t_{l-1},(2)} = \mathbb{0}_{l-1}, \mathcal{\textbf{L}}_{z}^{t_{l-1}}= \boldsymbol{\ell}^{t_{l-1}}, \mathcal{\textbf{G}}_{z_{*}}^{t_{l-1}} =\textbf{m}^{t_{l-1}}, \underline{L}^0= \underline{\ell}^0\big)
    \end{align*}

Finally, to identify the conditional probability $P\Bigl[T_i(\textbf{Z}^{t_{j-1}} = \textbf{z}^{t_{j-1}}, \mathcal{\textbf{G}}_{z_{*}}^{t_{j-1}}= \textbf{m}^{t_{j-1}}) \in  (t_{j-1},t_{j}], \mathcal{D}_i(\textbf{Z}^{t_{j-1}} = \textbf{z}^{t_{j-1}}, \mathcal{\textbf{G}}_{z_{*}}^{t_{j-1}}= \textbf{m}^{t_{j-1}}) =1 \big| 
    \mathcal{\textbf{Y}}_{z}^{t_{j-1},(1)} = \mathcal{\textbf{Y}}_{z}^{t_{j-1},(2)} = \mathbb{0}_{j-1}, \mathcal{\textbf{L}}_{z}^{t_{j-1}} = \boldsymbol{\ell}^{t_{j-1}}, \mathcal{\textbf{G}}_{z_{*}}^{t_{j-1}} = \textbf{m}^{t_{j-1}}, \underline{L}^0= \underline{\ell}^0 \Bigr]$
from the observed data, we demonstrate the steps for the second visit (since all participants are alive at the first visit) and the same argument applies for the subsequent visits by induction. 
    
Observe that:
\begin{align*}
        & P\Bigl[T_i(Z^{t_{1}} = z^{t_{1}}, \mathcal{G}_{z_{*}}^{t_{1}} = m^{t_{1}} ) \in   (t_{1},t_{2}], D_i(Z^{t_{1}} = z^{t_{1}}, \mathcal{G}_{z_{*}}^{t_{1}} = m^{t_{1}} ) = 1  | \mathcal{Y}_{z}^{t_{1},(1)} = \mathcal{Y}_{z}^{t_{1},(2)} = 0,\mathcal{L}_{z}^{t_{1}} = \ell^{t_{1}}, \mathcal{G}_{z_{*}}^{t_{1}} = m^{t_{1}}, \underline{L}^0= \underline{\ell}^0 \Bigr]
        \\
        &= P\Bigl[T_i(Z^{t_{1}} = z^{t_{1}}, \mathcal{G}_{z_{*}}^{t_{1}} = m^{t_{1}} ) \in   (t_{1},t_{2}], D_i(Z^{t_{1}} = z^{t_{1}}, \mathcal{G}_{z_{*}}^{t_{1}} = m^{t_{1}} ) = 1  |  
         Y^{t_{1},(1)}(z^{t_{0}}, m^{t_{0}}) = Y^{t_{1},(2)}(z^{t_{0}}, m^{t_{0}}) = 0,
          \\&  
          L^{t_{1}}(z^{t_{1}}, m^{t_{0}}) = \ell^{t_{1}},
        \mathcal{G}_{z_{*}}^{t_{1}} = m^{t_{1}}, \underline{L}^0= \underline{\ell}^0 \Bigr]
         \quad \quad \{\text{By the definition of } \mathcal{L}_{z}^{t_{1}}  \text{ and } \mathcal{Y}_{z}^{t_{1},(k)}\}
        \\
        &= P\Bigl[T_i(Z^{t_{1}} = z^{t_{1}}, \mathcal{G}_{z_{*}}^{t_{1}} = m^{t_{1}} ) \in   (t_{1},t_{2}], D_i(Z^{t_{1}} = z^{t_{1}}, \mathcal{G}_{z_{*}}^{t_{1}} = m^{t_{1}} ) = 1  |Y^{t_{1},(1)}(z^{t_{0}}, m^{t_{0}}) = Y^{t_{1},(2)}(z^{t_{0}}, m^{t_{0}}) = 0,
          \\&
          L^{t_{1}}(z^{t_{1}}, m^{t_{0}}) = \ell^{t_{1}}, 
        \underline{L}^0= \underline{\ell}^0 \Bigr] 
        \quad
        \big\{ \text{conditional on }  \ell^{t_{1}},  \text{ and } \underline{L}^0= \underline{\ell}^0 , m^{ t_{1}} \text{ is a random draw from  } \mathcal{G}_{z_{*}}^{t_{1}} 
        \text{(randomized intervention)} \\&
        \text{and does not affect }  T_i(Z^{t_{1}} = z^{t_{1}}, \mathcal{G}_{z_{*}}^{t_{1}} = m^{t_{1}} )  \text{ and } D_i(Z^{t_{1}} = z^{t_{1}}, \mathcal{G}_{z_{*}}^{t_{1}} = m^{t_{1}} )\big\}
        \\
        &= P\Bigl[T_i(Z^{t_{1}} = z^{t_{1}}, \mathcal{G}_{z_{*}}^{t_{1}} = m^{t_{1}} ) \in   (t_{1},t_{2}], D_i(Z^{t_{1}} = z^{t_{1}}, \mathcal{G}_{z_{*}}^{t_{1}} = m^{t_{1}} ) = 1  |Y^{t_{1},(1)}(z^{t_{0}}, m^{t_{0}}) = Y^{t_{1},(2)}(z^{t_{0}}, m^{t_{0}}) = 0, Z^{t_{0}} =  z^{t_{0}}, \\ &
        L^{t_{1}}(z^{t_{1}}, m^{t_{0}}) = \ell^{t_{1}},  M^{t_{0}} =  m^{t_{0}}, \underline{L}^0= \underline{\ell}^0 \Bigr] \quad \{ \text{By Assumptions 6 (i) and  (ii), and  }  Z^{t_{-1}} = L^{t_{-1}} = M^{t_{-1}}= \emptyset\} 
        \\
        &= P\Bigl[T_i(Z^{t_{1}} = z^{t_{1}}, \mathcal{G}_{z_{*}}^{t_{1}} = m^{t_{1}} ) \in   (t_{1},t_{2}] , D_i(Z^{t_{1}} = z^{t_{1}}, \mathcal{G}_{z_{*}}^{t_{1}} = m^{t_{1}} ) = 1 |Y^{t_{1},(1)}= Y^{t_{1},(2)} = 0,  Z^{t_{0}} =  z^{t_{0}}, L^{t_{1}}(z^{t_{1}}, m^{t_{0}}) = \ell^{t_{1}}, \\ & M^{t_{0}} =  m^{t_{0}},  \underline{L}^0= \underline{\ell}^0 \Bigr]  \quad
        \{Y^{t_{1},(k)}(z^{t_{0}},m^{t_{0}}) =  Y^{t_{1},(k)}  \text{ by Assumption 4.1 } \} 
        \\
        &= P\Bigl[T_i(Z^{t_{1}} = z^{t_{1}}, \mathcal{G}_{z_{*}}^{t_{1}} = m^{t_{1}} ) \in   (t_{1},t_{2}], D_i(Z^{t_{1}} = z^{t_{1}}, \mathcal{G}_{z_{*}}^{t_{1}} = m^{t_{1}} ) = 1 |Y^{t_{1},(1)}= Y^{t_{1},(2)} = 0, Z^{t_{1}} = z^{t_{1}}, L^{t_{1}}(z^{t_{1}}, m^{t_{0}}) = \ell^{t_{1}}, \\ & M^{t_{0}} =  m^{t_{0}},  \underline{L}^0= \underline{\ell}^0 \Bigr]  \quad\{ \text{By Unmeasured Confounding Assumption 6 (i)   } \}
        \\
        &= P\Bigl[T_i(Z^{t_{1}} = z^{t_{1}}, \mathcal{G}_{z_{*}}^{t_{1}} = m^{t_{1}} ) \in   (t_{1},t_{2}] , D_i(Z^{t_{1}} = z^{t_{1}}, \mathcal{G}_{z_{*}}^{t_{1}} = m^{t_{1}} ) = 1 |Y^{t_{1},(1)}= Y^{t_{1},(2)} = 0, Z^{t_{1}} = z^{t_{1}}, L^{t_{1}} = \ell^{t_{1}},  \\ & 
        M^{t_{0}} =  m^{t_{0}}, \underline{L}^0= \underline{\ell}^0 \Bigr] \quad \quad
        \{L^{t_{1}}({z}^{t_{1}},m^{t_{0}}) = L^{t_{1}} \text{ by Consistency (iii))}\} 
        \\
        &= P\Bigl[T_i(Z^{t_{1}} = z^{t_{1}}, \mathcal{G}_{z_{*}}^{t_{1}} = m^{t_{1}} ) \in   (t_{1},t_{2}] , D_i(Z^{t_{1}} = z^{t_{1}}, \mathcal{G}_{z_{*}}^{t_{1}} = m^{t_{1}} ) = 1 |Y^{t_{1},(1)}= Y^{t_{1},(2)} = 0, Z^{t_{1}} = z^{t_{1}}, L^{t_{1}} = \ell^{t_{1}}, \\&
        M^{t_{1}} = m^{t_{1}}, \underline{L}^0= \underline{\ell}^0 \Bigr]  \quad \quad
        \{ \text{By Unmeasured Confounding Assumption 6 (ii)  } \} 
        \\
        &= P\Bigl[T_i \in  (t_{1},t_{2}], d_i = 1 |Y^{t_{1},(1)}= Y^{t_{1},(2)} = 0, Z^{t_{1}} = z^{t_{1}}, L^{t_{1}} = \ell^{t_{1}}, M^{t_{1}} = m^{t_{1}}, \underline{L}^0= \underline{\ell}^0 \Bigr]  \\ &  \{ \text{By \textbf{Consistency 1} assumption on } T_i(\cdot) \text{ and } D_i(\cdot) \}  
    \end{align*}
Hence, the result follows.   
\end{proof}

The positivity assumptions specified for the proposition above assure that the conditional densities used to identify the mediation parameter are well-defined.

\subsubsection{Indentification for the competing event}

Under the assumptions specified in Section 3.1, we can identify $\mathcal{P}^{(2)}_{\textbf{z},\textbf{z}_{*}}(t_{j})$ using the observed data distribution as: \\
\begin{align*}
      \mathcal{P}^{(2)}_{\textbf{z},\textbf{z}_{*}}(t_{j}) &=  \int_{\textbf{m}^{t_{j-1}}}\int_{\boldsymbol{\ell}^{t_{j-1}}}\int_{\textbf{y}^{t_{j},(1)},\textbf{y}^{t_{j-1},(2)}}\int_{\underline{\ell}^{0}} p\big(\underline{L}^0\big) \times
      \\& 
      P \Bigl[T_i \in  (t_{j-1},t_{j}], d_i =2\big| \textbf{Y}^{t_{j},(1)} = \mathbb{0}_{j},
       \textbf{Y}^{t_{j-1},(2)} = \mathbb{0}_{j-1},
      \textbf{Z}^{t_{j-1}} = \textbf{z}^{t_{j-1}}, 
      \textbf{L}^{t_{j-1}} = \boldsymbol{\ell}^{t_{j-1}},\textbf{M}^{t_{j-1}} = \textbf{m}^{t_{j-1}},  \underline{L}^0= \underline{\ell}^0 \Bigr] \times  
      \\&
      p\big(Y^{t_{j},(1)}\big|\textbf{Y}^{t_{j-1},(1)} 
      = \textbf{Y}^{t_{j-1},(2)} = \mathbb{0}_{j-1},
      \textbf{Z}^{t_{j-1}} = \textbf{z}^{t_{j-1}}, \textbf{L}^{t_{j-1}} = \boldsymbol{\ell}^{t_{j-1}},
      \textbf{M}^{t_{j-1}} = \textbf{m}^{t_{j-1}},  \underline{L}^0= \underline{\ell}^0 \big) \times
      \\& 
       \prod_{l=1}^{j-1} \Bigl\{  p\big(Y^{t_{l},(2)}, Y^{t_{l},(1)}\big|\textbf{Y}^{t_{l-1},(1)} = 
       \textbf{Y}^{t_{l-1},(2)} = \mathbb{0}_{l-1},
      \textbf{Z}^{t_{l-1}} = \textbf{z}^{t_{l-1}}, \textbf{L}^{t_{l-1}} = \boldsymbol{\ell}^{t_{l-1}},
      \textbf{M}^{t_{l-1}} = \textbf{m}^{t_{l-1}},  \underline{L}^0= \underline{\ell}^0 \big) \times 
      \\&
      p\big(L^{t_{l}}\big|\textbf{Y}^{t_{l},(1)} 
      = \textbf{Y}^{t_{l},(2)} = \mathbb{0}_{l},
      \textbf{Z}^{t_{l}} = \textbf{z}^{t_{l}},
      \textbf{L}^{t_{l-1}} = \boldsymbol{\ell}^{t_{l-1}},
      \textbf{M}^{t_{l-1}} = \textbf{m}^{t_{l-1}},  \underline{L}^0 = \underline{\ell}^0 \big) \times
      \\&
      p\big( M^{t_{l}}\big| \textbf{Y}^{t_{l},(1)} 
      = \textbf{Y}^{t_{l},(2)} = \mathbb{0}_{l},
      \textbf{Z}^{t_{l}} = \textbf{z}_{*}^{t_{l}},  \textbf{L}^{t_{l}} = \boldsymbol{\ell}^{t_{l}},
      \textbf{M}^{t_{l-1}} = \textbf{m}^{t_{l-1}}, \underline{L}^0= \underline{\ell}^0\big)    
          \Bigr\}    d\underline{\ell}^{0}d\textbf{y}^{t_{j-1},(k)}d\boldsymbol{\ell}^{t_{j-1}}d\textbf{m}^{t_{j-1}}
  \end{align*}

\begin{proof}
 By definition,
   \begin{equation}
       \mathcal{P}^{(2)}_{\textbf{z},\textbf{z}_{*}}(t_{j}) = P\Bigl[T_i(\textbf{Z}^{t_{j-1}} = \textbf{z}^{t_{j-1}}, \mathcal{\textbf{G}}_{z_{*}}^{t_{j-1}}= \textbf{m}^{t_{j-1}}) \in  (t_{j-1},t_{j}], \quad \mathcal{D}_i(\textbf{Z}^{t_{j-1}} = \textbf{z}^{t_{j-1}}, \mathcal{\textbf{G}}_{z_{*}}^{t_{j-1}}= \textbf{m}^{t_{j-1}}) =2\Bigr]
   \end{equation}

Using the law of total probability iteratively for all $t_{j} \in \{t_{1},  \ldots, t_{J} \}$, we get:
\begin{align*}
    \mathcal{P}^{(2)}_{\textbf{z},\textbf{z}_{*}}(t_{j}) 
    &= \int_{\textbf{m}^{t_{j-1}}}\int_{\boldsymbol{\ell}^{t_{j-1}}}\int_{\textbf{y}^{t_{j},(1)},\textbf{y}^{t_{j-1},(2)}}\int_{\underline{\ell}^{0}} \\&  P\Bigl[T_i(\textbf{Z}^{t_{j-1}} = \textbf{z}^{t_{j-1}}, \mathcal{\textbf{G}}_{z_{*}}^{t_{j-1}}= \textbf{m}^{t_{j-1}}) \in  (t_{j-1},t_{j}], \mathcal{D}_i(\textbf{Z}^{t_{j-1}} = \textbf{z}^{t_{j-1}}, \mathcal{\textbf{G}}_{z_{*}}^{t_{j-1}}= \textbf{m}^{t_{j-1}}) =2 \big| 
    \mathcal{\textbf{Y}}_{z}^{t_{j},(1)} = \mathbb{0}_{j}, \mathcal{\textbf{Y}}_{z}^{t_{j-1},(2)} = \mathbb{0}_{j-1}, 
    \\& 
    \mathcal{\textbf{L}}_{z}^{t_{j-1}} = \boldsymbol{\ell}^{t_{j-1}}, \mathcal{\textbf{G}}_{z_{*}}^{t_{j-1}} = \textbf{m}^{t_{j-1}}, \underline{L}^0= \underline{\ell}^0 \Bigr] \times 
    \\&
    p\big(\mathcal{Y}_{z}^{t_{j},(1)} = y^{t_{j},(1)}\big|\mathcal{\textbf{Y}}_{z}^{t_{j-1},(1)} = \mathcal{\textbf{Y}}_{z}^{t_{j-1},(2)} = \mathbb{0}_{j-1}, \mathcal{\textbf{L}}_{z}^{t_{j-1}}= \boldsymbol{\ell}^{t_{j-1}}, \mathcal{\textbf{G}}_{z_{*}}^{t_{j-1}} =\textbf{m}^{t_{j-1}}, \underline{L}^0= \underline{\ell}^0\big)\\&
        \prod_{l=1}^{j-1} \Big\{p\big(\mathcal{Y}_{z}^{t_{l},(1)} = y^{t_{l},(1)}, \mathcal{Y}_{z}^{t_{l},(2)} = y^{t_{l},(2)}\big|\mathcal{\textbf{Y}}_{z}^{t_{l-1},(1)} = \mathcal{\textbf{Y}}_{z}^{t_{l-1},(2)} = \mathbb{0}_{l-1}, \mathcal{\textbf{L}}_{z}^{t_{l-1}}= \boldsymbol{\ell}^{t_{l-1}}, \mathcal{\textbf{G}}_{z_{*}}^{t_{l-1}} =\textbf{m}^{t_{l-1}}, \underline{L}^0= \underline{\ell}^0\big) \times  
        \\&
        p\big(\mathcal{L}_{z}^{t_{l}} = \ell^{t_{l}}\big|\mathcal{\textbf{Y}}_{z}^{t_{l},(1)} = \mathcal{\textbf{Y}}_{z}^{t_{l},(2)} = \mathbb{0}_{l}, \mathcal{\textbf{L}}_{z}^{t_{l-1}}= \boldsymbol{\ell}^{t_{l-1}}, \mathcal{\textbf{G}}_{z_{*}}^{t_{l-1}} =\textbf{m}^{t_{l-1}}, \underline{L}^0= \underline{\ell}^0\big) \times  
        \\& 
         p\big(\mathcal{G}_{z_{*}}^{t_{l}} = m^{t_{l}}\big| \mathcal{\textbf{Y}}_{z}^{t_{l},(1)} = \mathcal{\textbf{Y}}_{z}^{t_{l},(2)} = \mathbb{0}_{l},\mathcal{\textbf{L}}_{z}^{t_{l}} = \boldsymbol{\ell}^{t_{l}}, \mathcal{\textbf{G}}_{z_{*}}^{t_{l-1}} =  \textbf{m}^{t_{l-1}},\underline{L}^0= \underline{\ell}^0\big)\Big\} \times  \\&
         p(\underline{L}^0) \text{ } d\underline{\ell}^{0}d\textbf{y}^{t_{j-1},(k)}d\boldsymbol{\ell}^{t_{j-1}}d\textbf{m}^{t_{j-1}}
    \end{align*}
\end{proof}
The rest of the proof follows the same reasoning as the proof for identification in the case of the main event. The only difference is that at each visit $j$, we integrate out an additional main event component $y^{t_{j},(1)}$. This is because the joint conditional mass function of the main event and the competing event decomposes into the conditional mass function of the main event multiplied by the conditional mass function of the competing event given the main event.

\section{Supplementary Section 2: Additional results}

To calculate the cumulative mean blood pressure of an individual over a specified age range, we compute the following integral:

    \begin{align*}
        &\int_{t_0}^{t_1} MBP(a_{ij})da_{ij} 
        \\&= \int_{t_0}^{t_1} (\beta_0 + \beta_1 RACE_{i}+ \beta_2 SEX_{i}+  \beta_3 BMI_{i}+ \beta_4  a_{ij}  +  \beta_5 a_{ij}^2 \\& +
         \beta_6 RACE_{i}* a_{ij} + \beta_7 SEX_{i}* a_{ij} + \beta_8 BMI_{i}* a_{ij} + b_{0i}[a_{ij}]_{t_0}^{t_1} +  b_{1i} a_{ij}) \quad                 da_{ij} \\
         &= \beta_0 [a_{ij}]_{t_0}^{t_1} + \beta_1 RACE_{i}[a_{ij}]_{t_0}^{t_1} + \beta_2 SEX_{i}[a_{ij}]_{t_0}^{t_1} +  \beta_3 BMI_{i}[a_{ij}]_{t_0}^{t_1} + \beta_4  \int_{t_0}^{t_1} a_{ij} da_{ij}  +  \beta_5 \int_{t_0}^{t_1} a_{ij}^2 da_{ij} \\& +
        \beta_6 RACE_{i} \int_{t_0}^{t_1} a_{ij} da_{ij} + \beta_7 SEX_{i} \int_{t_0}^{t_1} a_{ij} da_{ij} + \beta_8 BMI_{i} \int_{t_0}^{t_1} a_{ij} da_{ij} + b_{0i}[a_{ij}]_{t_0}^{t_1} +  b_{1i}\int_{t_0}^{t_1} a_{ij} da_{ij}\\
        &= \beta_0 [a_{ij}]_{t_0}^{t_1} + \beta_1 RACE_{i}[a_{ij}]_{t_0}^{t_1} + \beta_2 SEX_{i}[a_{ij}]_{t_0}^{t_1} +  \beta_3 BMI_{i}[a_{ij}]_{t_0}^{t_1} + \beta_4  [\frac{a_{ij}^2}{2}]_{t_0}^{t_1}  +  \beta_5 [\frac{a_{ij}^3}{3}]_{t_0}^{t_1} \\& +
        \beta_6 RACE_{i} [\frac{a_{ij}^2}{2}]_{t_0}^{t_1} + \beta_7 SEX_{i} [\frac{a_{ij}^2}{2}]_{t_0}^{t_1} + \beta_8 BMI_{i} [\frac{a_{ij}^2}{2}]_{t_0}^{t_1} + b_{0i}[a_{ij}]_{t_0}^{t_1} +  b_{1i}[\frac{a_{ij}^2}{2}]_{t_0}^{t_1}\\
        &= \beta_0 (t_1 - t_0) + \beta_1 RACE_{i}(t_1 - t_0) + \beta_2 SEX_{i}(t_1 - t_0) +  \beta_3 BMI_{i}(t_1 - t_0) + \beta_4  \frac{(t_1^2 - t_0^2)}{2}   +  \beta_5 \frac{(t_1^3 - t_0^3)}{3} \\& + 
        \beta_6 RACE_{i} \frac{(t_1^2 - t_0^2)}{2} + \beta_7 SEX_{i} \frac{(t_1^2 - t_0^2)}{2} + \beta_8 BMI_{i} \frac{(t_1^2 - t_0^2)}{2} + b_{0i} (t_1 - t_0) +  b_{1i}\frac{(t_1^2 - t_0^2)}{2}.
    \end{align*}

\section{Supplementary Section 3: Additional tables (Tables 7-10)}

\setcounter{table}{6}

\begin{center}
\begin{table}[!h]
\caption{Posterior means and 95 \% Bayesian credible intervals of causal effects in the absence of a  competing event. The reported results indicate the differences in survival probabilities between (hypertensive at baseline) men who were on BP medication and those who were not on BP medication in ARIC.}
\begin{tabular}{llccccc}
 \multicolumn{4}{c}{\bfseries \normalsize Hypertensive at baseline population in ARIC (Men only)}  \\
\cline{1-4}
&  &   Visit 2 & Visit 3  \\
\cline{1-4}
Age 45-49         & $IDE$                       &  -0.027(-0.12, 0.034)                & -0.036(-0.13, 0.029)   \\
at baseline        & $IIE$                        & 0.0003(-0.0056, 0.0068)                & 0.0001(-0.0079, 0.0084)   \\
& $TE$              &  -0.026(-0.13, 0.034)                & -0.036(-0.13, 0.028)   \\
\toprule
 Age 50-54        & $IDE$                        &  -0.026(-0.13, 0.036)                & -0.036(-0.13, 0.027)   \\
at baseline        & $IIE$                        & 0.0002(-0.0053, 0.0074)                & 0.0001(-0.0080, 0.0086)   \\
& $TE$              &  -0.026(-0.13, 0.037)                & -0.036(-0.13, 0.028)   \\
\toprule
 Age 55-59        & $IDE$                        &  -0.025(-0.12, 0.031)                & -0.035(-0.12, 0.026)   \\
at baseline        & $IIE$                        & 0.0001(-0.0058, 0.0068)                & -0.0001(-0.0088, 0.0086)   \\
& $TE$              &  -0.025(-0.12, 0.032)                & -0.035(-0.12, 0.027)   \\
\toprule
 Age 60-64        & $IDE$                        &  -0.025(-0.11, 0.031)                & -0.034(-0.12, 0.029)   \\
at baseline        & $IIE$                        & 0.0002(-0.0062, 0.0080)                & 0.0002(-0.0098, 0.0097)   \\
& $TE$              &  -0.024(-0.11, 0.031)                & -0.034(-0.12, 0.027)   \\

\toprule
\end{tabular}
\begin{tablenotes}%
\item \textbf{Note:} $IDE$ = Interventional Direct Effects, $IIE$ = Interventional Indirect Effects, and $TE$ = Total Effects.\vspace*{6pt}
\end{tablenotes}
\end{table}
\end{center}

\begin{center}
\begin{table}[!h]
\caption{Posterior means and 95 \% Bayesian credible intervals of causal effects in the presence of a  competing event. The reported results indicate the differences in potential cumulative incidence functions (CIFs) between (hypertensive at baseline) participants who were on BP medication and those who were not on BP medication in ARIC.}
\begin{tabular}{llccccc}
 \multicolumn{4}{c}{\bfseries \normalsize Hypertensive at baseline population in ARIC (Men only)}  \\
\cline{1-4}
&  &   Visit 2 & Visit 3  \\
\cline{1-4}
\toprule 
Age 45-49        & $IDE^{(1)}$                                   &  0.0062(-0.011, 0.033)             & 0.011(-0.013, 0.042)              \\
at baseline        & $IIE^{(1)}$                                   &  -0.0001(-0.0055, 0.0058)             & -0.0002(-0.0081, 0.0078)              \\
& $TE^{(1)}$      &  0.0061(-0.011, 0.032)             & 0.011(-0.014, 0.043)              \\
\toprule
 & $IDE^{(2)}$        &  0.0069(-0.027, 0.051)             & 0.0075(-0.028, 0.048)              \\
& $IIE^{(2)}$         &  0.0002(-0.0098, 0.0097)             & 0.0001(-0.011, 0.011)              \\
& $TE^{(2)}$        &  0.0071(-0.027, 0.053)             & 0.0076(-0.026, 0.050)              \\
\toprule
\toprule
Age 50-54        & $IDE^{(1)}$                                   &  0.0063(-0.011, 0.034)             & 0.011(-0.015, 0.043)              \\
at baseline        & $IIE^{(1)}$                                  &  0.0001(-0.0054, 0.0059)             & 0.0000(-0.0079, 0.0078)              \\
& $TE^{(1)}$      &  0.0063(-0.012, 0.034)             & 0.011(-0.015, 0.042)              \\
\toprule
& $IDE^{(2)}$      &  0.0073(-0.031, 0.050)             & 0.0077(-0.029, 0.047)              \\
 & $IIE^{(2)}$      &  0.0000(-0.0088, 0.0081)             & 0.0000(-0.010, 0.0097)              \\
& $TE^{(2)}$                         &  0.0072(-0.030, 0.051)             & 0.0077(-0.030, 0.046)              \\
\toprule
\toprule
Age 55-59        & $IDE^{(1)}$                                   &  0.0061(-0.010, 0.031)             & 0.011(-0.014, 0.041)              \\
at baseline        & $IIE^{(1)}$                                  &  0.0001(-0.0059, 0.0064)             & 0.0001(-0.0095, 0.0087)              \\
& $TE^{(1)}$          &  0.0062(-0.010, 0.032)             & 0.011(-0.014, 0.042)              \\
\toprule
    & $IDE^{(2)}$     &  0.0077(-0.026, 0.053)             & 0.0084(-0.025, 0.051)              \\
    & $IIE^{(2)}$     &  0.0000(-0.0095, 0.0098)             & 0.0000(-0.011, 0.010)              \\
& $TE^{(2)}$        &  0.0077(-0.026, 0.051)             & 0.0083(-0.027, 0.051)              \\
\toprule
\toprule
Age 60-64       & $IDE^{(1)}$                                   &  0.0061(-0.011, 0.031)             & 0.011(-0.014, 0.042)              \\
at baseline        & $IIE^{(1)}$                                  &  -0.0002(-0.0071, 0.0064)             & -0.0001(-0.0092, 0.0093)              \\
& $TE^{(1)}$     &  0.0059(-0.011, 0.033)             & 0.011(-0.014, 0.043)              \\
\toprule
& $IDE^{(2)}$        &  0.0081(-0.027, 0.053)             & 0.0089(-0.027, 0.052)              \\
 & $IIE^{(2)}$      &  0.0002(-0.0087, 0.011)             & 0.0002(-0.010, 0.012)              \\
    & $TE^{(2)}$     &  0.0084(-0.028, 0.051)             & 0.0091(-0.027, 0.051)              \\
\toprule
\end{tabular}
\begin{tablenotes}%
\item \textbf{Note:} $IDE^{(1)}$ and $IDE^{(2)}$ represent direct effects on the main event and the competing event, respectively. We similarly define the indirect effects ($IIE^{(1)}$, $IIE^{(2)}$) and the total effects  ($TE^{(1)}$, $TE^{(2)}$) for both events.\vspace*{6pt}
\end{tablenotes}
\end{table}
\end{center}

\begin{center}
\begin{table}[!h]
\caption{Posterior means and 95 \% Bayesian credible intervals of causal effects in the absence of a  competing event. The reported results indicate the differences in survival probabilities between (hypertensive at baseline) women who were on BP medication and those who were not on BP medication in ARIC.}
\begin{tabular}{llccccc}
 \multicolumn{4}{c}{\bfseries \normalsize Hypertensive at baseline population in ARIC (Women only)}  \\
\cline{1-4}
&  &   Visit 2 & Visit 3  \\
\cline{1-4}
\toprule 
Age 45-49         & $IDE$                       &  0.023(-0.062, 0.14)                & 0.035(-0.11, 0.20)   \\
 at baseline        & $IIE$                        &  0.0012(-0.0072, 0.014)                & 0.0031(-0.024, 0.031)   \\
          & $TE$              &  0.024(-0.058, 0.14)                & 0.039(-0.10, 0.21)   \\
\toprule
 Age 50-54        & $IDE$                        &  0.021(-0.052, 0.13)                & 0.032(-0.10, 0.18)   \\
 at baseline        & $IIE$                        &  0.0009(-0.0082, 0.012)                & 0.0026(-0.022, 0.030)   \\
           & $TE$              &  0.022(-0.051, 0.13)                & 0.035(-0.094, 0.19)   \\
\toprule
 Age 55-59        & $IDE$                        &  0.020(-0.053, 0.13)                & 0.031(-0.10, 0.18)   \\
 at baseline        & $IIE$                        &  0.0011(-0.0066, 0.012)                & 0.0033(-0.020, 0.028)   \\
           & $TE$              &  0.021(-0.053, 0.13)                & 0.034(-0.094, 0.19)   \\
\toprule
 Age 60-64        & $IDE$                        &  0.020(-0.050, 0.12)                & 0.029(-0.092, 0.17)   \\
 at baseline        & $IIE$                        &  0.0010(-0.0072, 0.012)                & 0.0037(-0.021, 0.030)   \\
           & $TE$              &  0.021(-0.051, 0.13)                & 0.033(-0.095, 0.18)   \\

\toprule
\end{tabular}
\begin{tablenotes}%
\item \textbf{Note:} $IDE$ = Interventional Direct Effects, $IIE$ = Interventional Indirect Effects, and $TE$ = Total Effects.\vspace*{6pt}
\end{tablenotes}
\end{table}
\end{center}

\begin{center}
\begin{table}[!h]
\caption{Posterior means and 95 \% Bayesian credible intervals of causal effects in the presence of a  competing event. The reported results indicate the differences in potential cumulative incidence functions (CIFs) between (hypertensive at baseline) participants who were on BP medication and those who were not on BP medication in ARIC. )}
\begin{tabular}{llccccc}
 \multicolumn{4}{c}{\bfseries \normalsize Hypertensive at baseline population in ARIC (Women only)}  \\
\cline{1-4}
&  &   Visit 2 & Visit 3  \\
\cline{1-4}
\toprule 
Age 45-49        & $IDE^{(1)}$                                   &  0.011(-0.0086, 0.042)             & 0.016(-0.017, 0.051)              \\
 at baseline        & $IIE^{(1)}$                                   &  0.0000(-0.0058, 0.0056)             & 0.0000(-0.0076, 0.0072)              \\
    & $TE^{(1)}$                         &  0.011(-0.0087, 0.041)             & 0.016(-0.016, 0.053)              \\
\toprule
                    & $IDE^{(2)}$                                  &  0.0053(-0.017, 0.032)             & 0.025(-0.011, 0.072)              \\
                    & $IIE^{(2)}$                                  &  -0.0001(-0.0038, 0.0034)             & -0.0001(-0.0073, 0.0061)              \\
                    & $TE^{(2)}$                         &  0.0052(-0.016, 0.032)             & 0.025(-0.010, 0.071)              \\
\toprule
\toprule
 Age 50-54        & $IDE^{(1)}$                                   &  0.0096(-0.0097, 0.039)             & 0.014(-0.016, 0.049)              \\
 at baseline        & $IIE^{(1)}$                                  &  0.0002( -0.0060, 0.0070)             & 0.0002(-0.0089, 0.0098)              \\
          & $TE^{(1)}$                         &  0.0099(-0.0095, 0.039)             & 0.015(-0.016, 0.048)              \\
\toprule
    & $IDE^{(2)}$                                  &  0.0049(-0.015, 0.031)             & 0.024(-0.010, 0.068)              \\ 
    & $IIE^{(2)}$                                  &  0.0000(-0.0042, 0.0045)             & 0.0000(-0.0095, 0.0087)              \\
    & $TE^{(2)}$                         &  0.0049(-0.015, 0.031)             & 0.024(-0.0099, 0.068)              \\
\toprule
\toprule
 Age 55-59        & $IDE^{(1)}$                                   &  0.0098(-0.0086, 0.039)             & 0.015(-0.016, 0.049)              \\
 at baseline        & $IIE^{(1)}$                                  &  0.0000(-0.0062, 0.0065)             & 0.0000(-0.0085, 0.0085)              \\
           & $TE^{(1)}$                         &  0.0098(-0.0092, 0.040)             & 0.015(-0.017, 0.049)              \\
\toprule
 & $IDE^{(2)}$                                  &  0.0051(-0.015, 0.031)             & 0.025(-0.010, 0.069)              \\ 
& $IIE^{(2)}$    &  0.0000(-0.0045, 0.0047)             & -0.0001(-0.0095, 0.0092)              \\
& $TE^{(2)}$    &  0.0051(-0.015, 0.031)             & 0.025(-0.010, 0.068)              \\
\toprule
\toprule
 Age 60-64       & $IDE^{(1)}$                                   &  0.0093(-0.0086, 0.038)             & 0.014(-0.016, 0.048)              \\
 at baseline        & $IIE^{(1)}$                                  &  0.0001(-0.0066, 0.0072)             & 0.0001(-0.0082, 0.0092)              \\
    & $TE^{(1)}$       &  0.0094(-0.0090, 0.036)             & 0.014(-0.016, 0.050)              \\
\toprule
 & $IDE^{(2)}$         &  0.0052(-0.015, 0.031)             & 0.025(-0.010, 0.071)              \\
& $IIE^{(2)}$           &  0.0000(-0.0047, 0.0042)             & 0.0000(-0.0093, 0.0079)              \\
& $TE^{(2)}$          &  0.0052(-0.016, 0.031)             & 0.025(-0.0090, 0.070)              \\

\toprule
\end{tabular}
\begin{tablenotes}%
\item \textbf{Note:} $IDE^{(1)}$ and $IDE^{(2)}$ represent direct effects on the main event and the competing event, respectively. We similarly define the indirect effects ($IIE^{(1)}$, $IIE^{(2)}$) and the total effects  ($TE^{(1)}$, $TE^{(2)}$) for both events.\vspace*{6pt}
\end{tablenotes}
\end{table}
\end{center}


\end{document}